\documentclass[twocolumn, tighten, twocolappendix]{aastex63}

\newcommand{\logten}{$\log_{10}$}
\newcommand{\hahb}{$\rm H{\alpha} / H{\beta}$}
\newcommand{\ha}{$\rm H{\alpha}$}
\newcommand{\hb}{$\rm H{\beta}$}
\newcommand{\haew}{$\rm {EW}_{H\alpha}$}

\newcommand{\hii}{H{\sc ii}}
\newcommand{\hiiexplorer}{{\tt HII{\scriptsize EXPLORER}}}
\newcommand{\ebv}{$E(B-V)$}
\newcommand{\ebvstar}{$E(B-V)_{\rm star}$}
\newcommand{\ebvgas}{$E(B-V)_{\rm gas}$}
\newcommand{\ebvratio}{$E(B-V)_{\rm star}/E(B-V)_{\rm gas}$}
\newcommand{\ebvdelta}{$E(B-V)_{\rm gas}-E(B-V)_{\rm star}$}
\newcommand{\hasb}{$\Sigma_{\rm H\alpha}$}

\newcommand{\hasbmin}{$\Sigma_{\rm H\alpha, min}$}

\newcommand{\hasbunit}{$\rm erg\;s^{-1}\;kpc^{-2}$}
\newcommand{\f}{$f$}
\newcommand{\ba}{$b/a$}
\newcommand{\oii}{[O{\sc ii}]}

\newcommand{\oiill}{[O{\sc ii}]$\;\lambda\lambda3726,3729$}
\newcommand{\oiii}{[O{\sc iii}]}
\newcommand{\oiiil}{[O{\sc iii}]$\;\lambda4959$}
\newcommand{\oiiir}{[O{\sc iii}]$\;\lambda5007$}
\newcommand{\oiiill}{[O{\sc iii}]$\;\lambda\lambda4959,5007$}
\newcommand{\nii}{[N{\sc ii}]}
\newcommand{\niil}{[N{\sc ii}]$\;\lambda6548$}
\newcommand{\niir}{[N{\sc ii}]$\;\lambda6583$}
\newcommand{\niill}{[N{\sc ii}]$\;\lambda\lambda6548,6583$}
\newcommand{\sii}{[S{\sc ii}]}
\newcommand{\siil}{[S{\sc ii}]$\;\lambda6717$}
\newcommand{\siir}{[S{\sc ii}]$\;\lambda6731$}
\newcommand{\siill}{[S{\sc ii}]$\;\lambda\lambda6717,6731$}
\newcommand{\age}{$t_L$}

\newcommand{\metgas}{$12+\log_{10}(\rm O/H)$}
\newcommand{\sigmamass}{$\Sigma_*$}

\newcommand{\mass}{$M_*$}
\newcommand{\niioii}{[N{\sc ii}]/[O{\sc ii}]}
\newcommand{\oiiioii}{[O{\sc iii}]/[O{\sc ii}]}
\newcommand{\niisii}{[N{\sc ii}]/[S{\sc ii}]}

\defcitealias{2020ApJ...896...38L}{Paper I}
\received{}
\revised{}
\accepted{}
\submitjournal{ApJ}

\shorttitle{Stellar and ionized gas dust attenuation}
\shortauthors{Li et al.}
\begin{document}

\title{Estimating dust attenuation from galactic 
spectra. II. Stellar and gas attenuation in star-forming 
and diffuse ionized gas regions in MaNGA}

\correspondingauthor{Niu Li \& Cheng Li}
%\email{cli2015@tsinghua.edu.cn}

\author[0000-0002-0656-075X]{Niu Li}
\affiliation{Department of Astronomy, Tsinghua University, Beijing 100084, China}
\email{liniu@tsinghua.edu.cn}

\author[0000-0002-8711-8970]{Cheng Li}
\affiliation{Department of Astronomy, Tsinghua University, Beijing 100084, China}
\email{cli2015@tsinghua.edu.cn}

\author{Houjun Mo}
\affiliation{Department of Astronomy, University of Massachusetts Amherst, MA 01003, USA}

\author{Shuang Zhou}
\affiliation{Department of Astronomy, Tsinghua University, Beijing 100084, China}

\author[0000-0003-2496-1247]{Fu-heng Liang}
\affiliation{Department of Astronomy, Tsinghua University, Beijing 100084, China}

\author{M{\'e}d{\'e}ric Boquien}
\affiliation{Centro de Astronom{\'i}a, Universidad de Antofagasta, 
Avenida Angamos 601, Antofagasta 1270300, Chile}

\author{Niv Drory}
\affiliation{McDonald Observatory, University of Texas at Austin,
1 University Station, Austin, TX 78712, USA}

\author{Jos{\'e} G. Fern{\'a}ndez-Trincado}
\affiliation{Instituto de Astronom{\'i}a y Ciencias Planetarias, 
Universidad de Atacama, Copayapu 485, Copiap{\'o}, Chile}

\author{Michael Greener}
\affiliation{School of Physics and Astronomy, 
University of Nottingham, University Park, Nottingham, NG7 2RD, UK}

\author[0000-0002-1321-1320]{Rog{\'e}rio Riffel}
\affiliation{Instituto de F{\'i}sica, Universidade Federal do Rio Grande do Sul,
Campus do Vale, Porto Alegre, RS, Brasil, 91501-970}
\affiliation{Laborat\'orio Interinstitucional de e-Astronomia - LIneA, 
Rua Gal. Jos\'e Cristino 77, Rio de Janeiro, RJ - 20921-400, Brazil}



\begin{abstract}
We investigate the dust attenuation in both stellar populations 
and ionized gas in kpc-scale regions in nearby galaxies, using 
integral field spectroscopy data from MaNGA MPL-9. We identify 
star-forming (\hii) and diffuse ionized gas (DIG) regions from 
MaNGA datacubes. From the stacked spectrum of each region, 
we measure the stellar attenuation, \ebvstar, using 
the technique developed by \citet{2020ApJ...896...38L}, 
as well as the gas attenuation, \ebvgas, from the 
Balmer decrement. We then examine the correlation 
of \ebvstar, \ebvgas, \ebvdelta\ and \ebvratio\ with  
16 regional/global properties, and for regions with different 
\ha\ surface brightnesses (\hasb). We find a stronger correlation 
between \ebvstar\ and \ebvgas\ in regions of higher \hasb. 
Luminosity-weighted age (\age) is found to be the property that 
is the most strongly correlated with \ebvstar, and consequently 
with \ebvdelta\ and \ebvratio. At fixed \hasb, \logten\age\ is linearly 
and negatively correlated with \ebvratio\ at all ages. 
Gas-phase metallicity and ionization level are important for 
the attenuation in the gas. Our results indicate that 
the ionizing source for DIG regions is likely distributed in 
the outer-skirt of galaxies, while for \hii\ regions our results 
can be well explained by the two-component dust model of 
\citet{2000ApJ...539..718C}.
\end{abstract}



\section{Introduction}
\label{sec:intro}

Dust accounts for $\lesssim 1\%$ of the interstellar medium (ISM) mass
in a typical galaxy, but has an 
important influence on the spectral energy distribution (SED) of 
the galaxy through absorbing and scattering the starlight, 
an effect known as dust attenuation or dust extinction
\citep[see reviews by][]{2018ARA&A..56..673G, 2020ARA&A..58..529S}.
Dust is produced from the ejecta of asymptotic giant branch (AGB) stars and 
supernovae \citep[e.g.,][]{1998ApJ...501..643D,2013MNRAS.434.2390N,
2014MNRAS.442.1440S,2017MNRAS.471.3152P,2017MNRAS.466..105A},
and grows in the ISM by accreting gas-phase metals \citep[e.g.,][]{
1997ApJ...480..647D,1998ApJ...501..643D,2011MNRAS.416.1340H,
2014A&A...562A..76Z}. Dust can be destroyed by supernova shocks, thermal
sputtering and grain collisions, or be incorporated into new-born stars 
\citep[e.g.,][]{1998ApJ...501..643D,2005MNRAS.358..379B,2007ApJ...666..955N}.
As dust attenuation can cause changes in the overall shape of 
galactic spectra and SEDs,
one has to consider and correct the effect of attenuation 
in order to reliably measure the intrinsic
properties of galaxies from their observed spectra and SEDs. 

A variety of methods have been used to estimate the stellar continuum dust 
attenuation, \ebvstar. Dust extinction can be probed by observing  
individual stars along different lines of sight in the Milky Way or very nearby 
galaxies \citep[e.g.,][]{1984A&A...132..389P,1989ApJ...345..245C,
1999PASP..111...63F,2003ApJ...594..279G}. Shorter wavelength photons are more 
susceptible to dust attenuation, and dust can re-emit photons in the infrared.
Thus, the $L_{\rm IR}/L_{\rm UV}$ ratio \citep[known as IRX; e.g.,][]{1999ApJ...521...64M,
2000ApJ...533..236G} is often used to estimate the dust attenuation.
A galactic spectrum contains a variety of information about the physical
properties of the galaxy. A simple approach to estimate dust attenuation
through a galactic spectrum is to match the attenuated spectrum with those 
of unattenuated galaxies that have similar stellar populations \citep[e.g.,][]{
2000ApJ...533..682C,2011MNRAS.417.1760W,2015ApJ...806..259R,
2017ApJ...840..109B,2017ApJ...851...90B}.
Alternatively, fitting the SED (or full spectrum) with a stellar population
synthesis model is also widely adopted to obtain the stellar attenuation 
\citep[e.g.,][]{2005MNRAS.358..363C,2007MNRAS.381..263A,2009MNRAS.400..273R,
2013ARA&A..51..393C,2015MNRAS.449..328W,2018MNRAS.478.2633G,
2019A&A...622A.103B,2020ApJ...896...38L,2021MNRAS.501.4064R}.
In ionized gas regions, the dust attenuation on emission lines, \ebvgas,
is commonly estimated from the Balmer decrement \hahb. The intrinsic line 
ratio can be calculated by atomic physics applied to a given environment
\citep{2006agna.book.....O}.

However, previous studies found that \ebvgas\ is not consistent with \ebvstar.
\cite{1988ApJ...334..665F} found that \ebvgas\ is significantly higher than 
\ebvstar. \cite{1994ApJ...429..582C,2000ApJ...533..682C} further confirmed 
the result and found that the typical value of the ratio $f=\;$\ebvratio\ 
is about 0.44. Moreover, studies based on both the local and high redshift galaxies 
show that \ebvgas\ tends to be larger than \ebvstar\ \citep[e.g.,][]{
2011ApJ...738..106W,2013ApJ...779..135W,2013ApJ...771...62K,
2014ApJ...788...86P,2015ApJ...807..141P,2017ApJ...847...18Z,
2018A&A...619A.135B,2019PASJ...71....8K},
which is also seen in the near infrared \citep[e.g.][]{2008MNRAS.388..803R}.
The value of \f\ found in the literature varies over a wide range, 
from 0.44 to $\sim 1$
\citep[e.g.,][]{1994ApJ...429..582C,2010ApJ...712.1070R,2011ApJ...738..106W,
2013ApJ...777L...8K,2014ApJ...788...86P,2015ApJ...807..141P,
2015ApJ...801..132V,2016A&A...586A..83P}, and 
the variation is found to be correlated with physical properties
of galaxies, such as total stellar mass \citep[e.g.,][]{2017ApJ...847...18Z,
2019PASJ...71....8K}, specific star formation rate \citep[sSFR; e.g.,][]{
2011MNRAS.417.1760W,2014ApJ...788...86P,2019PASJ...71....8K,
2019ApJ...886...28Q}, and axial ratio \citep[\ba;][]{2011MNRAS.417.1760W}.

In general, the discrepancy between the two attenuations may be explained by 
a two-component dust model \citep[e.g.,][]{2000ApJ...539..718C,
2011MNRAS.417.1760W,2013MNRAS.432.2061C}, which includes an optically thin, diffuse 
component distributed throughout the ISM and an optically thick, 
dense component (the birth clouds) where young stars are born. The typical 
lifetime of a birth cloud is about $10^7$yr \citep{1980ApJ...238..148B,
2000ApJ...539..718C}. In this model, the emission lines are produced in the 
\hii\ regions of the birth clouds and stars younger than $3\times 10^6$yr
produce most of the ionizing photons. Thus the emission lines and the 
continuum radiation of young stars in birth clouds are attenuated by both the 
dust in the ambient ISM and the dust in the birth clouds, while the continuum
radiation of older stars are attenuated only by the diffuse dust in the ambient
ISM \citep{2000ApJ...539..718C}. Consequently, emission lines suffer larger 
attenuation than the stellar continuum if the observed region is dominated by 
older stars. Only in idealized cases where one can resolve individual birth 
clouds or regions dominated by young stars in birth clouds, the two 
attenuations are expected to be roughly the same \citep[e.g.,][]{2007ApJS..173..457B}.

However, previous studies of dust attenuation have been mostly based on
global properties of galaxies. As dust attenuations depend on the 
geometrical distribution of the dust relative to stars 
\citep[e.g.,][]{2000ApJ...539..718C,
2011MNRAS.417.1760W}, the relation between \ebvstar\ and \ebvgas\ may be driven
by local properties within galaxies.
With the advent of new integral field units (IFU) facilities,  
such as the Calar Alto Legacy Integral Field Area survey 
\citep[CALIFA;][]{2012A&A...538A...8S}, the Sydney Australian Astronomical 
Observatory Multi-object Integral Field Spectrograph survey 
\citep[SAMI;][]{2012MNRAS.421..872C}, the Multi Unit Spectroscopic Explorer
Wide survey \citep[MUSE-Wide;][]{2019A&A...624A.141U}, and the Mapping Nearby 
Galaxies at Apache Point Observatory survey \citep[MaNGA;][]{
2015ApJ...798....7B}, spatially resolved spectroscopy can be used to 
provide large samples to study properties of regions within individual 
galaxies. In addition, most of the previous studies focused on 
galaxies or regions that have high star formation rate (SFR), 
ignoring the diffuse ionized gas \citep[DIG;][]{2009RvMP...81..969H} regions. 
As the mechanisms of producing emission lines are quite different 
between DIG and \hii\ regions
\citep[e.g.,][]{2017MNRAS.466.3217Z}, the relation between \ebvstar\ and 
\ebvgas\ may also be different between the two.
Spatially resolved spectroscopy provides a way to divide a galaxy 
into DIG-dominated and \hii-dominated regions, whereby making it 
possible to investigate dust attenuation in different types of regions
\citep[e.g.,][]{2017MNRAS.466.3217Z, 2017ApJ...844..155T, 2018MNRAS.474.3727L}.

Recently, using data of MaNGA galaxies in Sloan Digital Sky Survey Data 
Release 15 \citep[SDSS DR15;][]{2019ApJS..240...23A} and MaNGA value-added 
catalog (VAC) of the \texttt{Pipe3D} pipeline \citep{2018RMxAA..54..217S},
\cite{2020ApJ...888...88L} investigated the variations of 
\f\ from sub-galactic to galactic scales. They found that \ebvstar\ 
and \ebvgas\ have a stronger correlation for more active \hii\ regions, 
while \f\ is found to have a moderate correlation with tracers of DIG regions. 
Their results suggest that local physical conditions, such as metallicity and 
ionization level, play an important role in determining  \f, in that 
metal-poor regions with higher ionized level have larger \f. 
Using a sample of 232 star-forming spiral galaxies 
from MaNGA and the full spectral fitting code \texttt{STARLIGHT}
\citep{2005MNRAS.358..363C}, \cite{2020MNRAS.495.2305G} also found that 
the variation in dust attenuation properties is likely driven by 
local physical properties of galaxies, such as the SFR surface density. 
\cite{2021MNRAS.501.4064R} analyzed a sample of 170 active galactic nuclei (AGN) hosts with 
strong star formation and a control sample of 291 star-forming 
galaxies from MaNGA. They also found a strong correlation between \ebvstar\ 
(derived with \texttt{STARLIGHT}) and \ebvgas, with $f\sim0.37$,  
which is close to, but slightly smaller than the typical value of $f=0.44$ 
found for the general population of star-forming regions.

In this paper, we use the latest internal data release of MaNGA, the MaNGA 
Product Launch 9 (MPL-9) of about 8,000 unique galaxies 
(4,621 in SDSS DR15), to explore the correlation between \ebvstar\ and \ebvgas\ 
in DIG and \hii\ regions, to study how \f\ correlates with physical properties of 
galaxies and whether or not the two-component model works on spatially resolved regions.
Our analysis improves upon earlier ones by using a much larger sample
and by adopting a new method to estimate the dust attenuation
from full optical spectral fitting \citep[][hereafter Paper I]{2020ApJ...896...38L}.
In addition, following the Wolf-Rayet searching procedure of \cite{2020ApJ...896..121L},
we perform spatial binning for each data cube 
according to the \hasb\ map 
instead of the continuum signal-to-noise ratio (S/N) map. 
The ionized gas regions identified in this way are expected to 
be more uniform and smoother in terms of physical properties, 
compared to those binned by S/N. 

This paper is organized as follows. Section \ref{sec:data} describes the 
observational data, quantities measured from the data, and samples for our analysis.
In Section \ref{sec:results}, we present our main 
results. We discuss and compare our results with those obtained previously in 
Section \ref{sec:dis}. Finally, we summarize 
our results in Section \ref{sec:summary}.
Throughout this paper we assume a $\Lambda$ cold dark matter cosmology model 
with $\Omega_{\rm m}=0.3$, $\Omega_\Lambda=0.7$, and $H_0=70 \;{\rm km\;s^{-1}\;Mpc^{-1}}$, 
and a \cite{2003PASP..115..763C} initial mass function (IMF).

\section{Data}
\label{sec:data}

\subsection{Overview of MaNGA}
\label{sec:MaNGA}

The MaNGA survey is one of the three core programs of the 
fourth-generation Sloan Digital Sky Survey project 
\citep[SDSS-IV;][]{2017AJ....154...28B}. During the past six 
years from July 2014 through August 2020, MaNGA has obtained 
integral field spectroscopy (IFS) for 10,010 nearby galaxies
\citep{2015ApJ...798....7B}.  A total of 29 integral field 
units (IFUs) with various sizes are used to obtain the IFS
data, including 17 science fiber bundles with five different 
field of views (FoVs) ranging from 12$\mbox{\arcsec}$ 
(19 fibers) to 32$\mbox{\arcsec}$ (127 fibers), and 12 
seven-fiber mini-bundles for flux calibration. 
Obtained with the two dual-channel 
BOSS spectrographs at the 2.5-meter Sloan telescope 
\citep{Gunn-06, Smee-13} and a typical exposure time 
of three hours, the MaNGA spectra cover 
a wavelength range from 3622 \AA\ to 10354 \AA\ with 
a spectral resolution of $R\sim 2000$, and reach a target 
$r$-band signal-to-noise (S/N) of 
$4-8$ (\AA$\rm ^{-1}\;per\;2\mbox{\arcsec}\;fiber$)
at $1-2$ effective radius ($R_e$) of galaxies. A detailed description of 
the MaNGA instrumentation can be found in \citet{2015AJ....149...77D}.

Targets of the MaNGA survey are selected from the NASA Sloan Atlas
(NSA)\footnote{\url{http://www.nsatlas.org}},
a catalog of $\sim640,000$ low-redshift galaxies constructed 
by \citet{Blanton-11} based on the SDSS, GALEX and 2MASS. 
As detailed in \citet{2017AJ....154...86W}, the MaNGA sample 
selection is designed so as to simultaneously optimize 
the IFU size distribution, the IFU allocation strategy and 
the number density of targets. The Primary 
and Secondary samples, which are the main samples of the 
survey, are selected to effectively have a flat distribution 
of the $K$-corrected $i$-band absolute magnitude ($M_i$),
covering out to 1.5 and 2.5$R_e$, respectively. In addition, 
the Color-Enhanced sample selects galaxies that are not 
well sampled by the main samples in the  
$NUV-r$ versus $M_i$ diagram. Overall the MaNGA targets 
cover a wide range of stellar mass, 
$10^9M_\odot\lesssim M_\ast\lesssim 6\times10^{11}M_\odot$, 
and a redshift range, $0.01<z<0.15$, with a median redshift 
of $z\sim0.03$. 

MaNGA raw data are reduced using the Data Reduction Pipeline 
\citep[DRP;][]{2016AJ....152...83L}, and provide a datacube 
for each galaxy with a spaxel size of $0.5\mbox{\arcsec}\times 0.5\mbox{\arcsec}$ 
with an effective spatial resolution that can be described by 
a Gaussian with a full width at half maximum 
(FWHM) of $\sim 2.5$\arcsec. For more than 80\% of the wavelength 
range, the absolute flux calibration of the MaNGA spectra 
is better than 5\%. Details about the flux calibration, survey 
strategy and data quality tests are provided in 
\citet{2016AJ....151....8Y, 2016AJ....152..197Y}.
In addition, MaNGA also provides products of the Data Analysis Pipeline 
\citep[DAP;][]{2019AJ....158..231W} developed by the MaNGA 
collaboration, which performs full spectral fitting to the {\tt DRP} datacubes 
and obtains measurements of stellar kinematics, emission lines 
and spectral indices. The latest data release from the MaNGA 
was made in SDSS DR15 \citep[][]{2019ApJS..240...23A},
including {\tt DRP} and {\tt DAP} products of 4824 data cubes 
for 4621 unique galaxies.
In this paper, we make use of the latest internal sample, the 
MPL-9, which contains 8113 datacubes 
for 8000 unique galaxies.

\subsection{Identifying ionized gas regions}
\label{sec:gas_regions}

Our study aims to examine the spatially resolved dust attenuation 
in both starlight and ionized gas. To this end, we first identify 
ionized gas regions in the MaNGA galaxies, and then measure 
the stellar and gas attenuation as well as other properties for 
each region.  We make use of the \ha\ flux maps obtained by 
MaNGA {\tt DAP} for the identification of ionized gas regions. Given the 
\ha\ flux map of a galaxy, we calculate the surface density of 
\ha\ emission \hasb\ in units of \hasbunit\ for each spaxel 
in the map, and select all the spaxels with \hasb\ exceeding a 
threshold of \hasbmin\ for the identification. We have corrected 
the effect of dust attenuation on the \ha\ fluxes using the observed 
Balmer decrement, i.e. the \ha-to-H$\beta$ flux ratio 
(\hahb)$_{\rm obs}$ with the H$\beta$ flux also from the {\tt DAP}, 
assuming case-B recombination with an intrinsic 
Balmer decrement (\hahb)$_{\rm int}=2.86$ \citep{2006agna.book.....O}. 

We apply the public pipeline \hiiexplorer
\footnote{\url{http://www.astroscu.unam.mx/~sfsanchez/HII_explorer/index.html}}
developed by \citet{2012A&A...546A...2S} to the \hasb\ map to 
identify ionized gas regions.
The \hiiexplorer\ begins by picking up the {\em peak} spaxel 
in the map, i.e. the one with the highest \hasb, as the center of 
an ionized gas region. The spaxels in the vicinity are then appended 
to the region if their distance from the center is less than $r_p^{\rm max}$ 
and if their \hasb\ is higher than 10\% of the central \hasb. 
The latter requirement aims to make the region roughly coherent. 
The region is then removed from the map, and the pipeline 
moves on to the {\em peak} spaxel of the remaining map. 
This process is repeated until every spaxel with \hasb$\;>\;$\hasbmin\
is assigned to a region. 

The \hiiexplorer\ was initially designed for identifying \hii\ regions, 
and so the threshold density \hasbmin\ is usually
set to be a relatively high value, e.g. \hasbmin$\;=10^{39.5}$\hasbunit\ 
in a recent MaNGA-based study by \cite{2020ApJ...896..121L}.
Here we adopt a much lower threshold, 
\hasbmin$\;=10^{37.5}$\hasbunit, in order to extend to DIG regions.
Following \citet{2020ApJ...896..121L} we set 
$r_p^{\rm max}=1.5^{\prime\prime}$, which is comparable 
to but slightly larger than half of the MaNGA spatial resolution
($2.5^{\prime\prime}$, corresponding to $\sim1.5$ kpc at 
the MaNGA median redshift $z=0.03$). Due to the limited 
resolution one cannot resolve individual \hii\ regions whose 
sizes typically range from a few to hundreds of parsecs
\citep[e.g.,][]{1984ApJ...287..116K,2001ApJ...549..979K,2009A&A...507.1327H,
2011ApJ...731...91L,2019ApJ...871..145A}. Therefore, 
the ionized gas regions identified from MaNGA may contain 
a few to hundreds of individual \hii\ regions, or be a mixture of DIG 
and \hii\ regions. According to \cite{2017MNRAS.466.3217Z}, 
\hasb\ can be used to effectively separate 
DIG-dominated regions from \hii-dominated 
regions in the MaNGA galaxies, with an empirical dividing 
value of $\sim 10^{39}$\hasbunit. In what follows, 
we refer \hii-dominated regions as \hii\ regions and 
DIG-dominated regions as DIG regions, for simplicity. 

The above process results in a total of $\sim 5\times 10^5$ 
ionized gas regions down to \hasbmin$\;=10^{37.5}$\hasbunit, 
out of 8000 galaxies from the MPL-9.

\subsection{Measuring stellar attenuation}
\label{sec:measure_dust}

In our sample each ionized gas region contains a few to tens of spaxels 
that have similar \hasb\ by definition. We stack the spectra within each 
region to obtain an average spectrum with high S/N, from which we 
then measure the stellar and gas attenuation, as well as other properties
necessary for this work. We adopt the ``weighted mean''  estimator 
for the spectral stacking, which is described in detail in 
\citet{2020ApJ...896..121L}. Briefly, the spectra in a given region are 
corrected to the rest-frame considering both the 
redshift of the host galaxy and the relative velocities of each spaxel,
using the stellar velocity map from {\tt DAP}, and the fluxes at 
each wavelength point are weighted by their spectral error provided 
by the MaNGA {\tt DRP}. Error of the stacked spectrum is derived so 
as to correct for the effect of covariance, by following the formula given 
in figure 16 of \citet[][]{2016AJ....152...83L}. We find that
the stacking process effectively reduces the noise in the original spectra:
the S/N of the stacked spectra is higher than that of the original spectra 
by $\sim 20\%$ on average, ranging from about 15\% at the highest S/N
up to 50\% at the lowest S/N.
On average, one region contains $\sim 10$ spaxels and the
typical final S/N is $\sim 12$.

For each region we then estimate the relative attenuation 
curve  ($A_\lambda - A_V$) and the stellar color excess \ebvstar, 
by applying the method developed in \citetalias{2020ApJ...896...38L} to the stacked spectrum. 
In this method, the small-scale features ($S$) in the observed spectrum 
is firstly separated from the large-scale spectral shape ($L$) using a 
moving average filter.  The same separation is also performed 
for the spectrum of model templates, and the intrinsic dust-free 
model spectrum of the stellar component is then derived by fitting 
the observed ratio of the small- to large-scale spectral components 
($S/L$)$_{\rm obs}$ with the same ratio of the model spectra ($S/L$)$_{\rm model}$.
As shown in \citetalias{2020ApJ...896...38L} (see \S2.1 and Eqns. 3-8 in that 
paper), the small- and large-scale components are attenuated by 
dust in the same way so that their ratio $S/L$ is dust free, 
as long as the dust attenuation curves are similar for different 
stellar populations in a galactic region.
Finally, $A_\lambda - A_V$ is derived by comparing the observed spectrum 
with the best-fit model spectrum. The value of \ebvstar\ ($=A_B-A_V$) 
is then directly calculated from the $A_\lambda - A_V$. 

As shown in \citetalias{2020ApJ...896...38L},
one important advantage of this method is that the relative dust 
attenuation curve can be directly obtained without the need to assume 
a functional form for the curve. Furthermore, extensive tests on mock 
spectra have shown that the method is able to recover the input dust 
attenuation curve accurately from an observed spectrum with 
S/N$\;>5$, as long as the underlying stellar populations have similar 
dust attenuation or the optical depths are smaller than unity. 
This method should be well valid in the current study, where the 
ionized gas regions are selected to each span a limited range of 
\hasb, thus expected to have limited variations in the underlying 
stellar population properties.

\subsection{Measuring stellar populations}
\label{sec:spectral_fitting}

Using the measurements of relative dust attenuation curves obtained above, 
we correct the effect of dust attenuation for the stacked spectra of
the ionized gas regions in our sample. We then perform full spectral fitting 
to the dust-free spectrum, and measure both stellar 
population parameters from the best-fit stellar component and emission line 
parameters from the starlight-subtracted component.  We use the 
simple stellar populations (SSPs) given by 
\citet[][BC03]{2003MNRAS.344.1000B} to fit our spectra. 
BC03 provides the spectra for a set of 1,326 SSPs at a spectral resolution 
of 3\AA, covering 221 ages from $t=0$Gyr 
to $t=20$Gyr and  six metallicities from $Z=0.005Z_\odot$ 
to $Z=2.5Z_\odot$, where the solar metallicity $Z_\odot=0.02$.
The SSPs are computed using the Padova evolutionary track 
\citep{1994A&AS..106..275B} and the \citet{2003PASP..115..763C} IMF. 
We select a subset of 150 SSPs that cover 25 ages at each of the six 
metallicities, ranging from $t=0.001$Gyr to $t=13$Gyr with approximately 
equal intervals in $\log_{10}t$. For a given region, we fit its spectrum 
$F(\lambda)$ with a linear combination of the SSPs:
\begin{equation}\label{eqn:fit}
    F(\lambda) = \sum_{j=1}^{j=N_\ast} x_j f^j_{SSP}(\lambda),
\end{equation}
where $N_\ast=150$, $f^j_{SSP}(\lambda)$ is the spectrum of the $j$-th 
SSP, and $x_j$ is the coefficient of the $j$-th SSP to be determined 
by the fitting. 
The effect of stellar velocity dispersion 
is taken into account by convolving the SSPs in 
\autoref{eqn:fit} with a Gaussian.
We have carefully masked out all the detected 
emission lines in the spectra following the scheme described 
in \citet{2005AJ....129..669L}. 

We then obtain the following parameters to quantify the stellar 
populations in each region, based on the coefficients $x_j$ of the 
best-fit stellar spectrum or the best-fit spectrum itself. 
\begin{itemize}
    \item $\log_{10}\Sigma_\ast$ --- logarithm of the surface density of stellar mass given by
          $\Sigma_\ast=(\sum^{N_*}_{j=1}x_jM_j)/A$ in units of $M_\odot\,{\rm kpc}^{-2}$, 
          where $M_j$ is the stellar mass of the $j$-th SSP and $A$ is the area of the region.
    \item $\log_{10}t_L$ --- logarithm of the light-weighted stellar age, given by
          $\log_{10} t_L=\sum^{N_*}_{j=1} x_j\log_{10}t_j$, where $t_j$ is the age of the $j$-th SSP
          in units of yr. 
    \item $Z_L$ --- light-weighted stellar metallicity, given by
          $Z_L=\sum^{N_*}_{j=1} x_j Z_j$, where $Z_j$ is the metallicity of the $j$-th SSP
          in units of solar metallicity. 
    \item $\log_{10}t_M$ --- logarithm of the mass-weighted stellar age, given by
          $\log_{10} t_M=(\sum^{N_*}_{j=1} x_jM_j\log_{10}t_j)/(\sum^{N_*}_{j=1}x_jM_j)$.
    \item $Z_M$ --- mass-weighted stellar metallicity, given by 
          $Z_M=(\sum^{N_*}_{j=1} x_j M_j Z_j) / (\sum^{N_*}_{j=1}x_jM_j)$.
    \item $D_n4000$ --- the narrow-band version of the 4000\AA\ break, defined by 
    \citet{1999ApJ...527...54B}. We measure this parameter from the best-fit stellar spectrum 
    rather than the observed spectrum in order to minimize the effect of noise. 
\end{itemize}

Since the spectra are corrected for dust attenuation before they are used 
for the spectral fitting, the known degeneracy between dust attenuation and 
stellar populations should be largely alleviated. This provides more 
reliable estimates of the stellar age and metallicity,  
as shown in \citetalias{2020ApJ...896...38L} (Appendix B), where extensive tests are made 
on mock spectra that cover wide ranges in age and metallicity and include 
realistic star formation histories and emission lines. In these tests, Bayesian 
inferences of the stellar age and metallicity are derived by applying the 
spectral fitting code {\tt BIGS} \citep{2019MNRAS.485.5256Z} to 
the mock spectra with the same set of BC03 SSPs as used here, 
with or without correcting the effect of dust attenuation before the fitting. 
It is found that the inference uncertainties for both stellar parameters 
are significantly reduced  if the dust attenuation is corrected before the fitting
(see their Figure B1 and B2).

\subsection{Measuring emission lines and gas attenuations}
\label{sec:measure_lines}

We measure the flux, surface density and equivalent 
width (EW) for a number of emission lines for each region, 
including \oiill, \hb, \oiiill, \niill, \ha, and \siill. 
We subtract the best-fit stellar spectrum from the stacked spectrum, 
and fit the emission lines with Gaussian profiles. We use 
single Gaussian to fit \hb, \oiiil, \oiiir, \niil, \ha, \niir, \siil\ and \siir, 
and double Gaussian to fit \oiill. The line ratios of 
\niir/\niil\ and \oiiir/\oiiil\ are fixed to 3 during the fitting. 
The parameters of each line are then calculated from the 
Gaussian profiles, together with the best-fit stellar spectrum when needed. 
We have corrected the fluxes for the effect of attenuation using 
the observed \ha-to-\hb\ flux ratio as described in 
Section \ref{sec:gas_regions}. For this we have assumed the 
case-B recombination. An estimate of the dust attenuation 
in the gas, as quantified by \ebvgas, is obtained for each region:
\begin{equation}
    E(B-V)_{\rm gas} = \frac{2.5}{k(\lambda_{\rm H\beta}) - k(\lambda_{\rm H\alpha})}
    \log_{10}\left[\frac{(\rm H\alpha/\rm H\beta)
    _{\rm obs}}{2.86}\right],
    \label{eqn:def_ebvbd}
\end{equation}
where $k(\lambda_{\rm H\beta})$ and $k(\lambda_{\rm H\alpha})$ are the attenuation curves
evaluated at the wavelength of \ha\ and \hb. For each region, we use its attenuation curve
measured from our method, as described in Section \ref{sec:measure_dust}.

Based on the emission line measurements, we have estimated the following
parameters to quantify gas-related properties.
\begin{itemize}
    \item $\log_{10}$\hasb\ --- logarithm of the surface density of the \ha\ emission 
    in units of \hasbunit.
    \item $\log_{10}$\haew\ --- logarithm of the equivalent width 
    of the \ha\ emission line in units of \AA.
    \item sSFR --- specific star formation rate (SFR) defined by the ratio of star 
    formation rate to stellar mass in a region. The SFR is computed from the
    dust-corrected \ha\ luminosity following \cite{1998ARA&A..36..189K}
    with a \cite{2003PASP..115..763C} IMF:
    ${\rm SFR}(M_\odot \; {\rm yr}^{-1})=4.6\times 10^{-42}
    L_{\mbox{\ha}}({\rm erg\;s}^{-1})$. The stellar mass is obtained above 
    from the spectral fitting (Section \ref{sec:spectral_fitting}).
    \item 12+\logten(O/H) --- gas-phase metallicity. 
    We adopt the O3N2 estimator in which the metallicity is empirically estimated 
    from the parameter O3N2$\;\equiv\;$(\oiiir/\hb)/(\niir/\ha)
    \citep[e.g.,][]{1979A&A....78..200A,1994ApJ...420...87Z,
    2001A&A...369..594P,2002MNRAS.330...69D,2004MNRAS.348L..59P,2005ApJ...631..231P}.
    \item $\log_{10}$(\niioii) --- logarithm of the flux ratio between \niir\ and \oiill.
    \item $\log_{10}$(\oiiioii)--- logarithm of the flux ratio between \oiiir\ and \oiill.
    \item $\log_{10}$(\niisii) --- logarithm of the flux ratio between \niir\ and \siil.
\end{itemize}

\begin{figure*}
    \centering
    \fig{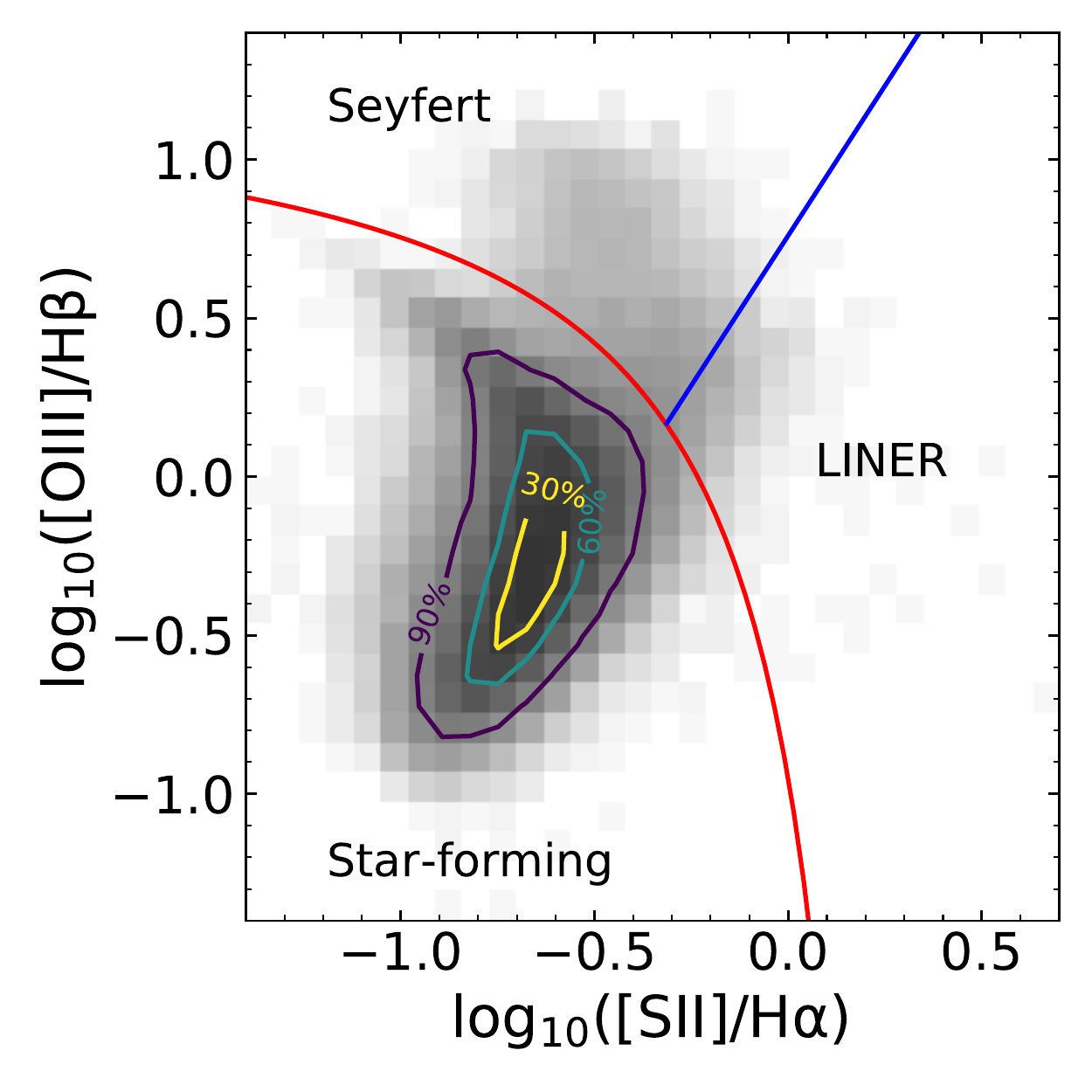}{0.24\textwidth}{(a1)}
    \fig{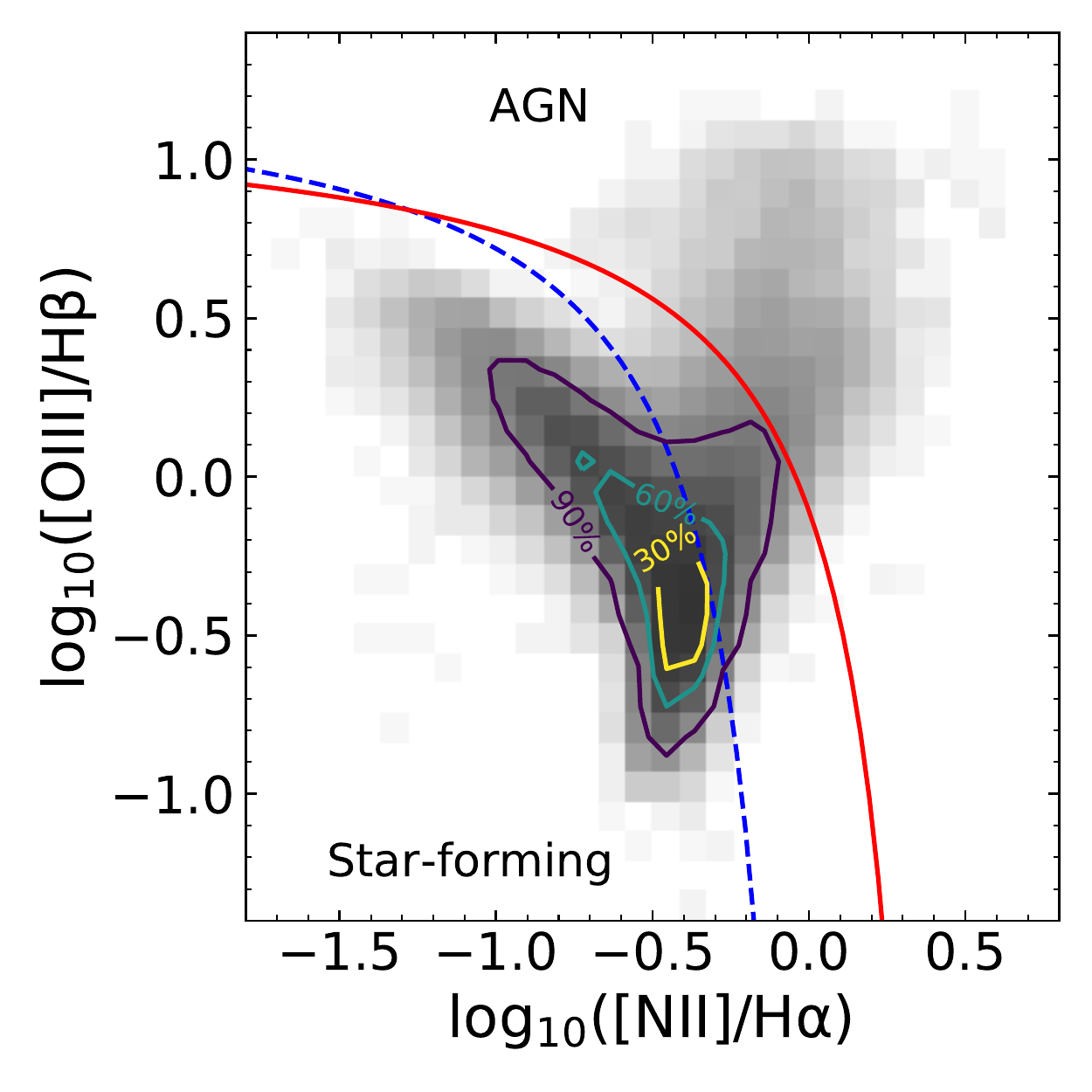}{0.24\textwidth}{(a2)}
    \fig{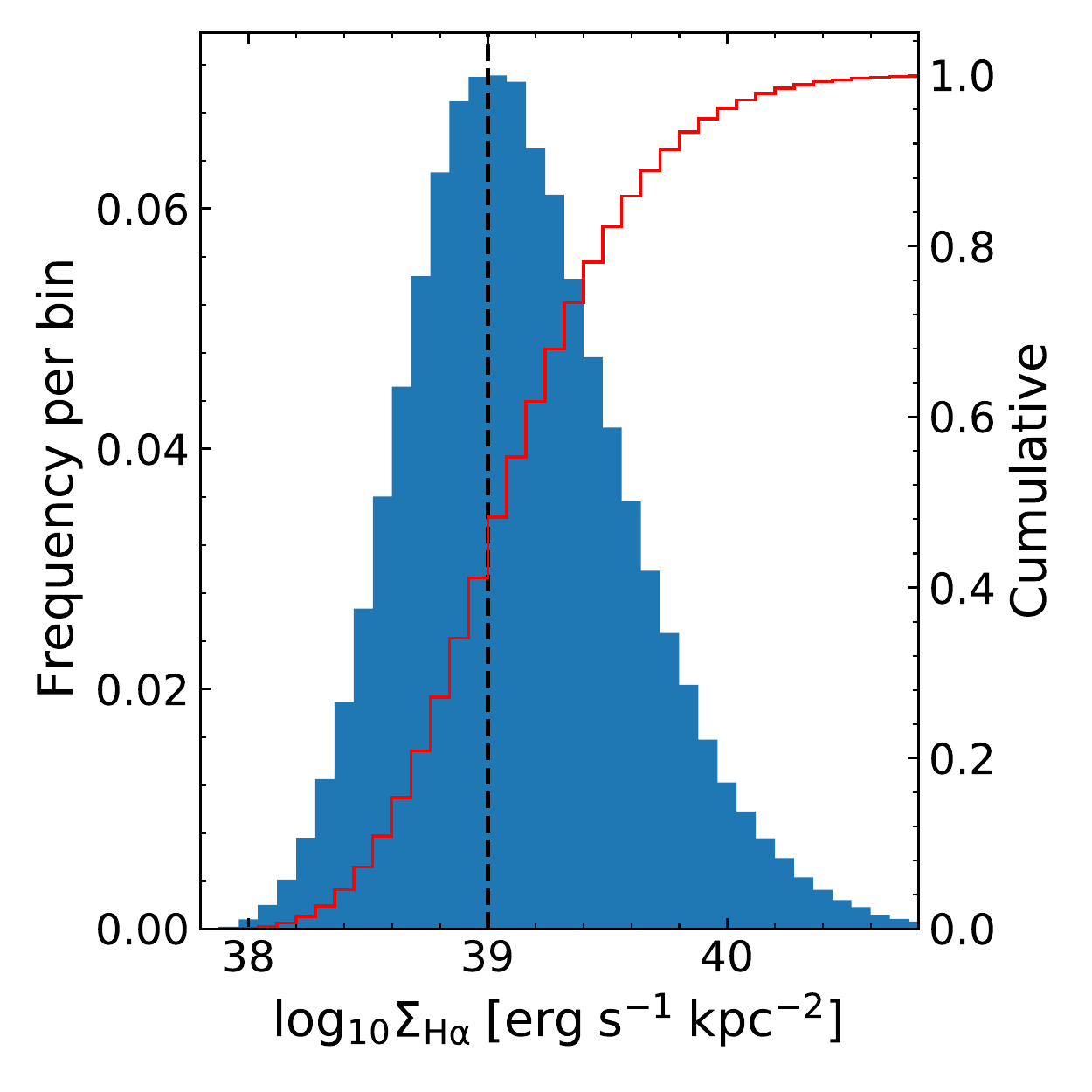}{0.24\textwidth}{(b)}
    \fig{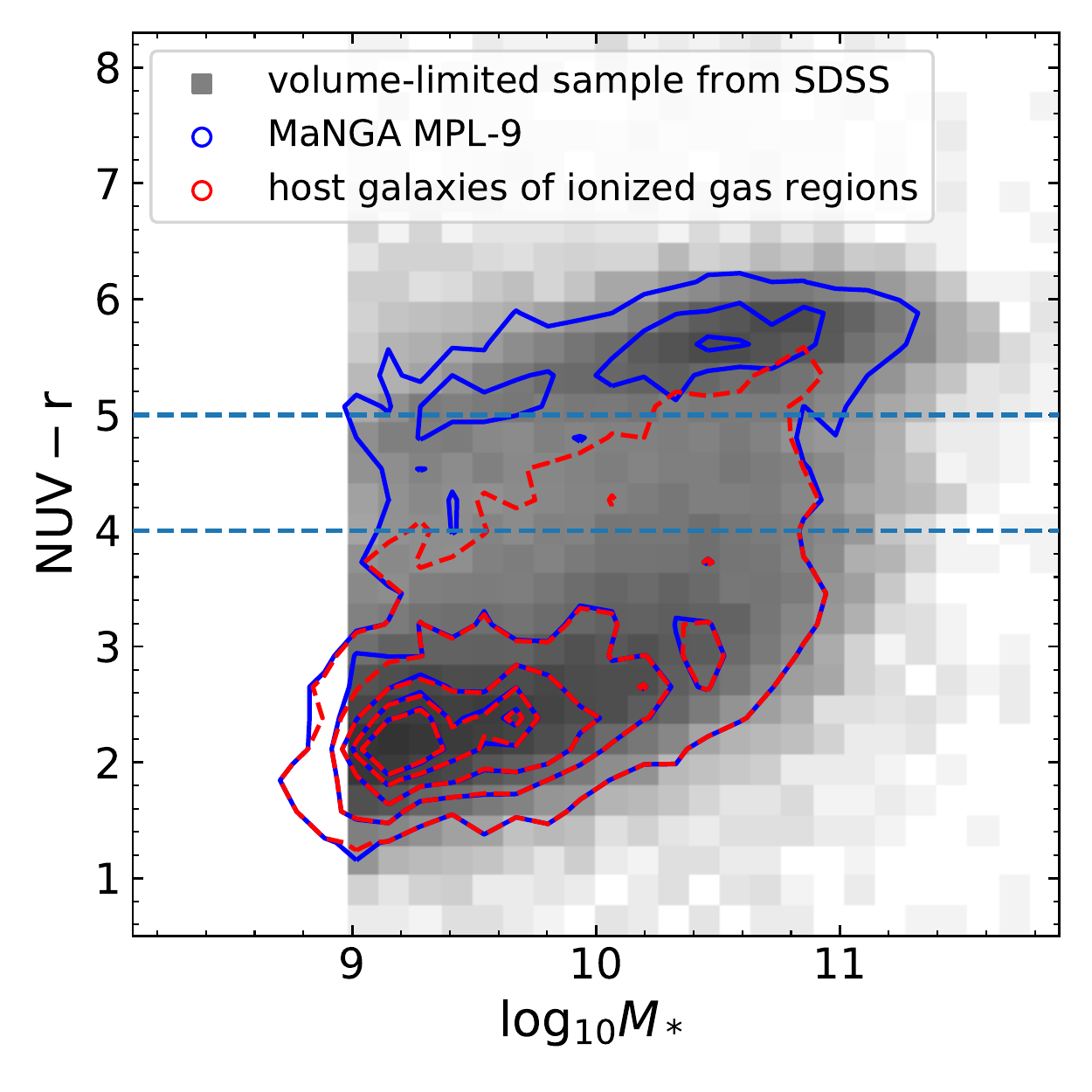}{0.24\textwidth}{(c)}
    \caption{Panel (a1): The BPT diagram of the ionized gas regions identified
    from the MaNGA MPL-9. The red line from \citet{2001ApJS..132...37K} 
    and the blue line from \citet{2006MNRAS.372..961K} jointly divide 
    the regions into three classes: star-forming regions (below the red line), 
    Syeferts (above the two lines) and LINERs (above the red but below the blue line).
     The gray scale indicates the number density of the ionized gas regions 
     in logarithm. The contours indicate the areas enclosing 30\%, 60\%, and 90\%
     of the total sample.
     Panel (a2): It's similar to panel (a1) but for \oiiir/\hb\ versus \niir/\ha\ diagram.
     The red line is from \citet{2001ApJS..132...37K} and 
     the blue line is from \citet{2003MNRAS.346.1055K}. Regions above the red line are
     classified as AGN.
     Panel (b): The histogram shows the distribution of 
     dust-corrected \hasb\ for the final sample of ionized gas regions.
     The red line shows the cumulative distribution of the final sample.
    Panel (c): Diagram of $NUV-r$ color versus stellar mass. The red dashed 
    contours show the distribution of the galaxies hosting the ionized 
    gas regions in our final sample. The blue contours show the distribution 
    of the parent MaNGA MPL-9 sample. The gray-scale distribution in the
    background is for a volume-limited sample of galaxies selected from the 
    SDSS with $M_\ast>10^{9}M_\odot$ and $0.01<z<0.03$. See the text 
    in Section \ref{sec:sample_selection}.}
    \label{fig:sample}
\end{figure*}

\subsection{Selection of ionized gas regions}
\label{sec:sample_selection}

For our analysis we further select a subset of the ionized gas regions with 
substantially higher S/N in both the stellar continuum and the relevant 
emission lines. Specifically, we require S/N$\;>5$ for the stellar 
continuum and the \ha\ and \hb\ lines, and S/N$\;>3$ for the 
\oiill, \oiiir, \niir\ and \siill. Furthermore, we exclude regions classified as 
AGN on the Baldwin-Phillips-Terlevich diagram 
\citep[BPT;][]{1981PASP...93....5B}.
\autoref{fig:sample} (panel (a1) and (a2)) displays the distribution of all the ionized gas 
regions in the planes of \oiii/\hb\ versus \sii/\ha\ and \oiii/\hb\ versus \nii/\ha. 
The majority of the
regions are located in the area classified as star-forming (or \hii). 
We first exclude all the regions that are classified as Seyfert
on \oiii/\hb\ versus \sii/\ha\ diagram. We also exclude the regions that are located
within 3\arcsec\ from their galactic center and are classified as either LINER
on the \oiii/\hb\ versus \sii/\ha\ diagram or AGN on the \oiii/\hb\ versus 
\nii/\ha\ diagram.

These restrictions exclude about 45\% of the ionized gas regions 
identified in Section \ref{sec:gas_regions}, which results in 
a final sample of $\sim 2.7\times 10^5$ regions. Panel (b) of \autoref{fig:sample}
displays the histogram of \hasb\ of the final sample. The distribution 
is peaked at around $10^{39}$\hasbunit, as indicated by the vertical dashed line. 
As mentioned above, \hasb$\;\sim10^{39}$\hasbunit\ can be used to effectively 
separate the low-\hasb\ DIG regions from the high-\hasb\ \hii\ regions.
Thus, about 40\% of our sample are DIG regions, while 60\% are \hii\ regions. 
Panel (c) of \autoref{fig:sample} displays 
the distribution of the galaxies that host our ionized gas regions 
in the $NUV-r$ versus $\log_{10}M_\ast$ plane, 
using data from the NSA \citep{Blanton-11}. 
For comparison, the distribution of the full sample of MaNGA MPL-9 
is plotted as blue contours. We have 
also selected from the NSA a volume-limited sample consisting 
of $\sim3.5\times10^4$ galaxies with stellar mass $M_\ast>10^{9}M_\odot$ 
and with redshift in the range $0.01<z<0.03$. The distribution of this sample 
is plotted as the gray background in the figure. Overall, galaxies in the MaNGA 
MPL-9 follow the general population. The host galaxies of our 
ionized gas regions are mostly found as the `blue cloud' population 
($NUV-r<4$), with a considerable fraction at $M_\ast\ga 10^{10}M_\odot$
extending to the green valley with $4<NUV<5$, and with a small fraction 
at the highest mass end ($M_\ast\sim10^{11}M_\odot$) to the red sequence
($NUV-r>5$). 

\begin{figure}
    \epsscale{1.15}
    \plotone{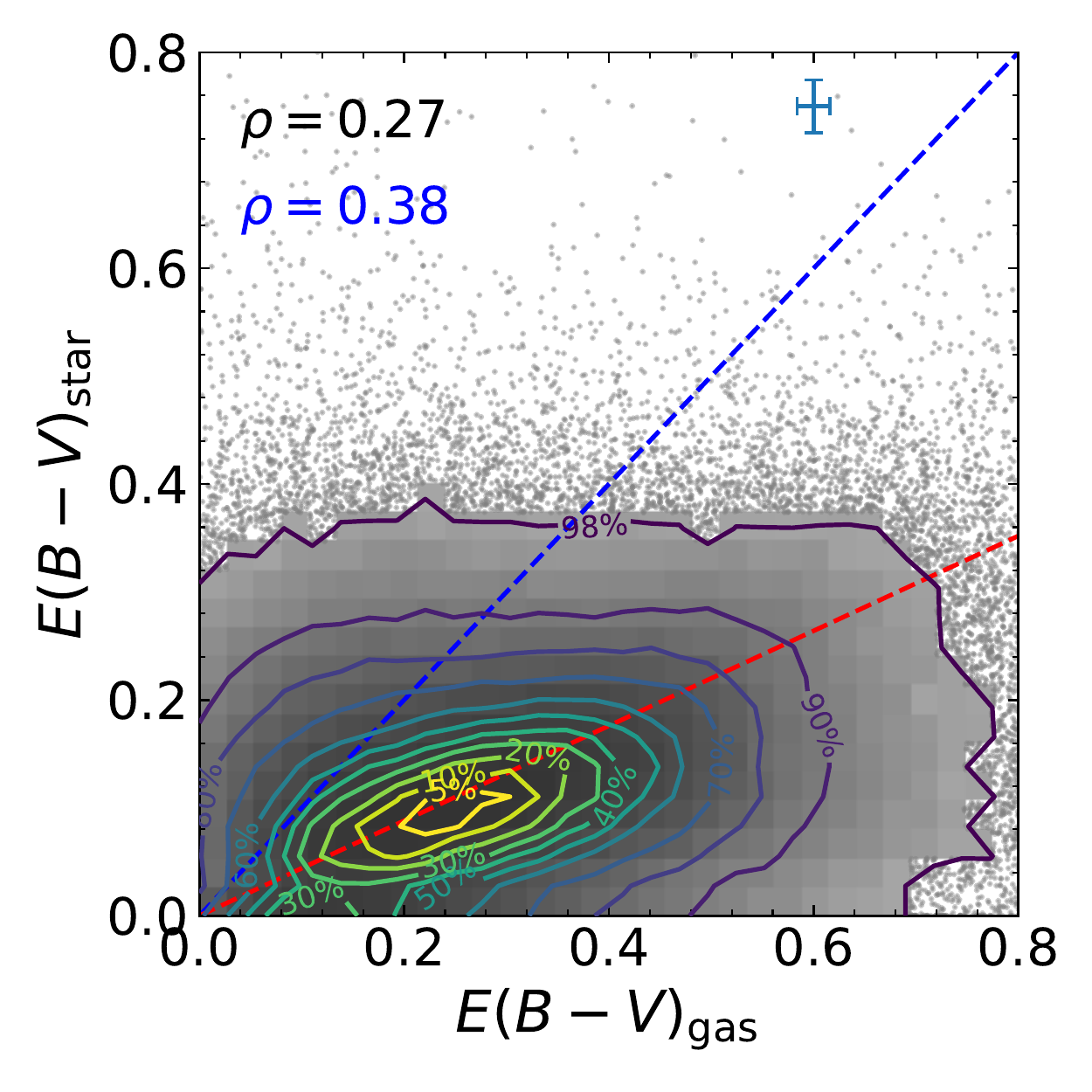}
    \caption{Distribution of the ionized gas regions in MaNGA galaxies
    on the diagram of \ebvstar\ versus \ebvgas. Contours of  
    sample fraction are plotted and the contour levels are indicated.
    Individual ionized gas regions are plotted as small dots outside the 98\% area.
    The Spearman rank correlation coefficients ($\rho$) are indicated
    for both the full sample (black) and  the \hii\ regions (blue) selected 
    by \logten\hasb$\;>39$. The blue dashed line 
    indicates the 1:1 relation, while the red dashed line represents 
    \ebvstar$\;=0.44$\ebvgas\ which is the average relation of star-forming 
    galaxies found by \citet{2000ApJ...533..682C}.
    The typical uncertainties are shown in the upper right corner.}
    \label{fig:ebv_all}
\end{figure}

\begin{figure*}
    \epsscale{1.15}
    \plotone{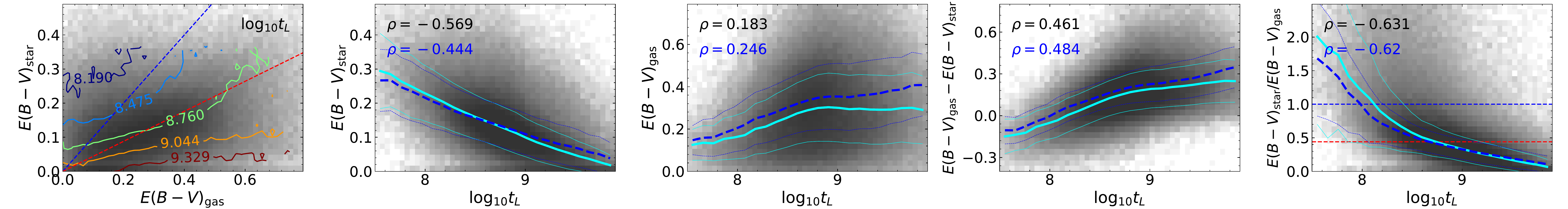}\\
    \plotone{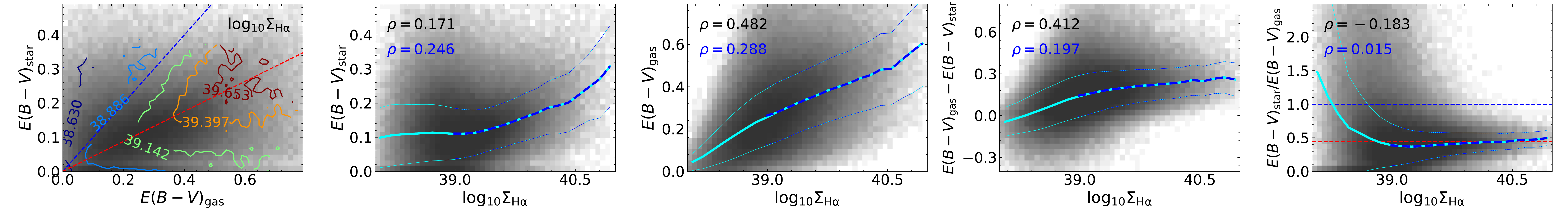}\\
    \plotone{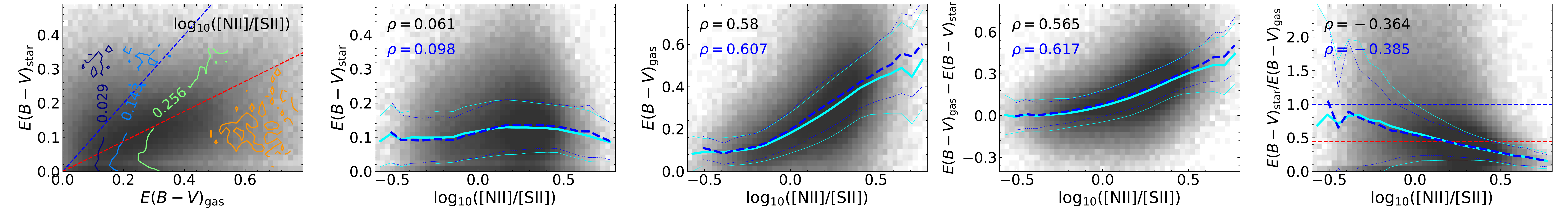}
    \caption{Correlations of \ebvstar\ and \ebvgas\ on 
    regional properties (panels from top to bottom):  
    \logten\age, \logten\hasb\ and \logten(\niisii). Panels in the left-most 
    column show the diagram of \ebvstar\ versus \ebvgas\ for all 
    the ionized gas regions in our sample, with 
    contours showing the distribution of the corresponding  
    property. Panels from the second to the last column show 
    the correlation of \ebvstar, \ebvgas, \ebvgas$-$\ebvstar\ and 
    \ebvstar/\ebvgas\ as functions of the property. 
    The distribution of all the ionized gas regions are plotted 
    as the gray-scale background in each panel, and the median 
    relation and the 1$\sigma$ scatter are plotted as the cyan
    lines. The median relation and the 1$\sigma$ scatter of the
    subset of \hii\ regions selected by \logten\hasb$\;>39$ are 
    additionally plotted as the blue dashed lines. The Spearman 
    rank correlation coefficients ($\rho$) for both  the full sample 
    (black) and the \hii\ regions (blue) are indicated. 
    }
    \label{fig:dependence_main}
\end{figure*}

\section{Results}
\label{sec:results}

\subsection{Correlations of dust attenuation with regional/global properties of galaxies}
\label{sec:res_drive}

\autoref{fig:ebv_all} displays the distribution of all the 
ionized gas regions in our sample in the  
\ebvstar\ versus \ebvgas\ plane. Overall, \ebvgas\ spans a wider
range than \ebvstar, and there is a weak 
correlation between the two parameters. 
The Spearman rank correlation coefficients for both the 
full sample of ionized gas regions ($\rho=0.27$) and the 
subset of \hii\ regions of \logten\hasb$\;>39$
($\rho=0.38$) are quite low. For reference, we also plot the
contours of constant sample fraction at levels ranging 
from 5\% to 98\%, as well as two linear relations: 
the 1:1 relation (blue dashed line) and 
\ebvstar$\;=0.44$\ebvgas\ (red dashed line), the previously found 
average relation of UV-bright starburst galaxies 
\citep[e.g.,][]{1994ApJ...429..582C,2000ApJ...533..682C}.
The majority of the ionized gas regions are a bit below the red line, with large scatter.
More interestingly, a significant population  
of the ionized gas regions is located above the 
1:1 line, where the stellar attenuation is larger than the 
gas attenuation. A key goal of our study is to understand 
what drives the scatter in this diagram, particularly the 
bimodal distribution divided (roughly) by the 1:1 line.

We examine the dependence of \ebvstar, \ebvgas, \ebvdelta,
and \ebvratio\ on a variety of physical properties, 
both regional and global, in order to find the driving 
factor(s) for the overall distribution shown in \autoref{fig:ebv_all}.
We consider 13 regional properties: 
\logten\hasb, \logten\haew, sSFR, \logten$\Sigma_\ast$,
\logten\age, \logten$t_M$, $Z_L$, $Z_M$, $D_n4000$, 12+\logten(O/H),
\logten\nii/\oii, \logten\oiii/\oii, \logten\nii/\sii, 
which are described in Section \ref{sec:spectral_fitting}
and Section \ref{sec:measure_lines}.
In addition, three global properties are considered: 
total stellar mass (\mass), morphological type 
($T$-type), and the $r$-band minor-to-major axial ratio (\ba).
The measurements of \mass\ and \ba\ are taken from the NSA catalog
\citep{2005AJ....129.2562B},
and the estimates of $T$-type come from \cite{2018MNRAS.476.3661D}. 
A negative $T$-type usually indicates early-type morphology, 
while a positive $T$-type indicates late-type.

We find that, out of the 16 regional/global properties, 
dust attenuation shows the strongest correlation with the following 
three: \age\ for the stellar attenuation while \hasb\ and \niisii\
for the gas attenuation. 
\autoref{fig:dependence_main} shows the distribution of the
ionized gas regions in the \ebvstar\ versus \ebvgas\ space, 
color-coded by the three properties (panels in the leftmost column), 
as well as the correlation of four dust attenuation parameters 
(\ebvstar, \ebvgas, and their difference and ratio)
with the three properties (panels in the 2nd to 5th columns). 
In each of the correlation panels, the median relation and the 
1$\sigma$ scatter around the median relation are plotted as 
solid cyan lines for all the ionized gas regions in our sample, 
and as blue dashed lines for the subset of \hii\ regions of 
\logten\hasb$\;>39$. The Spearman rank correlation coefficients 
($\rho$) are indicated for both the full sample and the subset. 
Results for the rest 13 properties, plotted in the same way, 
are presented in \autoref{sec:res_phys}. 

By comparing the Spearman rank correlation coefficients, one can see
that the stellar attenuation parameter \ebvstar\ shows the strongest 
(negative) correlation with stellar age. For the full sample, 
$\rho=-0.569$ for the luminosity-weighted age (\logten\age) and 
$\rho=-0.627$ for the mass-weighted age (\logten$t_M$; see 
\autoref{fig:dependence_other}). 
The stellar attenuation decreases with increasing \age,
from an average of \ebvstar$\sim0.3$ at \logten(\age/yr)$\;\sim7.5$ 
down to about zero at the oldest age, as can be seen from the first row 
of \autoref{fig:dependence_main}. The correlation of \ebvstar\ with 
the other 12 properties is much weaker, with $|\rho|<0.2$ 
for all properties except the sSFR for which $\rho=0.339$. 
Similar results are obtained when the analysis is limited to \hii\ regions. 
These results strongly suggest that stellar age is the dominant parameter 
in driving the stellar attenuation, for both \hii\ and DIG regions. 
Compared to the mass-weighted stellar age (\logten$t_M$),
the luminosity-weighted age (\logten\age) shows a more linear correlation
with stellar attenuation, although with a slightly smaller correlation 
coefficient. In addition, from the first row of \autoref{fig:dependence_main}, 
one can see that the stellar age is positively correlated with the gas attenuation,
although with a relatively small correlation coefficient ($\rho=0.183$). 
Consequently, \ebvdelta\ increases with \age, while 
\ebvratio\ decreases. Among all the properties and all the attenuation 
parameters, the strongest correlation is found between \ebvratio\ and 
\logten\age, with $\rho=-0.631$. Thus, the driving role of \logten\age\ 
in \ebvdelta\ and \ebvratio\  comes mainly from the strong correlation 
between \ebvstar\ and \logten\age\ , and the correlations of 
\ebvdelta\ and \ebvratio\ with other properties are largely produced 
by this strong correlation with age. 

Unlike \ebvstar, the attenuation in gas, \ebvgas, shows strong correlations 
with multiple properties, including \niisii\ ($\rho=0.58$)
and \hasb\ ($\rho=0.482$), as shown in \autoref{fig:dependence_main}, and 
\sigmamass\ ($\rho=0.432$), as shown in \autoref{fig:dependence_other}. 
From the second row of \autoref{fig:dependence_main}, we can see that 
both \ebvstar\ and \ebvgas\ are correlated with \hasb,
but in different ways at both the high and low end of \hasb.
For regions with \logten\hasb$\;\ga 39$, which are dominated by 
\hii\ regions, the two attenuation parameters behave quite similarly, 
showing positive correlations with \hasb.
As a result, both their difference and ratio show a weak correlation 
with \hasb. The \ebvdelta\ increases slightly from $\sim0.1$ mag 
at \logten\hasb$\;=39$ up to $\sim 0.25$ mag at the highest \hasb. 
The \ebvratio\ is almost constant at the value of 0.44 \citep{2000ApJ...533..682C},
ranging from $\sim0.4$ at \logten\hasb$\;=39$ to $\sim 0.5$ at 
\logten\hasb$\;=41$, with small scatter. At \logten\hasb$\;\lesssim 39$ 
where the regions are dominated by DIG,
\ebvstar\ shows no obvious correlation with \hasb, with an average 
of $\sim 0.1$ mag and large scatter, while \ebvgas\ rapidly increases 
with increasing \hasb. Consequently, the \ebvdelta\ increases 
and the \ebvratio\ decreases as the \hasb\ increases. 
At the lowest density, \logten\hasb$\;\lesssim 38.5$,  
the stellar attenuation is stronger than the gas attenuation, 
mainly because of the small amount of dust in the gas 
as seen from the rapid decrease of \ebvgas\ in the low-density end.
The negative correlation of \ebvratio\ with \hasb\ has recently 
been discussed in depth in \citet{2020ApJ...888...88L}, based on 
an earlier sample of MaNGA. Our results show in addition that the 
correlation is essentially a result of the dichotomy of the intrinsic 
properties of the ionized gas regions, \hii\ regions and DIG regions, 
divided at \hasb$\;\sim10^{39}$\hasbunit\ as suggested by \citet{2017MNRAS.466.3217Z}.
It is clear that \hii\ regions and DIG regions are distinct in terms 
of the dust attenuation, in the sense that \hii\ regions 
with higher \hasb\ tend to be more dusty and have tighter
correlation between the stellar and gas attenuation. It is thus 
necessary to consider the two types of regions separately when studying 
their dust properties.

\begin{figure*}
    \fig{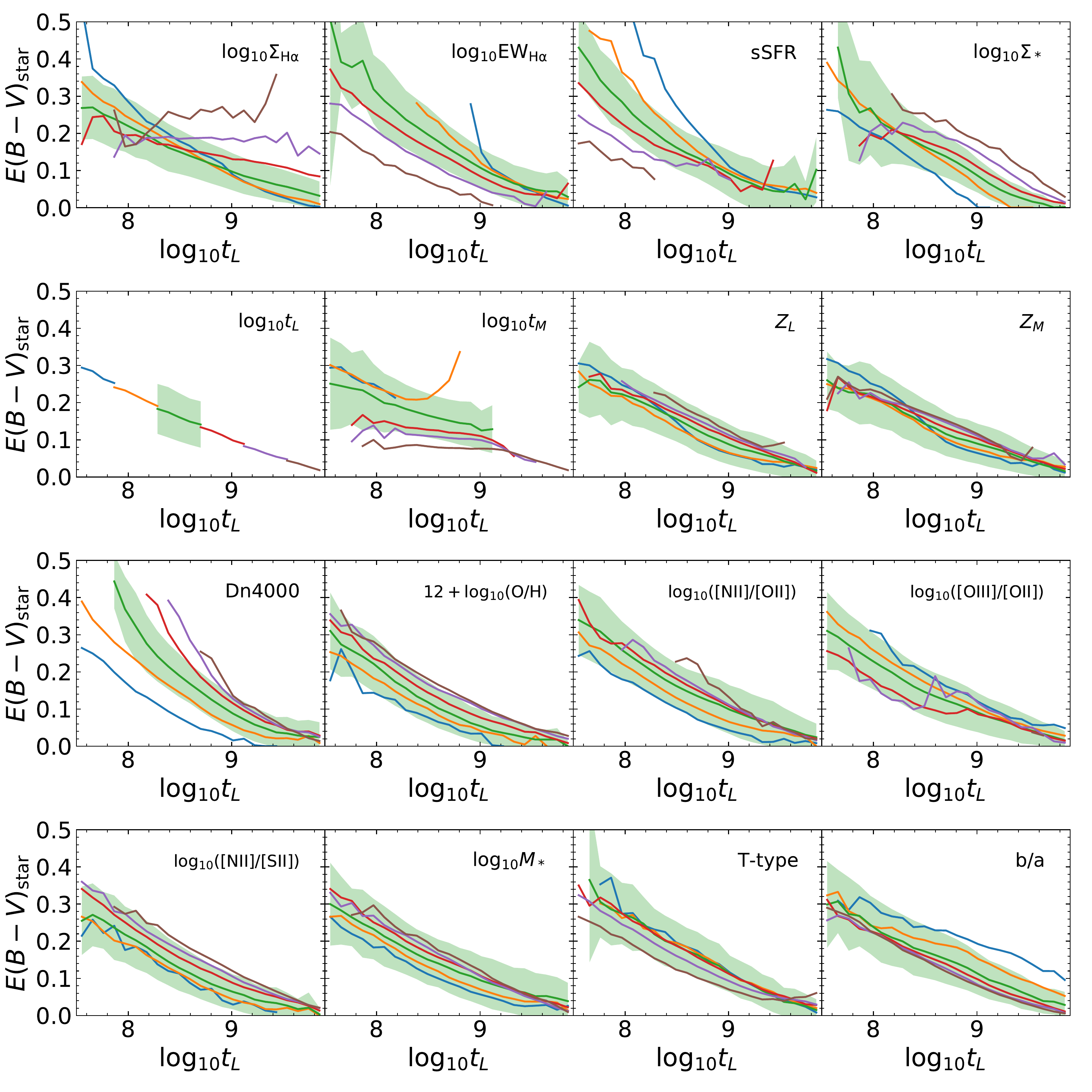}{0.48\textwidth}{(a) \ebvstar\ versus $\log_{10}$\age}
    \fig{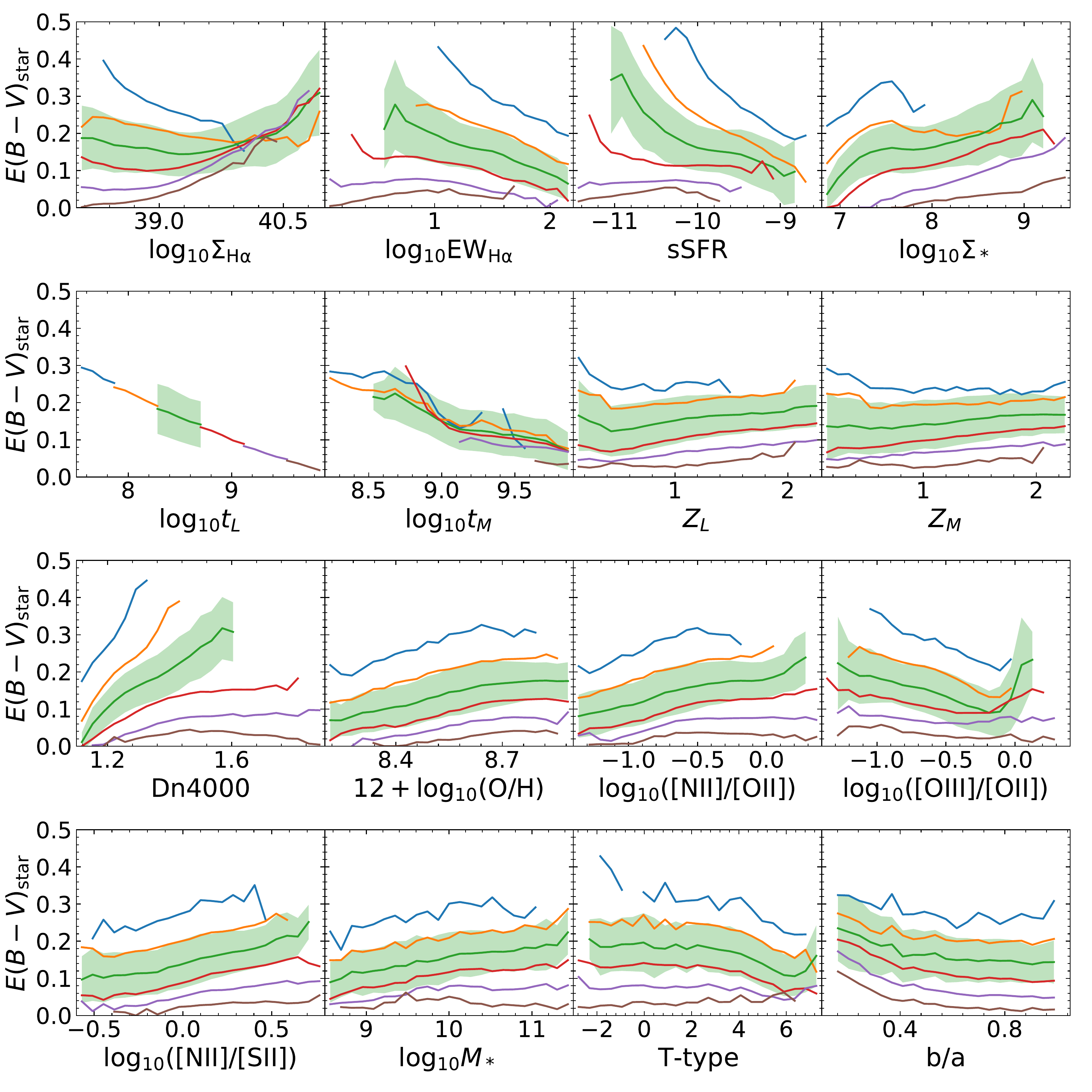}{0.48\textwidth}{(b) \ebvstar\ versus regional/global properties}
    \caption{Panels (a): \ebvstar\ as a function of \logten\age. In each panel the 
    different colors are for subsets of ionized gas regions selected by 
    one of the 16 regional/global properties as indicated.
    The colors of cyan, orange, green, red, purple and brown
    correspond to subsets with increasing values of the property.
    For clarity, we only show the scatter of the green line as the scatters of
    other lines are comparable to that of the green line.
    Panels (b): \ebvstar\ as a function of regional and global properties. In each 
    panel the different colors are for ionized gas regions in different intervals of 
    \logten\age. The stellar age binning can be seen from the left-most panel in the second row.}
    \label{fig:age_ebvstar}
\end{figure*}

The bottom row of \autoref{fig:dependence_main} shows the relationship 
between \niisii\ and the four dust attenuation parameters. Both a
strong correlation with the gas attenuation and the absence of 
correlation with the stellar attenuation can be clearly seen.  
For the gas attenuation, the strong correlation is seen only when 
\niisii\ exceeds $\sim -0.2$. In addition, noticeable correlations
between gas attenuation and the gas-phase metallicity or related parameters are also found: 
$\rho=0.445$ for \metgas\ and $\rho=-0.461$ for \oiiioii\ 
(see \autoref{fig:dependence_other}), which may share 
the same origin with the correlation with \sigmamass\ given 
the well-known mass-metallicity relation \citep{2004ApJ...613..898T}. 
Similar to \niisii, the \oiiioii\ also shows no correlation with 
the stellar attenuation, but a strong correlation with the gas attenuation 
(see~\autoref{fig:dependence_other}). 
Both \niisii\ and \oiiioii\ have been used as indicators of the 
ionization level of gas \citep{2013ApJS..208...10D, 2020ApJ...888...88L}.
In the ionization model of spherical \hii\ regions from 
\citet{2013ApJS..208...10D}, the line ratio of \oiiioii\ increases 
as the ionization parameter ($q$) increases, when the gas-phase metallicity 
diagnostics (e.g. \niioii) are fixed. Based on the same models and 
the MaNGA data, \citet{2020ApJ...888...88L} found the line ratio
\niisii\ to be sensitive to both the gas-phase metallicity and the ionization 
parameter, with smaller values of \niisii\ in metal-poorer regions 
at fixed $q$. Note that \ebvgas\ is also moderately correlated with 
$D_n4000$~($\rho=0.386$) and \logten\mass~($\rho=0.459$). 
The former may be related to its correlation with stellar 
age as mentioned above, while the latter may be a result of its 
correlation with \sigmamass. 

In conclusion, we find that the stellar attenuation is mostly driven by 
stellar age, while the gas attenuation is most strongly related to 
\hasb, \niisii\ and \oiiioii. In the latter case, \hasb\ divides 
ionized gas regions into two distinct types, DIG regions and \hii\ regions, 
and \niisii\ and \oiiioii\ are indicators of the gas-phase metallicity 
and ionization level.  
For the stellar-to-gas attenuation ratio (or difference), both stellar 
age and \hasb\ play important roles. In the following subsections, we 
will discuss these findings further. We will focus on the stellar 
age versus stellar attenuation relation in Section \ref{sec:res_t_L}, 
the role of \hasb, \niisii\ and \oiiioii\ for the gas attenuation 
in Section \ref{sec:role_hasb_niisii}, and the joint role of \logten\age\  
and \hasb\ for the stellar-to-gas attenuation ratio in 
Section \ref{sec:joint_dependence}. 

\subsection{The role of stellar age in driving stellar attenuation}
\label{sec:res_t_L}

In this subsection we concentrate on \age, which has the strongest 
correlation with \ebvstar,
when compared to any other regional/global properties. This implies the 
potential dominant role of the stellar age in driving  the dust 
attenuation parameters. In order to test this hypothesis, we divide the 
ionized gas regions into different intervals of \logten\age, and examine 
the correlations of \ebvstar\ with other regional/global properties. 
Similarly, for a given regional/global property, we divide the ionized 
regions into intervals of the property, and examine the correlations
of the dust attenuation parameters with \logten\age. The results 
are shown in \autoref{fig:age_ebvstar}.

Panels (a) of \autoref{fig:age_ebvstar} show the correlation of 
\ebvstar\ with \logten\age, but for the subsets of ionized gas regions 
selected by each of the regional/global properties. 
Panels (b) show the correlation of \ebvstar\ with different 
regional/global properties, but for the subsets of ionized gas regions 
selected by \logten\age. In most cases, \ebvstar\ shows tight 
correlations with \logten\age\ even when the sample is limited to 
a narrow range of a specific property, as can be seen from panels (a), 
while \ebvstar\ shows no or rather weak dependence on all other 
properties once the stellar age is limited to a narrow range, as 
shown in panels (b). This result reinforces the conclusion
from the previous subsection that the stellar attenuation is predominantly  
driven by the stellar age. 

One can identify two subsets of the ionized 
gas regions in which residual correlations are clearly seen. 
The subset, which consists of the regions with both high \ha\ surface brightness 
(\hasb$\;\ga 10^{39}$\hasbunit) and old stellar ages (\age$\;\ga 1$Gyr), 
shows that the \ebvstar\ is more strongly correlated with  
\hasb\ than with \logten\age. This can be seen from the top-left 
panel in both panels (a) and (b) of the figure. The second subset,
consisting of regions with very young ages (with \age\ less than a few 
$\times10^{8}$yr), shows residual correlations with $D_n4000$ 
(as well as with \haew\ and sSFR) at fixed \logten\age. 

We have done the same analysis for \ebvdelta\ and \ebvratio, and the 
results are presented in \autoref{sec:res_t_L_more}. We see that 
the stellar age is correlated with both \ebvdelta\ and \ebvratio\ 
in most cases when other properties are fixed, while at fixed \logten\age\
no strong correlation is seen between either of the attenuation parameters 
and other properties. These behaviors can be attributed largely  
to the correlation between \ebvstar\ and \logten\age, as shown 
in~\autoref{fig:age_ebvstar}. 

\begin{figure*}
    \centering
    \fig{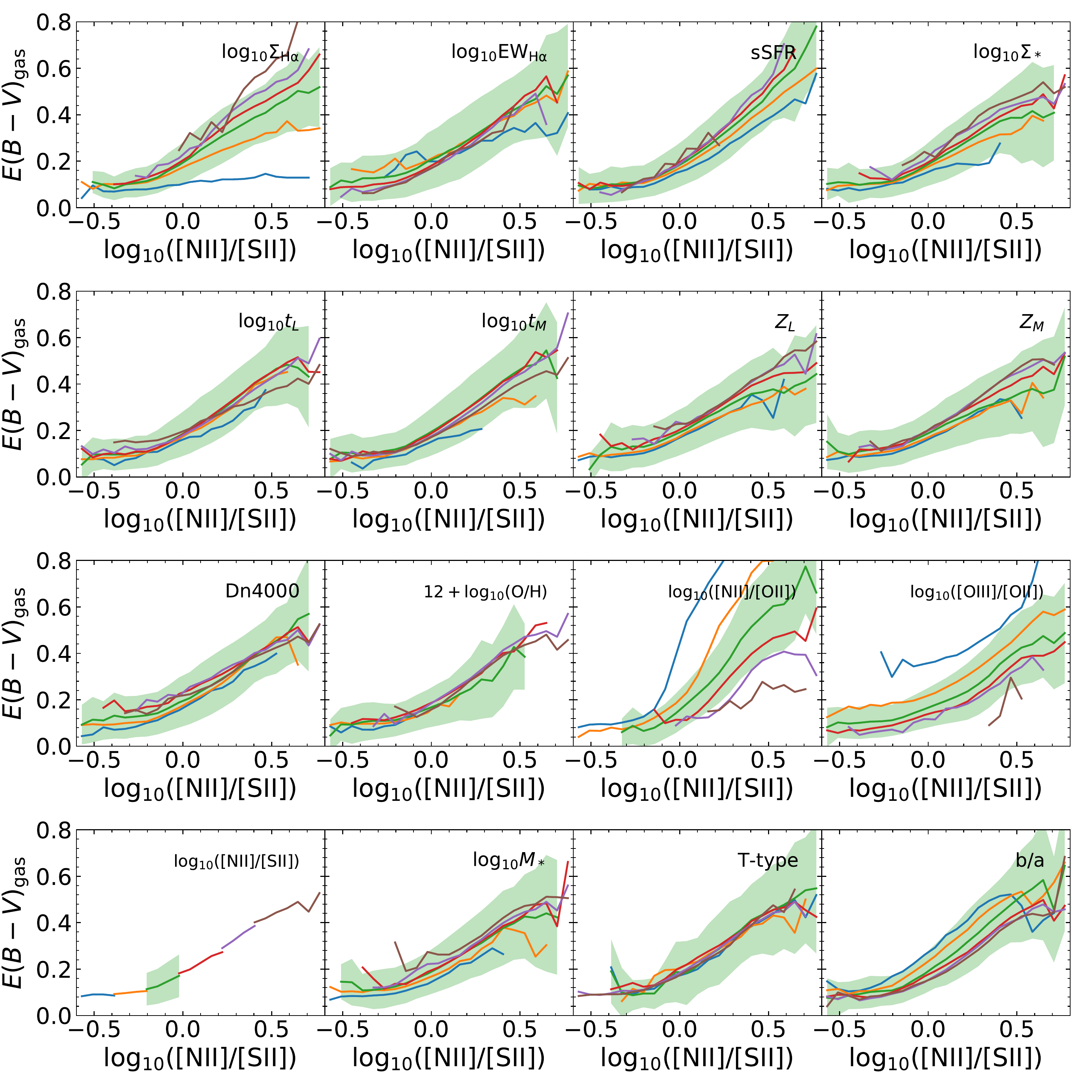}{0.48\textwidth}{(a) \ebvgas\ versus \logten(\niisii)}
    \fig{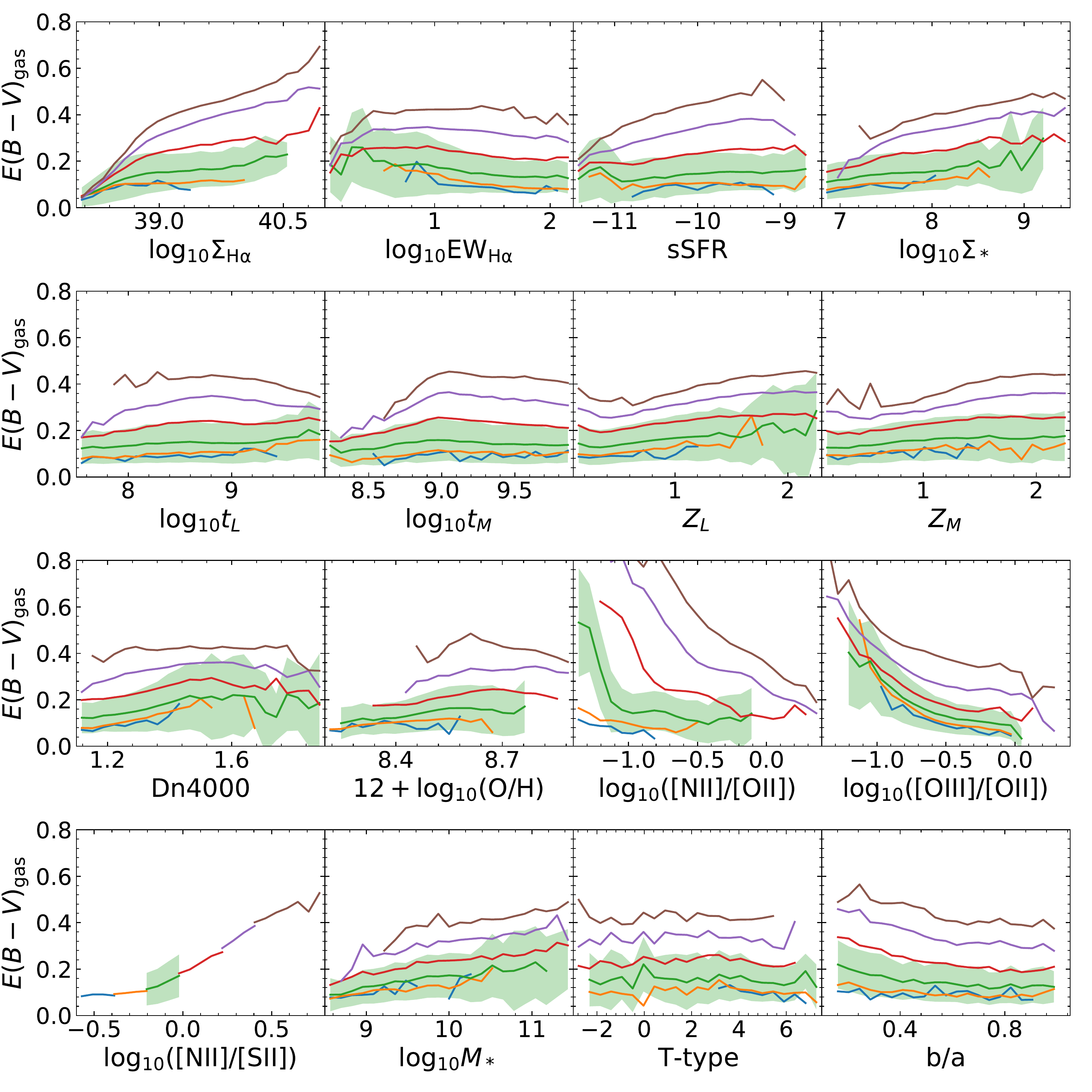}{0.48\textwidth}{(b) \ebvgas\ versus regional/global properties}
    \caption{\ebvgas\ as function of \logten(\niisii) for subsets of 
    ionized gas regions selected by each of the 16 regional/global properties 
    as indicated (panels a), and as functions of regional/global properties 
    for subsets of ionized gas regions selected by \logten(\niisii)
    (panels b). Symbols/lines are the same as in \autoref{fig:age_ebvstar}.}
    \label{fig:niisii_ebvgas}
\end{figure*}

\subsection{The role of \hasb, \niisii\ and \oiiioii\ in driving gas attenuation}
\label{sec:role_hasb_niisii}

Similarly, we have compared the dependence of \ebvgas\ on 
\hasb, \niisii\ and \oiiioii\ with that on other regional/global 
properties. We find that, among the three properties, \niisii\ 
shows stronger correlations with \ebvgas\ than both \hasb\ 
and \oiiioii\ when other properties are limited to a 
narrow range. This is consistent with the fact that 
the highest correlation efficient is between \ebvgas\ and \niisii\   
(see Section \ref{sec:res_drive}).   
We therefore only show the results for \niisii\ 
in~\autoref{fig:niisii_ebvgas}, and present the results for \hasb\ and
\oiiioii\ in \autoref{sec:res_t_L_more}. 

As can be seen from~\autoref{fig:niisii_ebvgas}, the gas attenuation 
is positively correlated with \niisii\ when other properties are 
fixed, but this is true only for \hii\ regions with \hasb$\;\ga 10^{39}$\hasbunit. 
This result echoes the \hasb\ dichotomy as discussed above.
For DIG regions with lower \hasb, the gas attenuation shows weak 
dependence on \niisii\ (see the top-left panel of panels (a)), 
and strong dependence on \hasb\ when \niisii\ is fixed 
(see the top-left panel of panels (b)).  
Apparently, \niisii\ plays a dominant role for \hii\ regions, 
showing strong correlations with \ebvgas\ in almost all cases when 
the regional/global properties, including \hasb\, are fixed
(panels (a)). No residual correlation of \ebvgas\ is seen for most of the 
regional/global properties if \logten(\niisii) is fixed (panels (b)).
Two properties, \niioii\ and \oiiioii,
show residual correlations with \ebvgas\ at fixed \logten(\niisii). 
This probably is not surprising. As found in \citet{2020ApJ...888...88L},
\niisii\ is sensitive to both gas-phase metallicity and 
the ionized level of the gas, and thus expected to correlate with 
both \niioii\, which is a metallicity indicator, and \oiiioii\, which 
is an indicator of the ionized level. 

\begin{figure}
    \epsscale{1.1}
    \plottwo{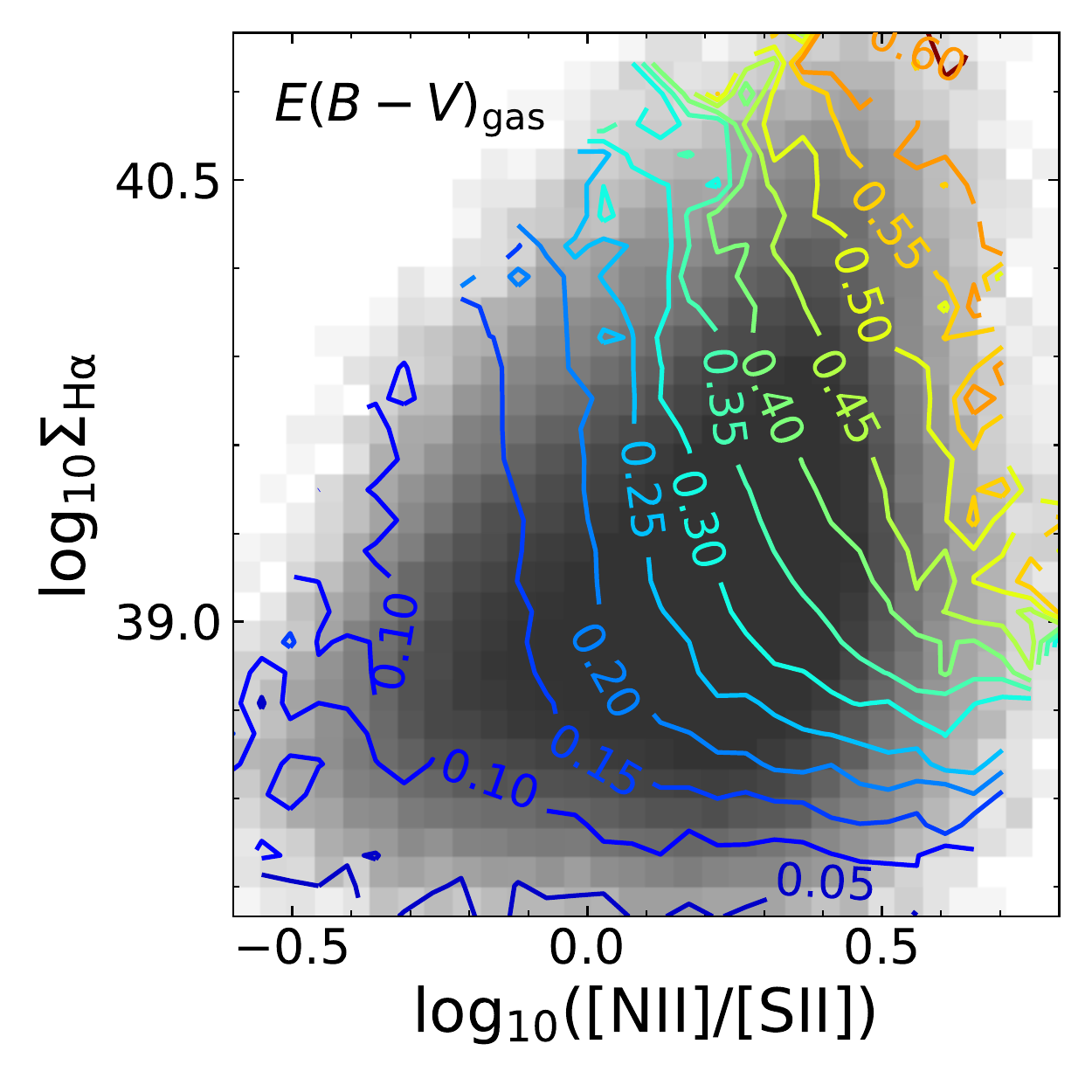}{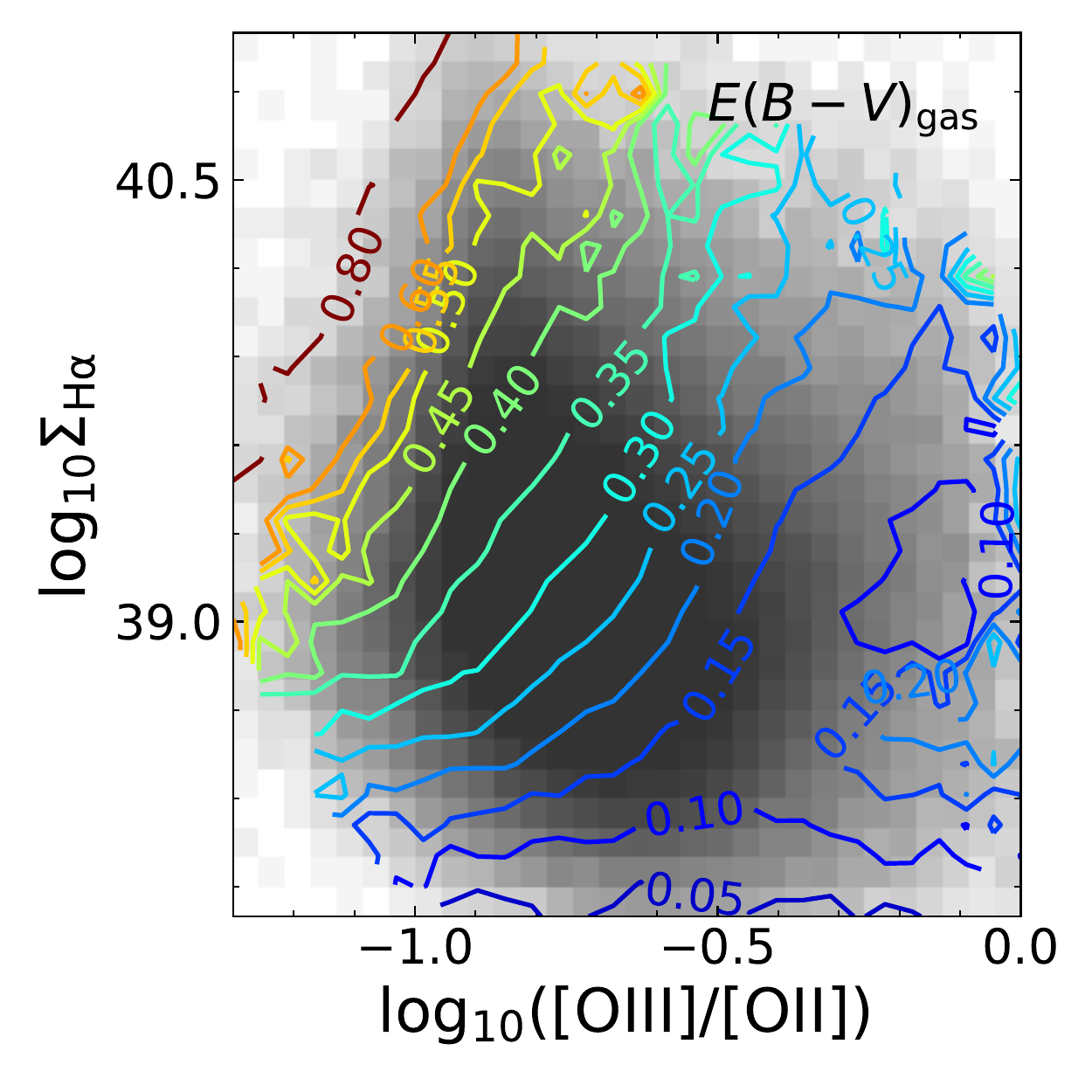}
    \caption{Distribution of all the ionized gas regions on the plane 
    of \logten\hasb\ versus \logten(\niisii) (left panel) and the 
    plane of \logten\hasb\ versus \logten(\oiiioii) (right panel)
    are plotted as the gray-scale background, while contours 
    of \ebvgas\ are plotted in the colored lines with contour 
    levels being indicated.}
    \label{fig:hasb_niisii_plane}
\end{figure}

\begin{figure}
    \epsscale{1.1}
    \plottwo{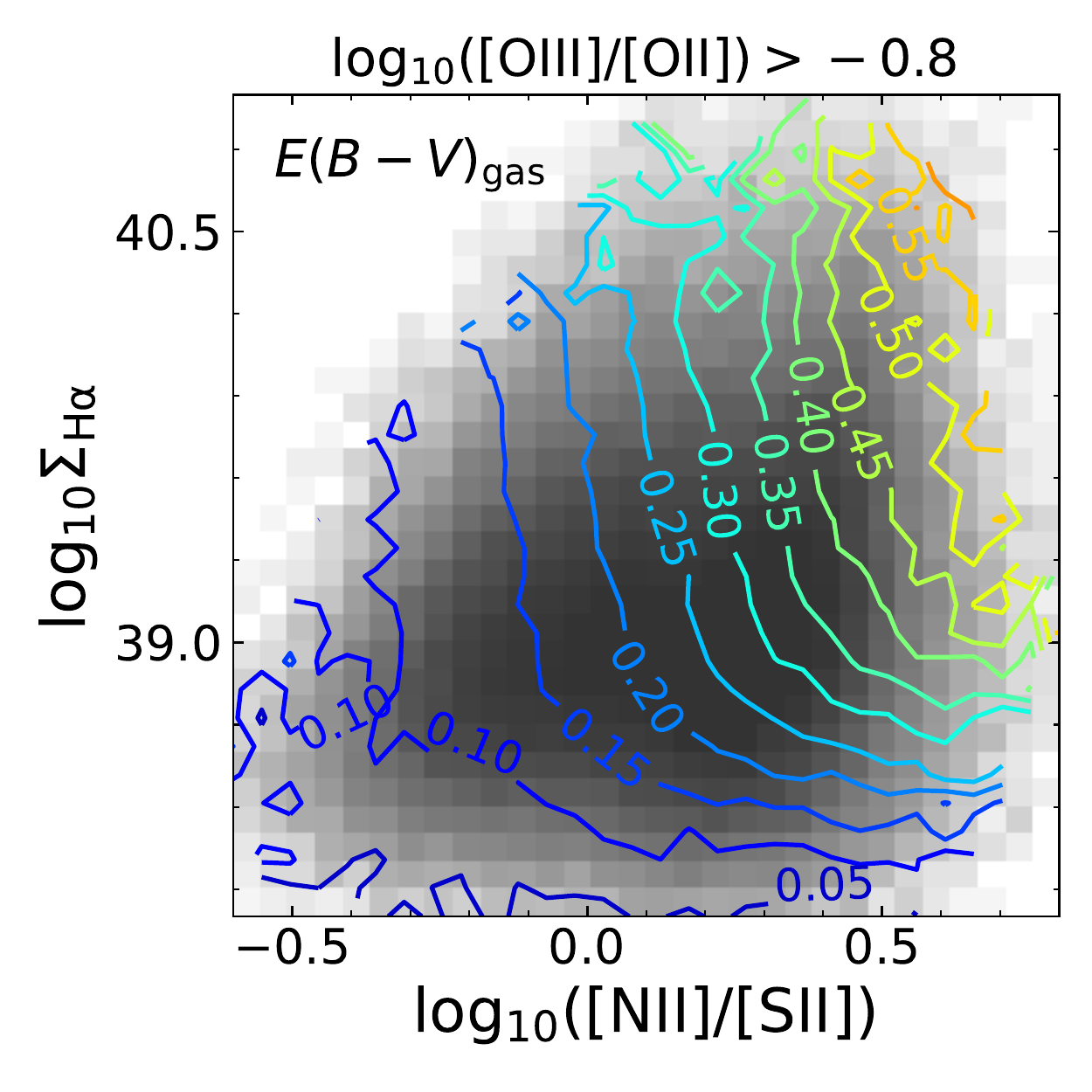}{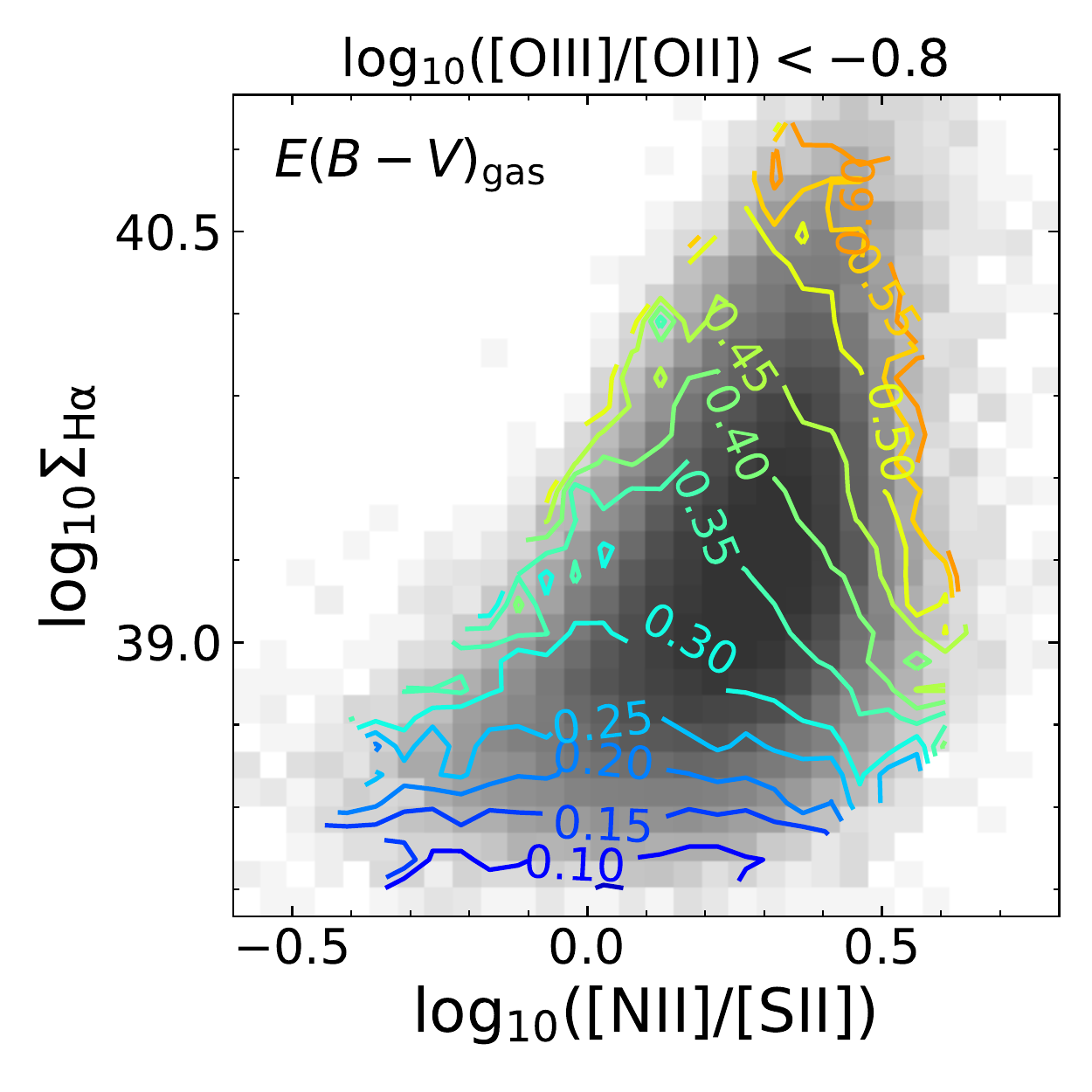}
    \caption{Distribution on the plane of \logten\hasb\ versus \logten(\niisii) 
    for ionized gas regions with \logten(\oiiioii)$\;>-0.8$ (left panel)
    and those with \logten(\oiiioii)$\;<-0.8$ (right panel). Symbols/lines 
    are the same as in \autoref{fig:hasb_niisii_plane}.}
    \label{fig:hasb_niisii_plane_oiiioii}
\end{figure}

\autoref{fig:niisii_ebvgas}
suggests that the \ebvgas\ is determined mainly by 
two parameters: \hasb\ for DIG regions with low \hasb, and 
\niisii\ for \hii\ regions with high \hasb. This finding is more 
clearly seen in \autoref{fig:hasb_niisii_plane} (left panel) where we 
show the contours of \ebvgas\ in the diagram of \logten\hasb\ 
versus \logten(\niisii). Below \logten\hasb$\;\sim 39$, the 
contours are almost horizontal, and so the change of \ebvgas\ 
is mainly driven by \hasb. Above \logten\hasb$\;\sim 39$, in contrast,  
the contours are largely vertical, indicating that \ebvgas\ is
driven mainly by \logten(\niisii). In the right panel of the same figure, we show 
the contours of \ebvgas\ in the \logten\hasb\ versus 
\logten(\oiiioii) plane. Below \logten\hasb$\;\sim 39$, again, the \ebvgas\ 
is mainly driven by \hasb\, with no correlation with \oiiioii. 
At higher \hasb, the contours are inclined, indicating that 
neither \hasb\ nor \oiiioii\ alone can dominate in the \ebvgas. 
At fixed \hasb, the correlation of \ebvgas\ with \oiiioii\ 
appears to be weaker in regions with 
\logten(\oiiioii)$\;\ga -0.8$ than in lower ionization regions.

\autoref{fig:hasb_niisii_plane_oiiioii} displays the \hasb--\niisii\ 
relation again, but separately for high and low ionization regions 
divided at \logten(\oiiioii)$\;=-0.8$.
The high ionization regions (left panel) cover almost the full range 
in the diagram, and the contours in 
\hii\ regions (high \hasb) and with high \niisii\ (high metallicity) 
become more vertical when compared to those for the full sample, 
suggesting that the diagnostic of \oiiioii\ in these regions may be negligible. 
This is consistent with the weak correlation of 
\ebvgas\ with \oiiioii\ at \logten(\oiiioii)$\;\ga -0.8$ seen 
in the right panel of \autoref{fig:hasb_niisii_plane}. 
Low ionization regions (right panel) are distributed differently: 
they are either DIG regions spanning all the range of  \niisii, 
or limited to \hii\ regions of high \niisii. In the latter case the \ebvgas\  
depends jointly on \hasb\ and \niisii, as indicated by the inclined 
contours. Apparently, the diagnostic of \oiiioii\ cannot be neglected 
for low ionization \hii\ regions.

\begin{figure*}
    \epsscale{1.1}
    \plotone{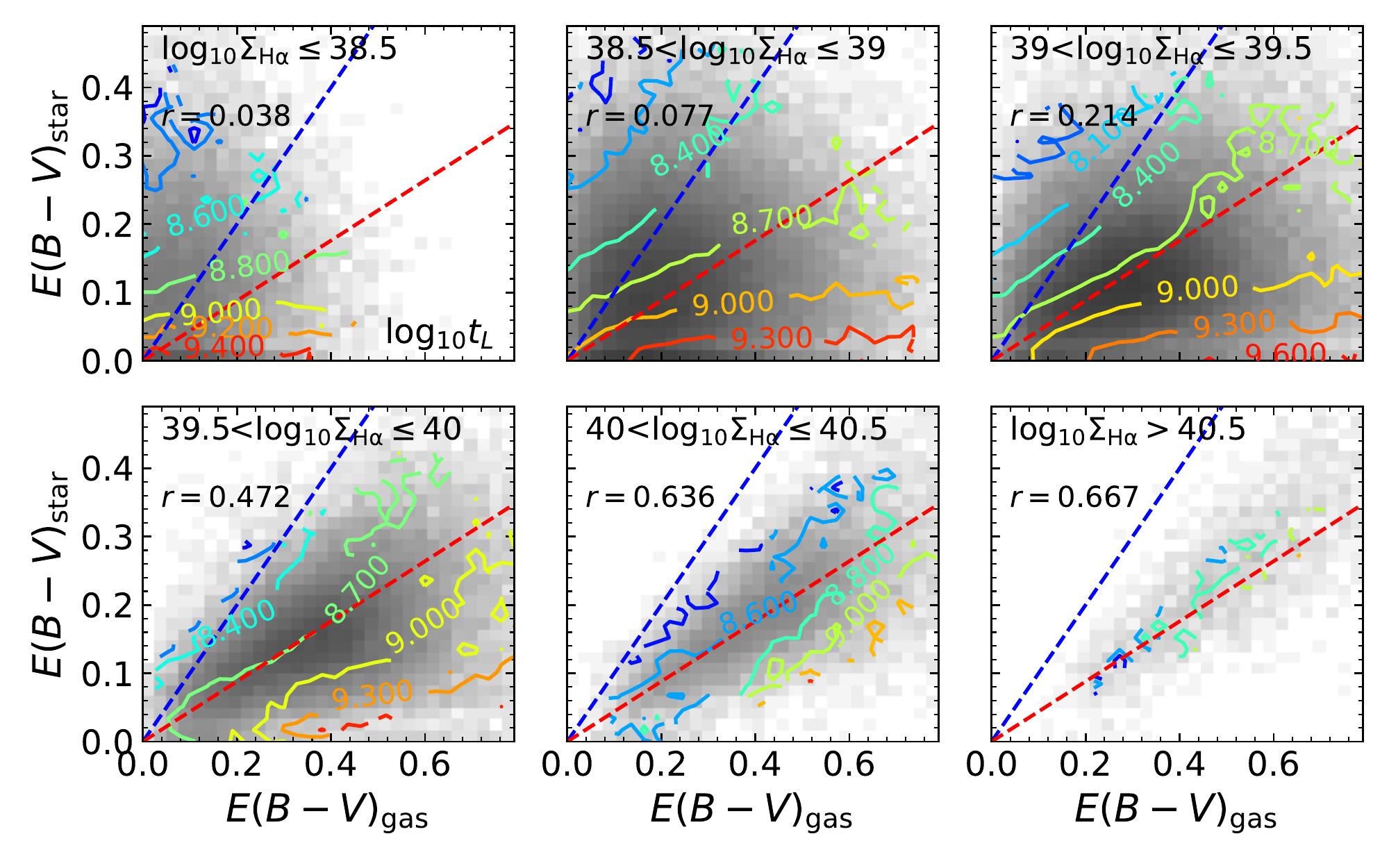}
    \caption{\ebvstar\ versus \ebvgas\ in different intervals of \hasb. 
    The colored lines are the contours of \logten\age\ as indicated, 
    and the gray-scale background shows the number density 
    distribution of the ionized gas regions in each \hasb\ bin.
    The blue dashed line is the 1:1 relation and the red dashed line 
    represents the relation of \ebvstar$\;=0.44$\ebvgas. 
    The Pearson linear correlation coefficient ($r$) between the two 
    attenuations is indicated in each panel.}
    \label{fig:ebvstar_ebvgas_habins}
\end{figure*}

\begin{figure*}
    \epsscale{1.1}
    \plotone{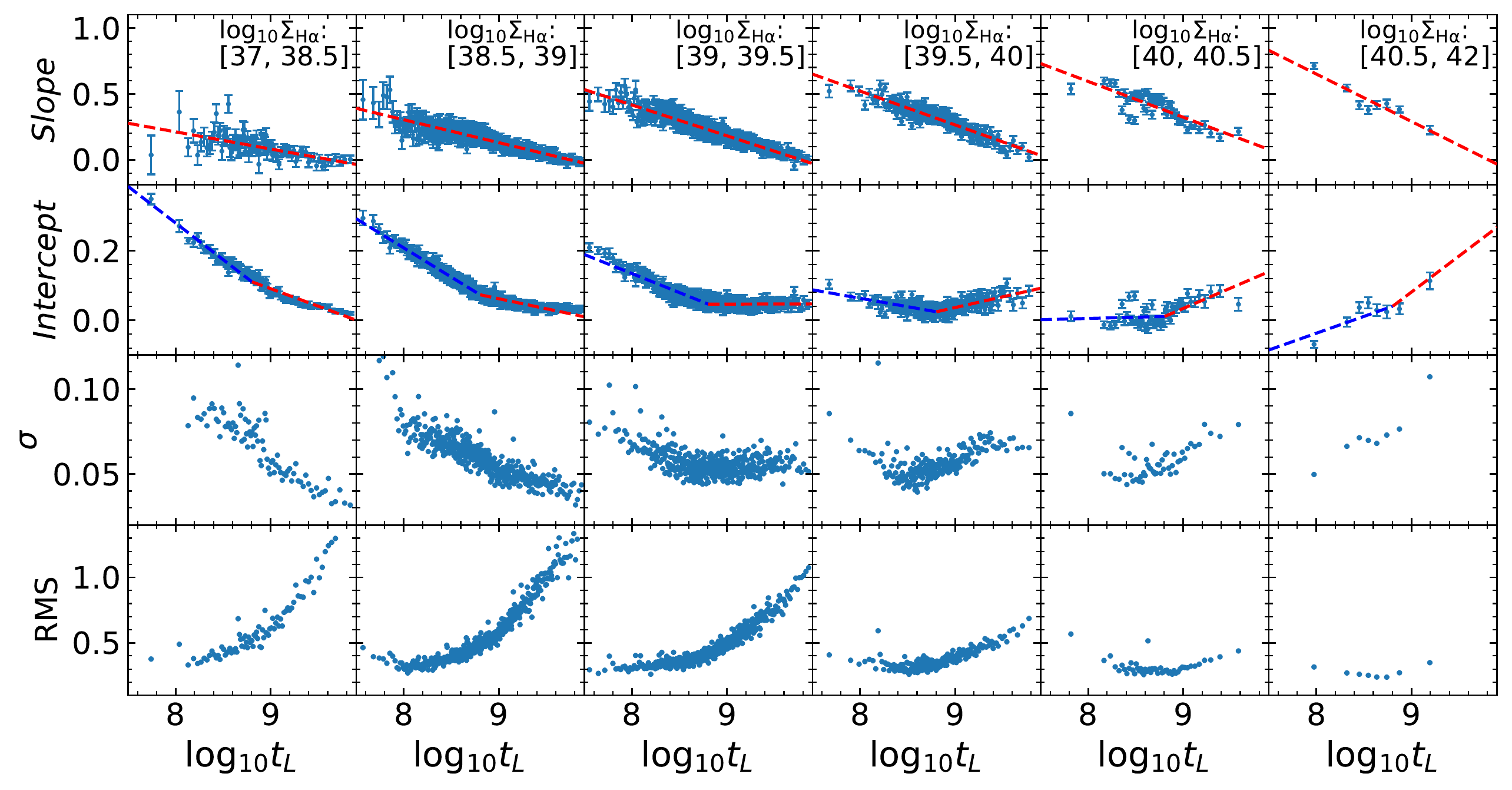}
    \caption{Upper two rows show the slope and intercept of the 
    relation in \autoref{eq:slope_inter} as a function of 
    \logten\age, for different intervals of \logten\hasb. 
    In each panel, the dashed red/blue lines are the best-fit 
    model. The third row shows the 
    standard deviation of ionized gas regions around the best-fit 
    relation ($\sigma$), while the bottom row shows the rms of the relative 
    difference of the ionized gas regions from the best-fit relation
    ($\sigma /$best-fit). See text for details.}
    \label{fig:slope_inter}
\end{figure*}

\subsection{Dependence of \ebvratio\ on both \hasb\ and \age}
\label{sec:joint_dependence}

In this subsection we focus on the stellar-to-gas attenuation ratio,
\ebvratio. The analysis so far has revealed that the 
stellar age is the property most strongly correlated 
with \ebvratio\ (see \autoref{fig:dependence_main}), 
and that the \hasb\ plays an important role as well. 
The residual dependence on \hasb\ at fixed \logten\age\ can largely 
be attributed to the dichotomy of the ionized gas regions. Thus, 
all the regions can be well divided into \hii\ regions with high \hasb\ 
and DIG regions with low \hasb. In this subsection 
we study the dependence of \ebvratio\ on \hasb\ and \age\ jointly.

\autoref{fig:ebvstar_ebvgas_habins} shows the distribution of the ionized 
regions in the \ebvstar\ versus \ebvgas\ plane, but for six 
successive, non-overlapping intervals of \logten\hasb. 
In each \logten\hasb\ interval, the distribution of the ionized 
gas regions is plotted in gray scales, overlaid 
with colored contours of \logten\age. The Pearson linear correlation 
coefficient is also indicated in each panel. For DIG regions with 
\logten\hasb$\;\lesssim 39$, the correlation coefficient is close 
to zero, indicative of no correlation. 
At fixed \ebvgas,  \ebvstar\ spans the full range from zero up to 
$\sim0.5$ mag, while \ebvgas\ is limited to relatively low values 
with the upper limit increasing from $\sim0.2$ mag at 
\logten\hasb$\;\le 38.5$ to $\sim0.6$ mag at 
$38.5<\;$\logten\hasb$\;\le39$. For \hii\ regions, as \hasb\ increases, 
the correlation between the two \ebv\ measurements becomes more and 
more evident and the scatter decreases steadily. The average relation 
of the \hii\ regions follows the relation \ebvstar$\;=0.44$\ebvgas, 
which is indicated by the red dashed line in each panel. When 
\logten\hasb\ exceeds $\sim 39.5$, almost all the regions fall below 
the 1:1 relation.

When \logten\age\ is fixed,
\ebvstar\ is linearly correlated with \ebvgas, and this is true 
in all the different \logten\hasb\ intervals, although the slope 
and intercept of the linear relation vary with both \logten\age\ 
and \logten\hasb. To quantify the joint dependence 
on \logten\hasb\ and \logten\age, we perform a linear fitting to 
the relation between \ebvstar\ and \ebvgas\ for given \logten\age\ 
and \logten\hasb. For each \logten\hasb\ bin, the ionized gas regions 
are sorted by increasing \logten\age, and are divided into a number 
of \logten\age\ bins by requiring each bin to contain 200 ionized gas regions. 
A linear relation is obtained by fitting the following equation to the data
in each bin:
\begin{equation}\label{eq:slope_inter}
    \mbox{{\ebvstar}} = Slope \times \mbox{\ebvgas} + Intercept.
\end{equation}
The upper two panels in \autoref{fig:slope_inter} show the best-fit 
$Slope$ and $Intercept$ as functions of \logten\age, for different \logten\hasb\ bins. 
Overall, the relations are quite flat at the lowest \hasb\ and 
the highest \age, with a slope close to zero for regions with \logten\hasb$\;<38.5$ 
and \age\ older than a few Gyr. The slope increases with \hasb\ at 
fixed \age, and decreases with \age\ at fixed \hasb. The intercept 
behaves differently for young regions with \age$\;\lesssim 1$Gyr and old regions 
with higher \age. For the young regions, the intercept decreases with both 
\age\ and \hasb, while for the old regions, the intercept decreases with \age\ 
in DIG regions, and increases with \age\ in \hii\ regions. 
For DIG regions, the small $Slope$ suggests a poor correlation 
between \ebvstar\ and \ebvgas.
In \hii\ regions, the $Intercept$ 
is relatively small, close to zero in \hii\ regions with \logten\hasb$\;>39.5$. 
The $Slope$ is thus approximately equal to \ebvratio\ for these \hii\ regions, 
with a value comparable to the average of \ebvratio$\;\approx 0.44$, 
but varying with both \age\ and \hasb.

The lower two panels of \autoref{fig:slope_inter} show the standard deviation of 
the ionized gas regions around the best-fit relation, and the rms of their relative 
difference from the best-fit relation.
The scatter of the ionized regions around the \ebvstar\ versus \ebvgas\ relations 
depends on \hasb\ and \age\ in a way similar to the dependence of the intercept
on these two parameters. Overall, the scatter is smaller than 0.1 mag in all cases, 
and is relatively large for DIG regions of young ages (\hasb$\;\lesssim 10^{39}$\hasbunit\ and 
\age$\;\lesssim 1$Gyr) and for high-\hasb\ \hii\ regions of old ages 
(\hasb$\;\ga10^{40}$\hasbunit\ and \age$\;\ga1$Gyr), ranging from 0.05 mag to 0.1 mag. 
The rms of the relative difference from the best-fit relations decreases 
with increasing \hasb, and it is smaller than 50\% at ages younger than $\sim1$Gyr 
with weak dependence on \age. At older ages, the rms is only 20-30\% 
at the highest \hasb\ ($>10^{40}$\hasbunit), but increases rapidly towards 
lower \hasb. The large rms at the low \hasb\ again reflects
the poor correlations between the stellar and gas attenuation in the DIG regions. 

\begin{deluxetable}{rcccccc}
    \label{tab:fits}
    \tablecaption{Best-fit coefficients in \autoref{eqn:slope_age} and 
    \autoref{eqn:intercept_age}
    which describe the slope and intercept of the stellar-to-gas attenuation linear relation
    as a function of luminosity-weighted stellar age.}
    \tablehead{
      \colhead{$\log_{10}$\hasb} & \colhead{$a_0$} & \colhead{$b_0$} & \colhead{$a_1$} & \colhead{$b_1$} &
      \colhead{$a_2$} & \colhead{$b_2$}
    }
    \startdata
    (, 38.5]     & -0.130 & 1.253 & -0.211 & 1.968 & -0.101 & 0.996 \\
    (38.5, 39.0] & -0.174 & 1.692 & -0.168 & 1.553 & -0.058 & 0.582 \\
    (39.0, 39.5] & -0.234 & 2.286 & -0.110 & 1.012 & 0.0006 & 0.041 \\
    (39.5, 40.0] & -0.257 & 2.576 & -0.049 & 0.452 & 0.062 & -0.520 \\
    (40.0, 40.5] & -0.271 & 2.766 & 0.007 & -0.052 & 0.118 & -1.024 \\
    (40.5, ]     & -0.360 & 3.527 & 0.097 & -0.812 & 0.207 & -1.783 \\
    \enddata
%    \tablecomments{aaa}
\end{deluxetable}

As can be seen from \autoref{fig:slope_inter}, both $Slope$ and $Intercept$ 
depend on \logten\age\ in a rather simple way. For a given \hasb\ range, 
we model $Slope$ and $Intercept$ as function of \logten\age\ using  
\begin{equation}\label{eqn:slope_age}
    Slope = a_0\times\log_{10}t_L + b_0,
\end{equation}
and 
\begin{eqnarray}\label{eqn:intercept_age}
    Intercept=
    \left\{
    \begin{array}{ll}
        a_1\times\log_{10}t_L + b_1, \mbox{if}\; \log_{10}t_L\leq8.8 \\
        a_2\times\log_{10}t_L + b_2, \mbox{if}\; \log_{10}t_L>8.8
    \end{array}
    \right.
\end{eqnarray}
The best-fit relations are plotted in \autoref{fig:slope_inter} as the dashed lines, 
and the best-fit values of $a_0$, $b_0$, $a_1$, $b_1$, $a_2$ and $b_2$ are listed 
in Table~\ref{tab:fits} for the different intervals of \hasb. 

\section{Discussion}
\label{sec:dis}

\subsection{Comparison with previous MaNGA-based studies}

Recently, \citet{2020ApJ...888...88L} and  \citet{2020MNRAS.495.2305G} have used the MaNGA data to
study the dust attenuation in nearby galaxies. In this 
subsection we compare our work with both studies. 
We note that, 
in a recent study based on MaNGA, \citet{2021MNRAS.501.4064R} 
also examined the correlation between \ebvstar\ and \ebvgas,
but focusing on AGN host galaxies and a control sample of 
star-forming galaxies. Here we will not make comparisons 
with this study, given the difference in sample selection.

\subsubsection{Comparison with \citet{2020ApJ...888...88L}}

\cite{2020ApJ...888...88L} used the dust attenuation parameters 
and spectroscopic properties from the MaNGA value-added catalog 
produced by applying the {\tt Pipe3D} pipeline \citep{2018RMxAA..54..217S} 
to the SDSS DR15 that contains $\sim 4,800$ datacubes from MaNGA. 
{\tt Pipe3D} performed spatial binning for each datacube 
and fitted the binned spectra with a set of simple stellar population 
models base \citep{2005MNRAS.357..945G,2010MNRAS.404.1639V,2013A&A...557A..86C}. 
The dust attenuation 
curve of \citet{1989ApJ...345..245C} and a selective extinction 
of $R_V=3.1$ were assumed, and a stellar attenuation was  
determined as one of the free parameters in the spectral fitting. 

The authors examined the correlation between \ebvstar\ and \ebvgas, 
and the dependence on a variety of galaxy properties, both regional 
and global. In particular, they found that both the Pearson and Spearman 
rank correlation coefficients increase rapidly with \logten\hasb, 
from $<0.1$ at \logten\hasb$\;<39$ up to $\sim0.8$ at \logten\hasb$\;>40.5$ 
(see their figure 2). This dependence is also seen in our results, 
e.g. in \autoref{fig:ebvstar_ebvgas_habins}.
We further find that the scatter in the relation between 
\ebvstar\ and \ebvgas\ at fixed \hasb\ can well be explained by 
the stellar age, which can also be seen from \autoref{fig:ebvstar_ebvgas_habins}. 
Once limited to a narrow range of \logten\age, the two attenuation 
parameters are linearly correlated, although the slope, intercept 
and scatter of the relation vary with \hasb\ and \age. We  
show that the variations can be described quantitatively by 
simple functions (see \autoref{fig:slope_inter} and Table~\ref{tab:fits}).

\citet{2020ApJ...888...88L} payed particular attention to \ebvratio, 
finding it to be larger in DIG regions than 
in \hii\ regions, which is also found from our data (see \autoref{fig:dependence_main}).
According to the Spearman rank correlation analysis, they found that 
\ebvstar/\ebvgas\ is the most strongly correlated with \logten(\niisii),
with a correlation coefficient close to $\rho=-0.5$, and is the least strongly 
correlated with $D_n4000$, with $\rho\sim-0.1$ (see their figure 3). 
In contrast, we find that, among other properties, the luminosity-weighted 
stellar age \age\ is the property that shows the strongest correlation with 
\ebvstar/\ebvgas, with $\rho=-0.63$ (see \autoref{fig:dependence_main}).
The Spearman correlation coefficient is $\rho=-0.38$ between \ebvstar/\ebvgas\ 
and $D_n4000$, not as weak as found in \citet{2020ApJ...888...88L}. 
The different correlation coefficient for $D_n4000$ may be attributed 
to the different data products used in the two studies. In fact, we 
have repeated the analysis of the correlation between \ebvstar/\ebvgas\  
and $D_n4000$ using measurements from the {\tt Pipe3D} VAC, and 
found a similarly weak correlation. 

In Section \ref{sec:res_drive} we have discussed the dominant role of 
the stellar age in driving \ebvstar, as well as its difference and ratio relative to
\ebvgas. This analysis was missing in \citet{2020ApJ...888...88L} for
two reasons. First, the authors did not consider the stellar age directly, but adopted 
$D_n4000$ as an indicator of it. The $D_n4000$ is indeed a good indicator
of stellar age, but sensitive only to populations younger than 1-2Gyr 
\citep{2003MNRAS.344.1000B, 2003MNRAS.341...33K}. Second, 
our measurements of \ebvstar\ are more reliable and accurate than 
in previous studies,  because we use the new technique of measuring stellar 
attenuation developed in \citetalias{2020ApJ...896...38L}. As shown in Appendix B of \citetalias{2020ApJ...896...38L}, 
our method can obtain \ebvstar\ before the stellar population 
is modeled, thus significantly alleviating the degeneracy between 
dust attenuation and stellar population properties, such as stellar age and metallicity. 
We have used data from the {\tt Pipe3D} VAC and found the Spearman 
rank correlation coefficient to be $\rho=-0.3$ between \ebvratio\ and \logten\age.
Thus, the dominant role of the stellar age would be missed 
in our analysis if the stellar attenuation and age measurements 
were not improved substantially. 

We find that the Spearman correlation coefficient between \ebvratio\ 
and \niisii\ is $\rho=-0.36$ (see \autoref{fig:dependence_main}), 
weaker than the coefficient of $\rho\sim -0.5$ found in \citet{2020ApJ...888...88L}. 
Again, using the {\tt Pipe3D} data we obtain $\rho=-0.49$, in good 
agreement with their result. Assuming the relevant emission lines are 
measured reasonably well in both studies, we conclude that the different 
correlation coefficients for the relation of \ebvratio with \niisii\ 
are caused mainly by the different measurements of \ebvstar. 
Our \ebvstar\ measurements are expected to be more reliable for the reasons  
mentioned above. On the other hand, we find that \niisii\ is indeed 
one of the dominant parameters for \ebvgas, 
as discussed in Section \ref{sec:res_drive}, and drives 
the gas attenuation in \hii\ regions.
Furthermore, we confirm the finding 
of \citet{2020ApJ...888...88L} that the ionization level of the gas as 
indicated by \oiiioii\ cannot be simply neglected, and we find that 
this parameter needs to be considered only in \hii\ regions with low 
ionization, \logten(\oiiioii)$\;\lesssim-0.8$.

To conclude, comparing with \citet{2020ApJ...888...88L}, 
we confirm the following results:
\begin{itemize}
\item The relation between \ebvstar\ and \ebvgas\ is stronger for 
more active \hii\ regions;
\item Local physical properties such as metallicity and ionization level 
play important roles in determining the dust attenuation. 
\end{itemize}
In addition we obtain the following new results:
\begin{itemize}
\item Stellar age is the driving factor for \ebvstar,
and consequently for \ebvdelta\ and \ebvratio; 
\item At fixed \hasb, the stellar age is linearly correlated with \ebvratio\ at all ages;
\item Gas-phase metallicity and ionization level are important, but only
for the attenuation in \ebvgas.  
\end{itemize}

\begin{figure}
    \epsscale{1.1}
    \plotone{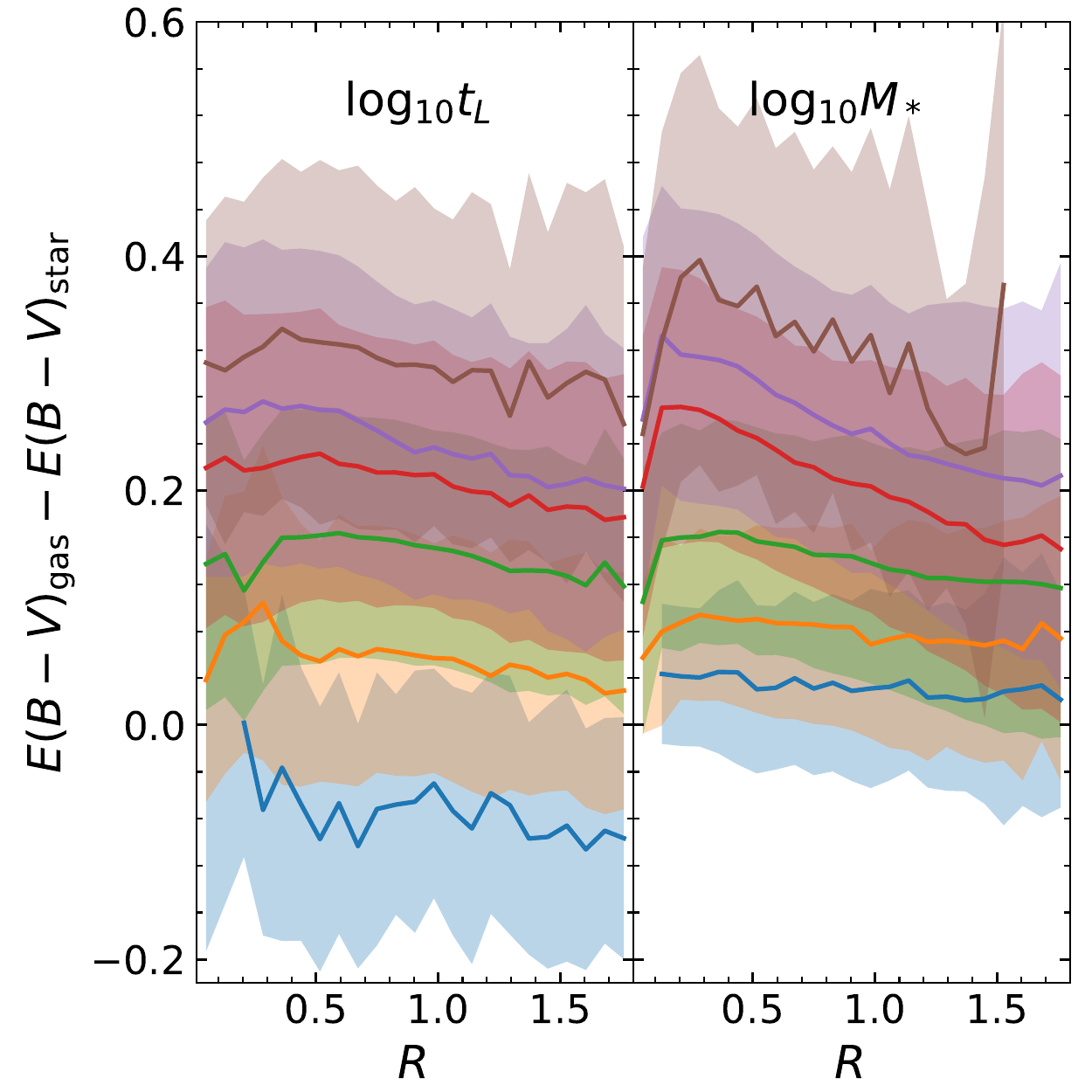}
    \caption{The median radial profiles of the gas-to-stellar attenuation 
    difference for ionized gas regions with different \logten\age\ 
    (left panel) or \logten$M_\ast$ (right panel). The shaded regions 
    indicate the $1\sigma$ scatter of the regions around the median 
    profile. }
    \label{fig:greener_extra}
\end{figure}

\subsubsection{Comparison with \citet{2020MNRAS.495.2305G}}

\cite{2020MNRAS.495.2305G} have recently studied the spatially resolved 
dust attenuation in 232 spiral galaxies from the eighth MaNGA Product Launch 
(MPL-8) with morphological classifications from Galaxy Zoo:3D \citep{2020IAUS..353..205M}. 
The stellar dust attenuation was measured using the full-spectrum stellar
population fitting code {\tt STARLIGHT} \citep{2005MNRAS.358..363C, 2018MNRAS.480.4480C}, 
with SSPs from the E-MILES \citep{2010MNRAS.404.1639V,2016MNRAS.463.3409V} templates 
and a \citet{2000ApJ...533..682C} attenuation curve 
with $R_V=4.05$, while the gas dust attenuation 
was measured from the Balmer decrement assuming $R_V=3.1$.

The authors found that dust attenuation increases with 
$\Sigma_{\rm SFR}$ (SFR surface density),
consistent with the positive correlations of both \ebvstar\ and 
\ebvgas\ with \hasb\ seen from \autoref{fig:dependence_main},
and that the ratio \ebvratio\ decreases with \mass, 
which is also seen from our data. 
They found that both \ebvstar\ and \ebvgas\ decrease as one goes 
from the galactic center outwards, which is also seen from our 
data. They also found that the ratio \ebvratio\ has no correlation 
with the distance from galactic center ($R$), while 
our data show a weak correlation between \ebvratio\ and $R$, with 
$\rho=0.25$ for all the ionized gas regions and $\rho=0.24$ for \hii\ regions. 
In addition, \cite{2020MNRAS.495.2305G}
found a high concentration of birth clouds near the galactic center, 
as indicated by a negative radial  gradient in \ebvdelta. 
As can be seen from \autoref{fig:greener_extra} (the right panel), our data
also show such negative radial profiles at fixed stellar mass and the 
gradient is stronger at higher mass. 
We find, however, that the gradient becomes rather weak 
when the ionized gas regions are limited to narrow ranges of 
\logten\age, as shown in the left panel of the same figure. 
Therefore, the higher \ebvgas$-$\ebvstar\ at smaller $R$ should be 
attributed to the positive correlation of \ebvgas$-$\ebvstar\ with 
stellar age (see \autoref{fig:dependence_main}), given that 
the stellar populations in the inner regions of galaxies are 
typically older than those in the outer regions. 
We notice from \autoref{fig:greener_extra} 
a weak upturn within $\sim0.5R_e$ in the 
\ebvdelta\ of the youngest regions. This may
be explained if birth clouds are concentrated in the 
central regions of galaxies, as suggested by \citet{2020MNRAS.495.2305G},
because very young stars are mostly embedded in birth clouds. 
However, this result should not be overemphasized, given that the 
upturn is weak and the noise level is high. 
We will come back to this in the future. 

\subsection{Implications of ionizing source for DIG}\label{sec:dis_dig}

We find that in DIG regions, \ebvstar\ is weakly 
correlated with \ebvgas\ and in many cases \ebvstar\ is larger than \ebvgas.
The behavior is very different from that for \hii\ regions.  
\hii\ regions are mostly located below the 1:1 relation in the \ebvstar\ versus \ebvgas\ 
diagram and follow the \ebvstar$\;=0.44$\ebvgas\ relation on average (\autoref{fig:ebvstar_ebvgas_habins}).
For DIG regions, most of the scatter in the relation between \ebvstar\ 
and \ebvgas\ can be explained by the variance in stellar age, as shown in 
Section \ref{sec:joint_dependence} and \autoref{fig:slope_inter}. 

The different behaviors of the dust attenuation between DIG and \hii\ regions
imply that the emission in the two classes are produced by different mechanisms. 
The physical origin of the DIG is not clear. A variety of ionizing sources 
have been proposed for the DIG, including supernova shocks, 
cosmic rays, leaky radiation from \hii\ regions, and hot evolved stars
\citep[e.g.,][]{1992ApJ...400L..33R,2009RvMP...81..969H,Yan-Blanton-2012,
2014MNRAS.440.3027B,2015MNRAS.447..559B,2017MNRAS.466.3217Z}.
Using MaNGA data, \cite{2017MNRAS.466.3217Z} examined the
ionization and metallicity diagnostics of DIG-dominated regions, finding 
that DIG has a lower ionization level than \hii\ regions, which can 
enhance $\rm [NII]/H\alpha$, $\rm [SII]/H\alpha$, $\rm [OII]/H\alpha$
and $\rm [OI]/H\alpha$, but reduce $\rm [OIII]/H\beta$. As the
leaky \hii\ region model cannot produce LINER-like emission commonly 
seen in DIG regions, the authors suggested that hot evolved stars could 
be a major ionization source for DIG, at least in the following two cases: 
(1) low-surface brightness regions that are located far from \hii\ regions, 
such as regions at large vertical height, and (2) post-starburst galaxies 
and quiescent galaxies.

\begin{figure}
    \epsscale{1.1}
    \plottwo{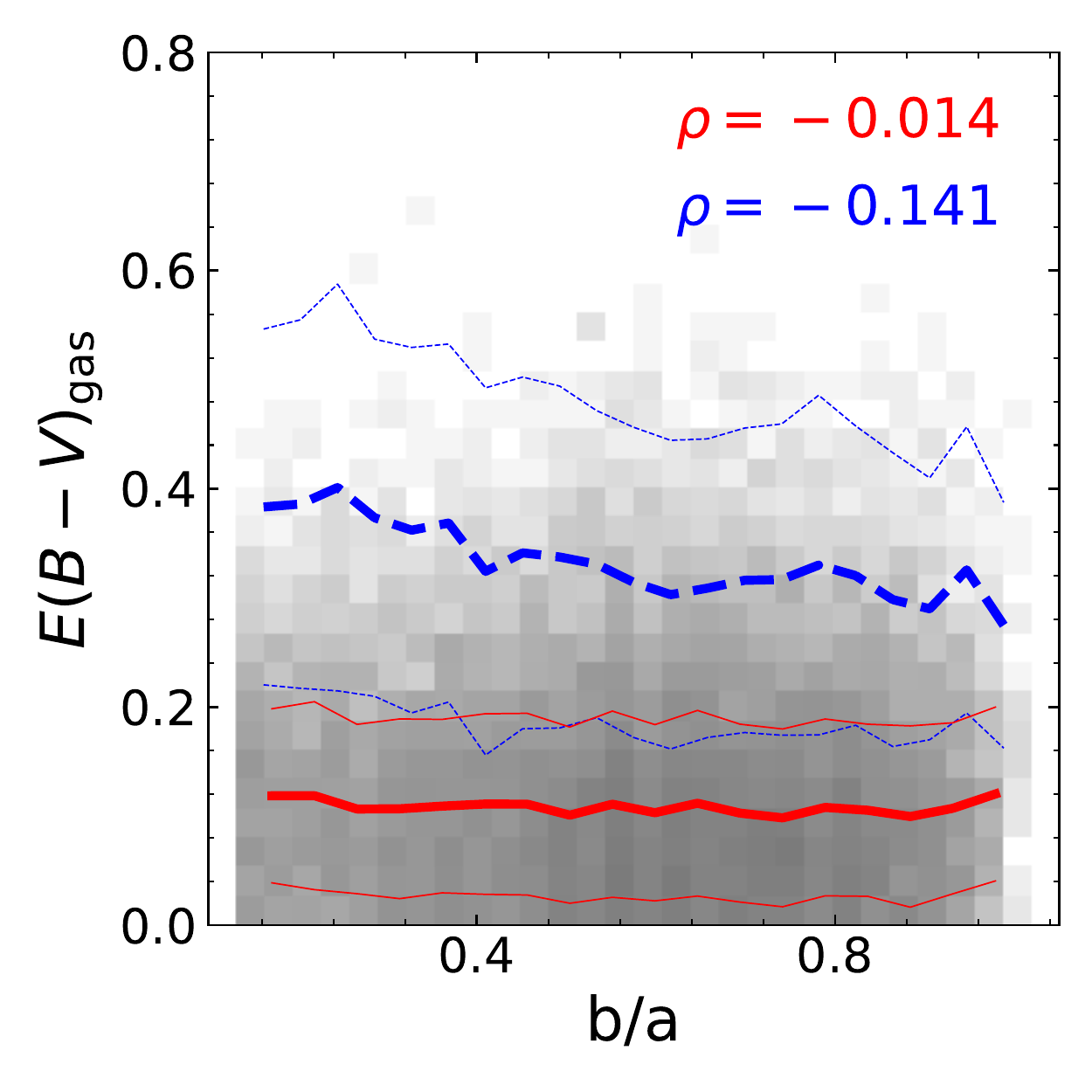}{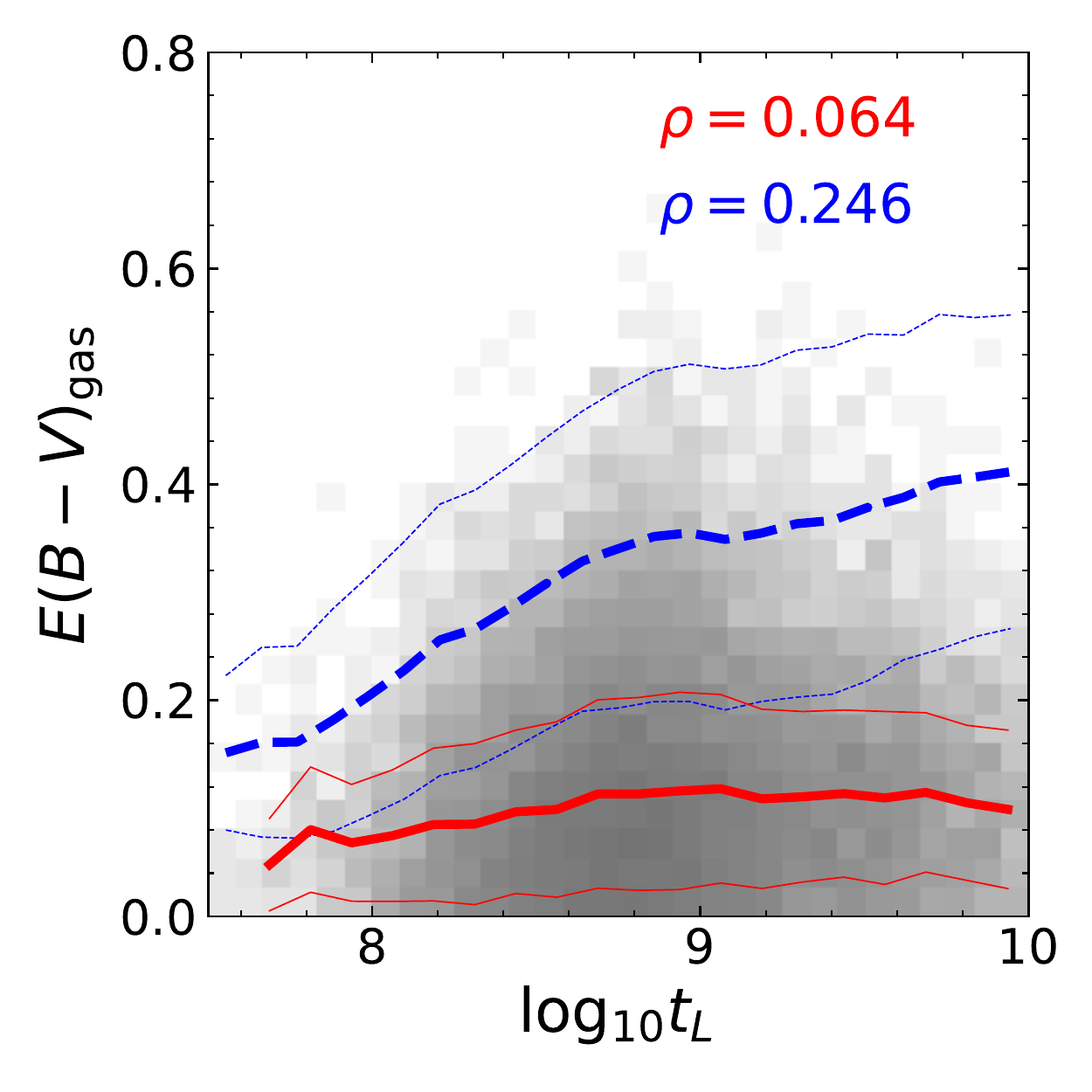}
    \caption{\ebvgas\ 
    as function of \ba\ (left) and \logten\age\ (right)
    for DIG regions with \hasb$\;<10^{38.5}$\hasbunit\ (red lines)
    and \hii\ regions with \hasb$\;>10^{39}$\hasbunit\ (blue lines).
    The number density distribution of the DIG regions is plotted as the 
    gray-scale background. The solid lines are the median 
    relations and the dashed lines indicate the $1\sigma$ scatter 
    of the DIG or \hii\ regions around the median relation.
    The Spearman rank correlation coefficients are indicated 
    for both types of regions, with red for DIG and blue for \hii\ regions.}
    \label{fig:dig_t_L_ba}
\end{figure}

Our data support the suggestion of \citet{2017MNRAS.466.3217Z}. 
In \autoref{fig:dig_t_L_ba} we plot \ebvgas\ as a function of 
both \ba\ and \logten\age, for DIG regions with \hasb$\;<10^{38.5}$\hasbunit. 
Here we use a \hasb\ limit 0.5 dex smaller than the usually adopted 
value to reduce the contamination of the \hii\ regions. 
For comparison, the results for the \hii\ regions with \hasb$\;>10^{39}$\hasbunit\ 
are also plotted, as blue solid/dashed lines. Overall, the DIG and \hii\ 
regions behave quite differently, in the sense that 
\ebvgas\ depends on both \ba\ and \logten\age\ in \hii\ regions but 
\ebvgas\ shows no dependence on either parameter in DIG regions. 
In addition, the median value of \ebvgas\ in DIG regions is pretty small, 
constantly at $\sim0.1$ mag for all values of \ba\ and \logten\age. 
This weak gas attenuation in DIG regions and its independence of \ba\ 
strongly indicates that the ionizing source for DIG is distributed in the 
outskirts of galaxies, either at large vertical heights or at large radial distances, 
or both. Ionizing photons emerging from these locations can easily escape 
the galaxy without being attenuated much because of the short propagation 
path through the ISM, thus leading to \ebvgas\ values that are small 
and similar in different viewing angles. 

\begin{figure}
    \epsscale{1.1}
    \plottwo{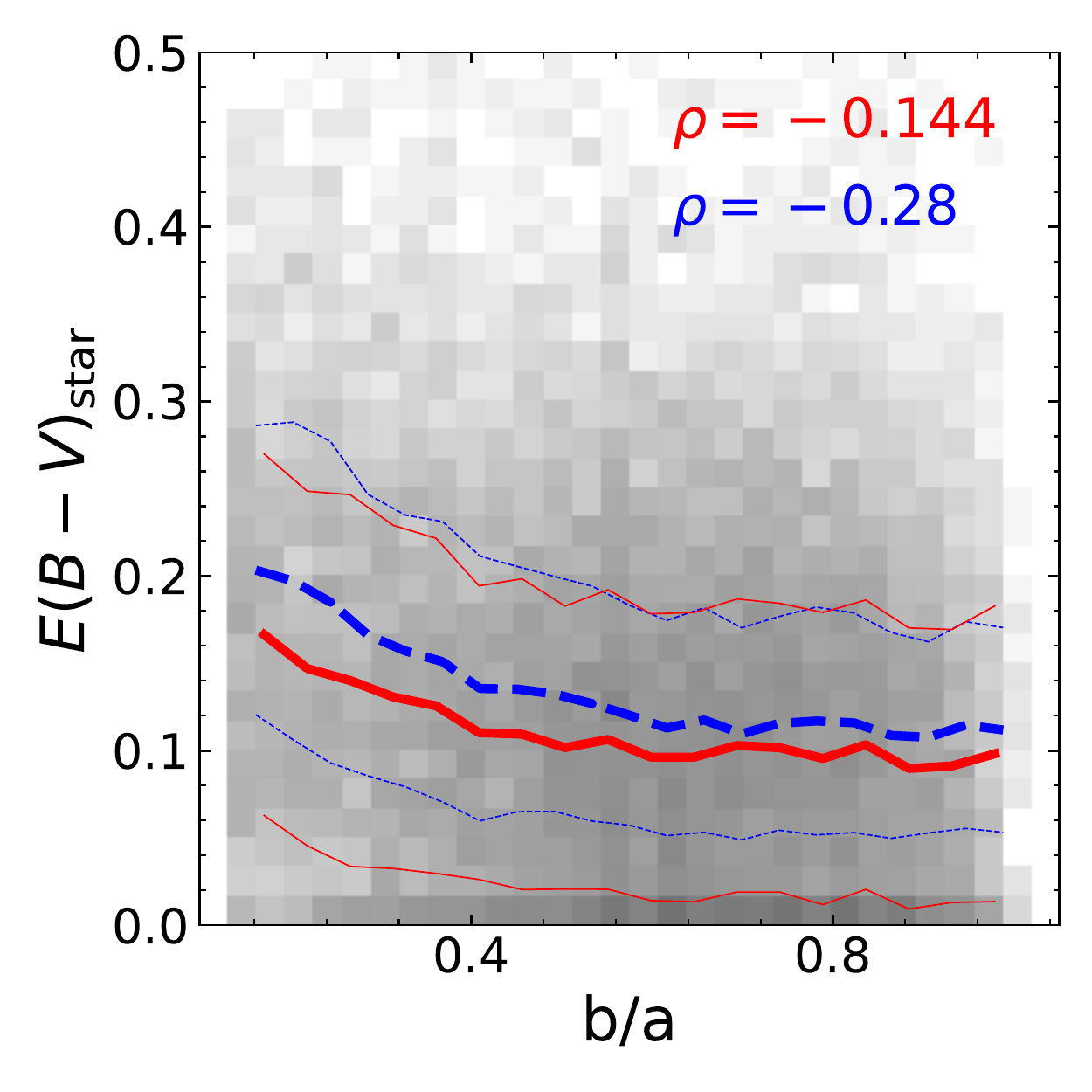}{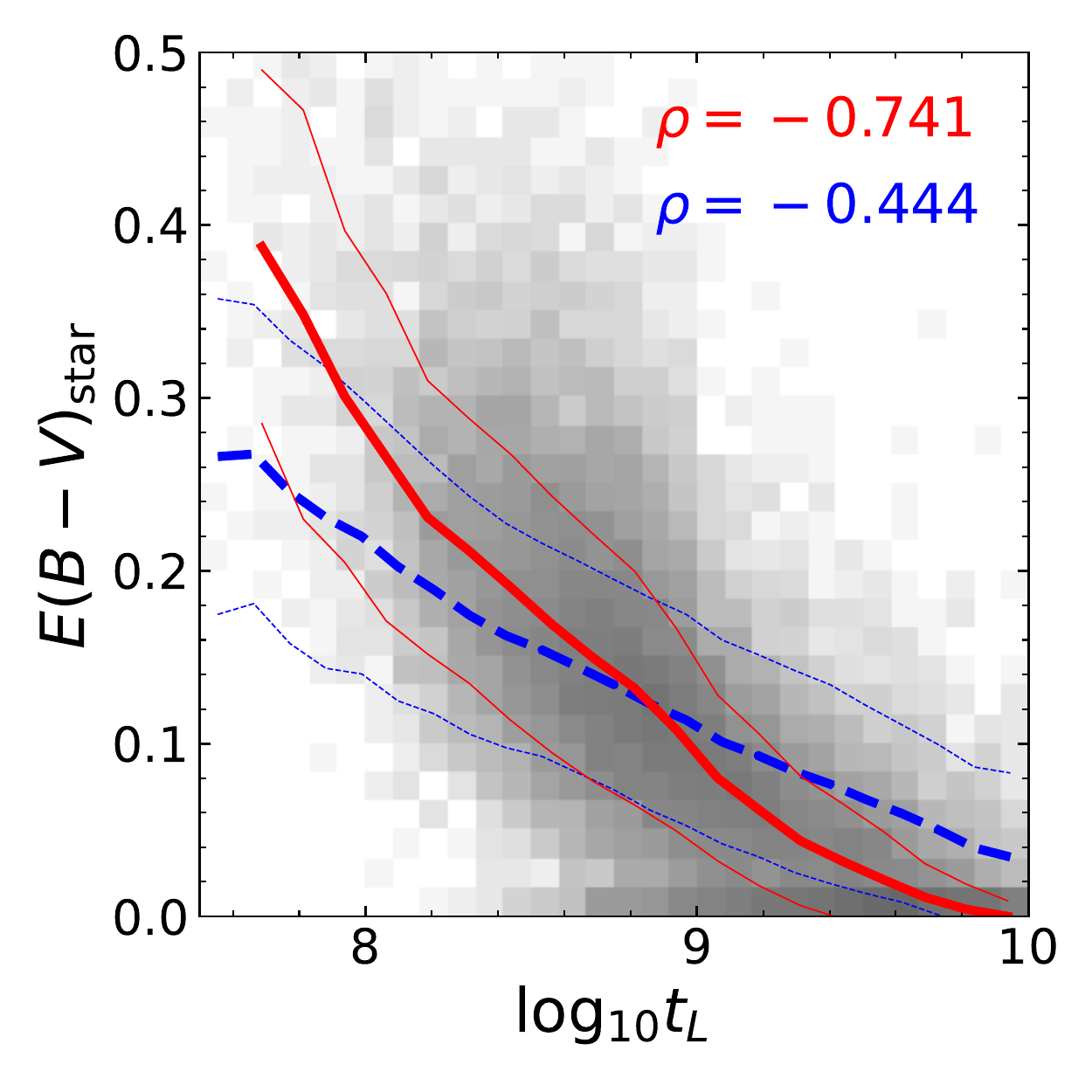}
    \caption{\ebvstar\ 
    as function of \ba\ (left) and \logten\age\ (right)
    for DIG regions with \hasb$\;<10^{38.5}$\hasbunit\ (red lines)
    and \hii\ regions with \hasb$\;>10^{39}$\hasbunit\ (blue lines).
    Symbols/lines are the same as in \autoref{fig:dig_t_L_ba}.}
    \label{fig:dig_t_L_ba_ebvstar}
\end{figure}

\autoref{fig:dig_t_L_ba_ebvstar} shows the stellar attenuation parameter 
\ebvstar\ as a function of \ba\ and \logten\age\ for the DIG and \hii\ regions. 
The two types of regions behave similarly, with \ebvstar\  
decreasing as both \ba\ and \logten\age\ increase. For DIG regions, 
the dependence of \ebvstar\ on \ba\ appears to be contradictory 
to the independence of \ebvgas\ on \ba. The stronger stellar attenuation   
at smaller \ba\ indicates that the non-ionizing continuum photons
emitted from stars must have propagated a longer path than 
the ionizing photons emerging in the outskirt of the galaxy. 
The ionized gas regions are selected from the two-dimensional map of \hasb, 
and so they include emission and attenuation along the line of sight. 
Therefore, the stellar continuum of each DIG region is not emitted from the 
DIG region itself, but is the integration of the attenuated continuum 
photons from all stars in the area covered by the region in question, 
while the ionizing photons are produced in the DIG region.
This conjecture is supported by the similar behaviors of DIG and 
\hii\ regions in the right-hand panel of \autoref{fig:dig_t_L_ba_ebvstar}, 
which shows again the driving role of the stellar age in the stellar attenuation.

\subsection{The dominant role of stellar age and the dust model of 
\citet{2000ApJ...539..718C}}

We find in this paper that the stellar age plays the dominant role in driving 
the correlation of stellar dust attenuation with many other regional/global 
properties of galaxies (see Section \ref{sec:res_drive}). As a result, the stellar 
age together with \hasb\ can explain the properties of \ebvratio\ 
in both \hii\ and DIG regions (see Section \ref{sec:joint_dependence}).  
Among all the correlations considered, the strongest is found between 
\ebvratio\ and \logten\age\, with a Spearman rank correlation coefficient 
$\rho=-0.63$ (\autoref{fig:dependence_main}). The 
stellar attenuation itself has a strong correlation 
also with the stellar age, with $\rho=-0.569$ for the \ebvstar$-$\logten\age\ relation 
(\autoref{fig:dependence_main}). 
The negative correlation of \ebvstar\ with stellar age is expected, 
because stars are born in environments with enhanced dust contents
\citep{2007MNRAS.375..640P}. Such an age-dependent stellar attenuation
has been considered in some earlier studies \citep[e.g.,][]{2009A&A...507.1793N,
2012A&A...545A.141B,2017MNRAS.472.1372L,2019MNRAS.488.2301T}.

In DIG regions, the ionizing source is distributed in the outskirts of galaxies
while the stellar continuum is dominated by the underlying stars of intermediate/old ages
in the ISM (see Section \ref{sec:dis_dig}). Consequently, ionizing photons are 
less affected by the attenuation than non-ionizing continuum photons. The 
strong correlation between \ebvratio\ and \logten\age\ actually reveals the 
expected correlation between \ebvstar\ and \logten\age\ in DIG regions.

In \hii\ regions, our results are broadly consistent with the dust model of 
\citet{2000ApJ...539..718C}. 
In this model, non-ionizing continuum photons emitted by young stars 
and ionizing photons propagate through both the H{\sc i} envelope of the 
``birth clouds'', where young stars are born and embedded, before 
propagating through the ambient ISM, while emission from long-lived 
(old) stars only propagate through the ISM because of the finite 
lifetime of the stellar birth clouds. As emphasized in \citet{2000ApJ...539..718C},
the finite lifetime of birth clouds is a key ingredient 
for resolving the discrepancy between the attenuation of line and 
continuum photons. In very young \hii\ regions, most of the stars are still 
embedded in ``birth clouds'', so that the stellar attenuation is comparable to 
the gas attenuation. \cite{2007ApJS..173..457B} studied a sample of 
Ultraviolet-luminous galaxies whose \ebvstar\ are comparable to \ebvgas. 
They found that galaxies whose stars formed only recently are 
very young and can be explained by a model where the majority
of stars are within birth clouds. In very old \hii\ regions, where the 
stellar continuum is dominated by old stars in the ISM, both \ebvstar\ 
and \ebvratio\ are low, as seen in \autoref{fig:dependence_main}.
The intermediate-age \hii\ regions may be considered as a mixture 
of both young \hii\ regions and old \hii\ regions, so that 
their \ebvratio\ cover the whole range from zero to unity. 
This is supported by the results shown in \autoref{fig:ebvstar_ebvgas_habins}, 
where younger \hii\ regions tend to have steeper slope, and almost all the 
regions are below the 1:1 relation at \logten\hasb$\;>39.5$.

\begin{figure*}
    \centering
    \fig{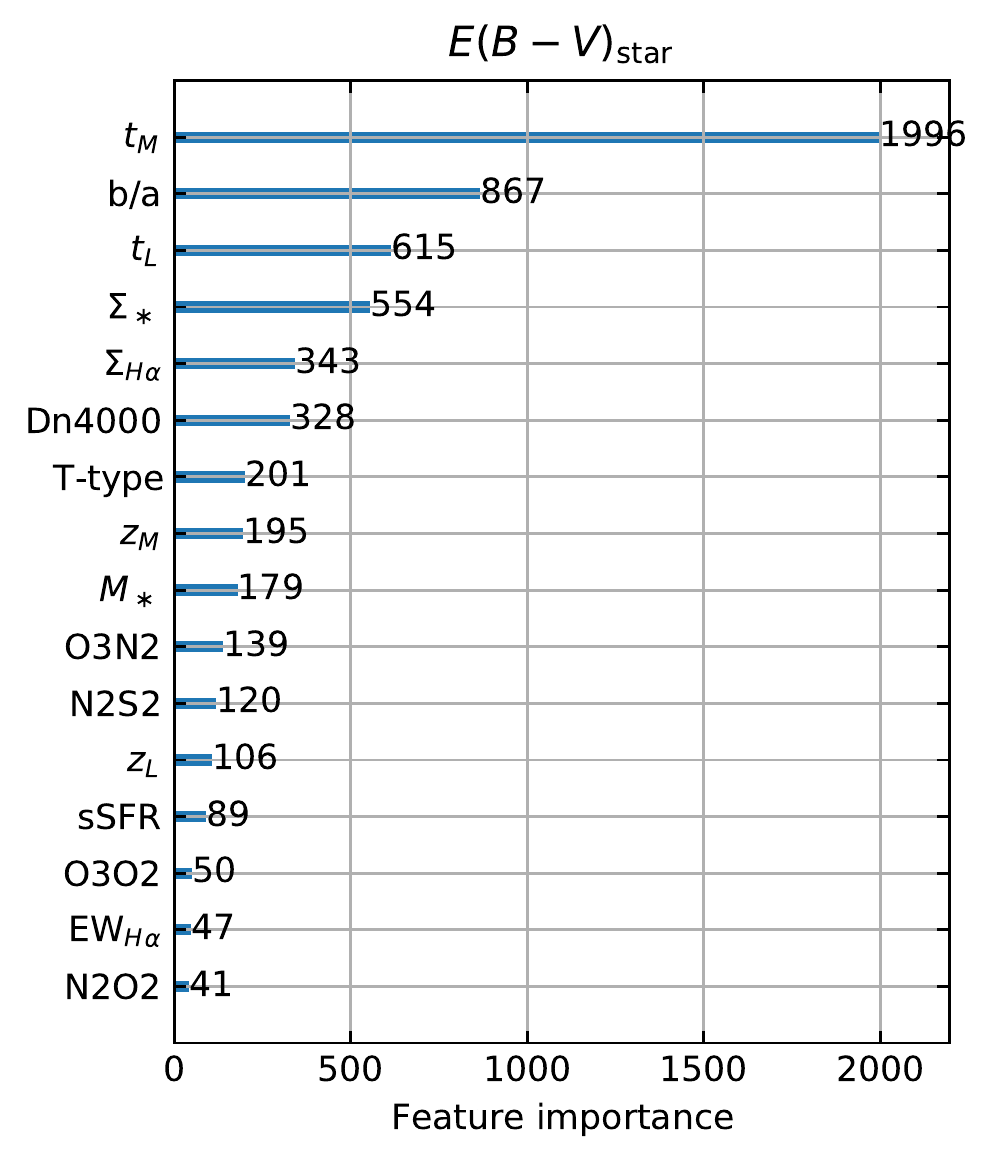}{0.48\textwidth}{(a)}
    \fig{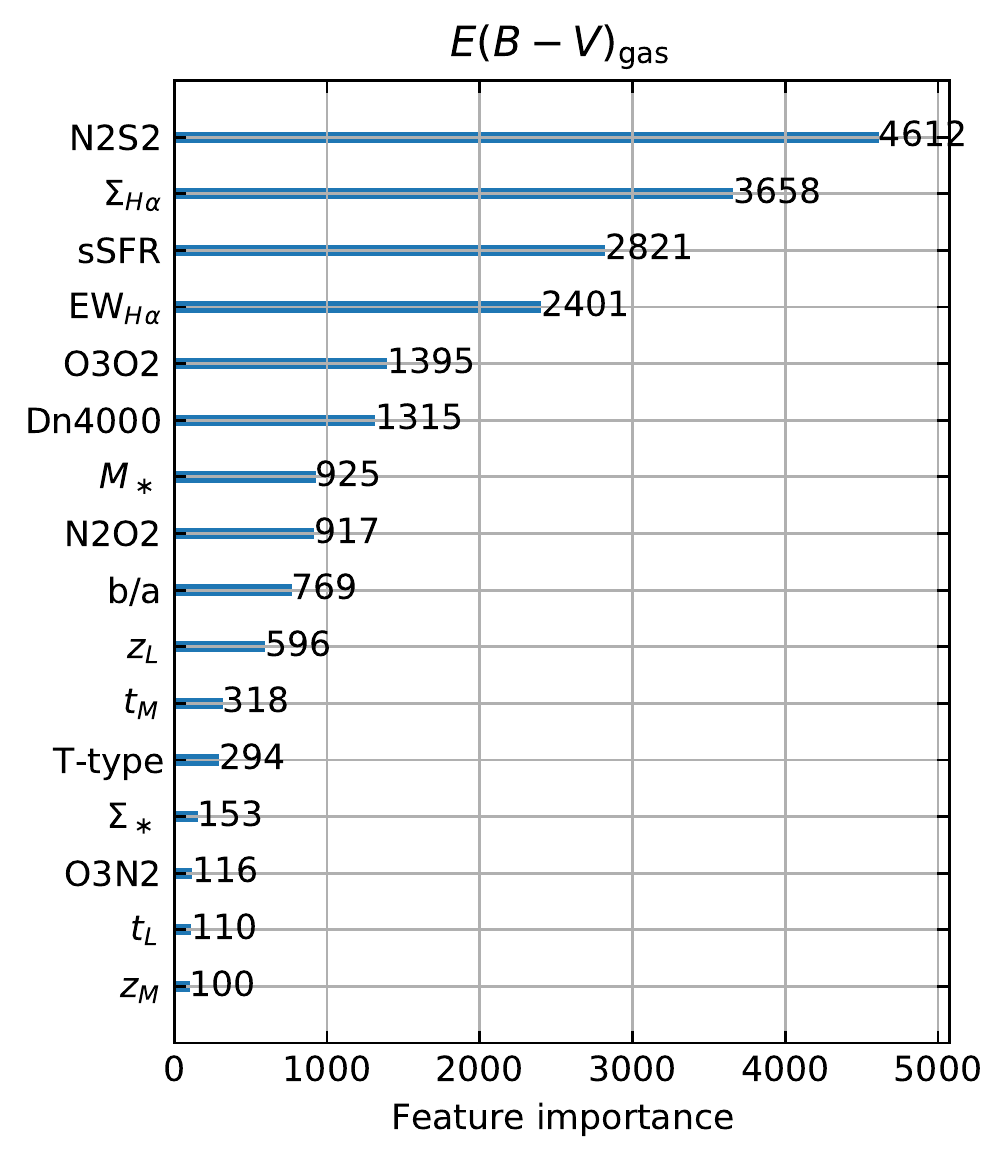}{0.48\textwidth}{(b)}\\
    \fig{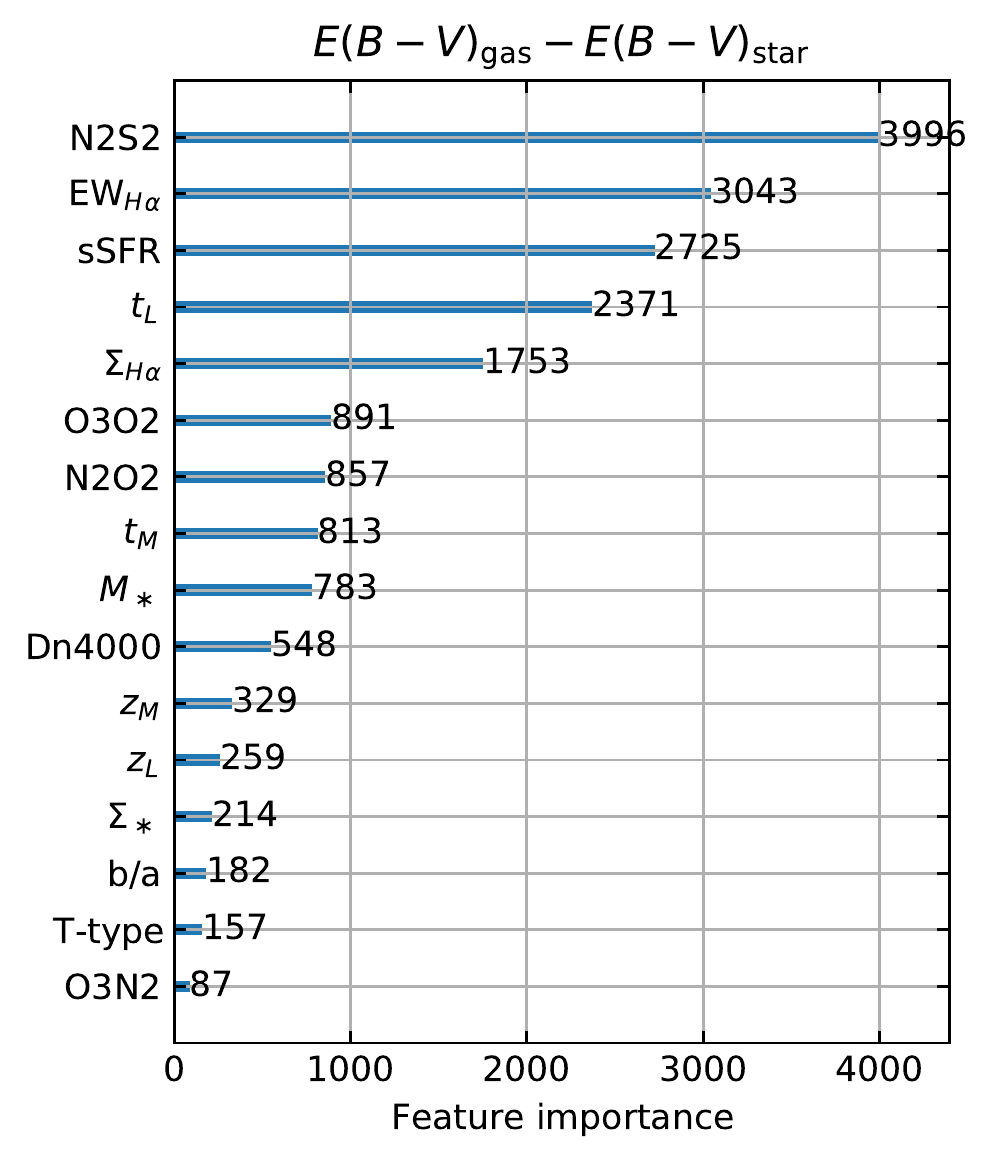}{0.48\textwidth}{(c)}
    \fig{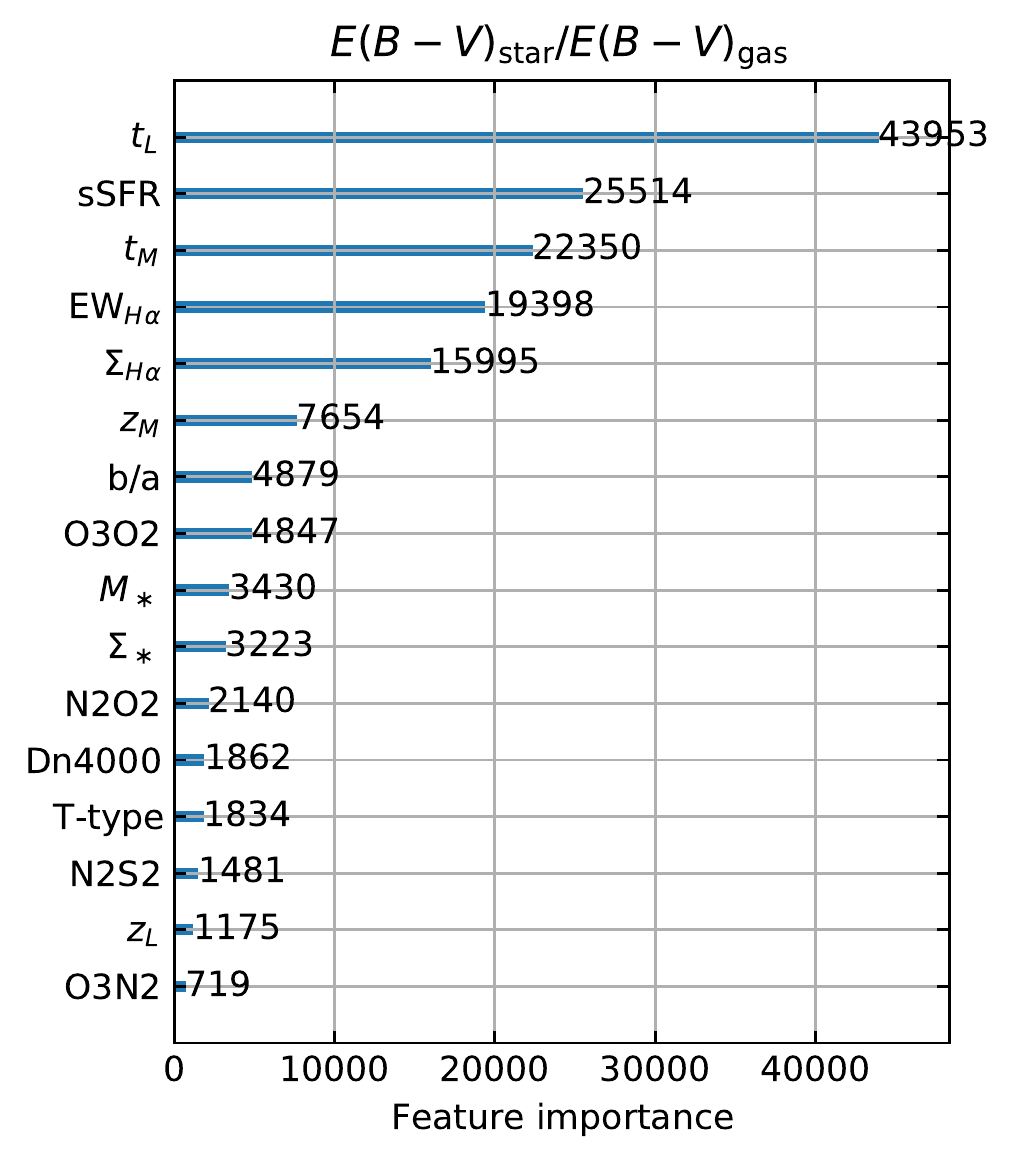}{0.48\textwidth}{(d)}
    \caption{
    Feature importance of the 16 regional/global properties is shown 
    respectively for the four dust attenuation parameters (panels a,b,c,d):
    \ebvstar, \ebvgas, \ebvdelta, \ebvratio. 
    N2S2, N2O2, O3O2 represent \niisii, \niioii, \oiiioii.}
    \label{fig:feature_importance}
\end{figure*}

\subsection{Feature importance of the properties based on machine learning}
In Section \ref{sec:res_drive} we have examined the correlation of dust attenuations
with a large number of regional/global properties. The relative
importance of different properties is quantified by the commonly
used Spearman rank correlation coefficient. Nowadays, the 
technique of machine learning has been widely applied to more 
efficiently deal with such big-data problems. Here we have 
also done a machine-learning analysis as a consistency check
on our results. We use \texttt{LightGBM}
\footnote{\url{https://lightgbm.readthedocs.io/en/latest/index.html}},
which is a gradient boosting framework using tree-based learning 
algorithms to evaluate feature importance of the regional/global
properties of our galaxies to the dust attenuation parameters.

\autoref{fig:feature_importance} shows the feature importance 
of all the 16 properties for \ebvstar, \ebvgas, \ebvdelta\ and 
\ebvratio, respectively. Overall, it is encouraging that the results 
are well consistent with our previous analysis based on Spearman
correlation coefficients: stellar age is indeed the most important
property to \ebvstar\ and \ebvratio, and \niisii\ and \hasb\ are the 
most important to \ebvgas. In particular, the mass-weighted age 
($t_M$) shows the highest feature importance to \ebvstar, while 
the luminosity-weighted age ($t_L$) is most important to 
\ebvratio. This is also in good agreement with the Spearman correlation
analysis, where \ebvstar\ has a correlation coefficient of 
$\rho=-0.627$ with $t_M$ (\autoref{fig:dependence_other}) and 
$\rho=-0.529$ with $t_L$ (\autoref{fig:dependence_main}), and
the strongest correlation is found between \ebvratio\ and $t_L$ 
with $\rho=-0.631$ (\autoref{fig:dependence_main}). It is interesting
to note that the $b/a$ shows a high feature importance to 
\ebvstar, falling in between $t_M$ and $t_L$ in panel (a) of the figure.
The Spearman correlation coefficient between \ebvstar\ and $b/a$ is 
relatively low ($\rho=-0.228$; \autoref{fig:dependence_other}), however.
This is probably because the $b/a$ is a global parameter, thus
more sensitive to the overall stellar dust attenuation of the whole 
galaxy, while the Spearman correlation coefficients simply reflect
the average behavior of individual regions. In this regard, the 
machine learning appears to be more powerful in comprehensively
revealing the intrinsic correlations in a complex, high-dimensional dataset.

\section{summary}
\label{sec:summary}

In this paper, we study the relationship between \ebvstar\ and \ebvgas\ 
in both \hii\ regions and DIG regions
of kpc scales, using data of integral field spectroscopy from the MaNGA MPL-9. 
For each region, we stack the original spectra and perform full 
spectral fitting to obtain \ebvstar\ and other properties of the 
underlying stellar populations. We then 
measure emission lines from the starlight-subtracted spectra and 
calculate \ebvgas\ from the Balmer decrement. With these measurements,
we examine the correlations of \ebvstar, \ebvgas, 
\ebvdelta\ and \ebvratio\ with 16 regional/global properties. 
Our main results can be summarized as follows. 
\begin{itemize}
    \item The relation between \ebvstar\ and \ebvgas\ is stronger for 
    more active \hii\ regions.
    \item Stellar age is the driving factor for 
    \ebvstar, and consequently for \ebvdelta\ and \ebvratio\ as 
    well. At fixed \hasb, the stellar age is linearly and negatively 
    correlated with \ebvratio\ at all ages. 
    \item Gas-phase metallicity and ionization level are important for 
    \ebvgas.
    \item The ionizing sources for DIG regions are likely distributed in 
    the outskirt of galaxies. 
    \item The attenuation in \hii\ regions can be well explained 
    by the two-component dust model of \citet{2000ApJ...539..718C}.
\end{itemize}

\acknowledgments

We're grateful to the anonymous referee whose comments have helped 
improve the paper.
This work is supported by the National Key R\&D Program of China
(grant No. 2018YFA0404502), and the National Science
Foundation of China (grant Nos. 11821303, 11973030, 11733004).
MB acknowledges FONDECYT regular grant 1170618.
MG acknowledges the funding from the STFC in the UK.
RR acknowledges the Brazilian funding agencies CNPq, CAPES and FAPERS.

Funding for the Sloan Digital Sky Survey IV has been provided by the
Alfred P. Sloan Foundation, the U.S. Department of Energy Office of
Science, and the Participating Institutions. SDSS-IV acknowledges
support and resources from the Center for High-Performance Computing
at the University of Utah. The SDSS web site is www.sdss.org.

SDSS-IV is managed by the Astrophysical Research Consortium for the
Participating Institutions of the SDSS Collaboration including the
Brazilian Participation Group, the Carnegie Institution for Science,
Carnegie Mellon University, the Chilean Participation Group, the
French Participation Group, Harvard-Smithsonian Center for
Astrophysics, Instituto de Astrof\'isica de Canarias, The Johns
Hopkins University, Kavli Institute for the Physics and Mathematics of
the Universe (IPMU) / University of Tokyo, Lawrence Berkeley National
Laboratory, Leibniz Institut f\"ur Astrophysik Potsdam (AIP),
Max-Planck-Institut f\"ur Astronomie (MPIA Heidelberg),
Max-Planck-Institut f\"ur Astrophysik (MPA Garching),
Max-Planck-Institut f\"ur Extraterrestrische Physik (MPE), National
Astronomical Observatories of China, New Mexico State University,New
York University, University of Notre Dame, Observat\'ario Nacional /
MCTI, The Ohio State University, Pennsylvania State University,
Shanghai Astronomical Observatory, United Kingdom Participation Group,
Universidad Nacional Aut\'onoma de M\'exico, University of Arizona,
University of Colorado Boulder, University of Oxford, University of
Portsmouth, University of Utah, University of Virginia, University of
Washington, University of Wisconsin, Vanderbilt University, and Yale
University.

\appendix

\begin{figure*}
    \epsscale{1.15}
    \plotone{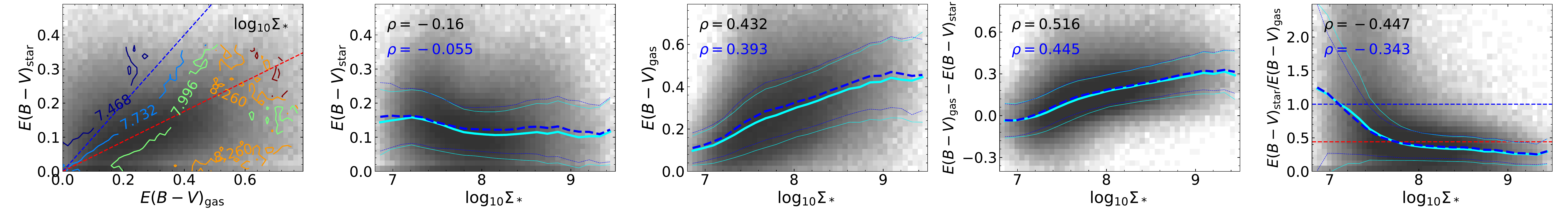}\\
    \plotone{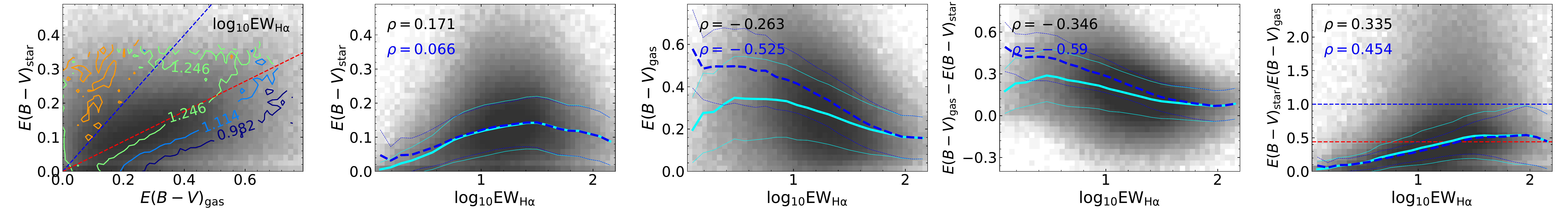}\\
    \plotone{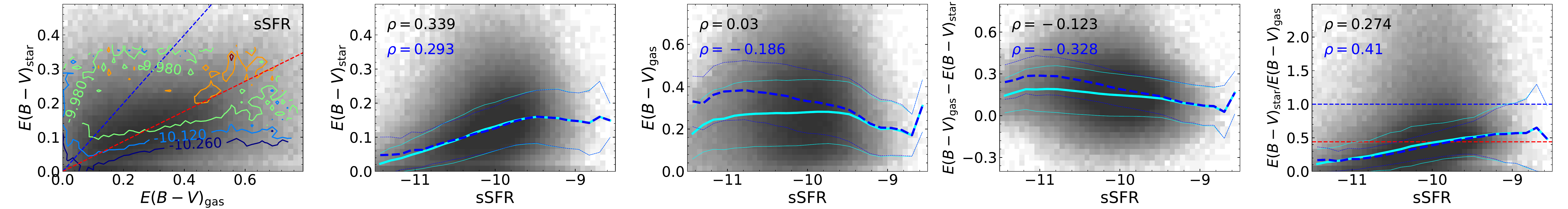}\\
    \plotone{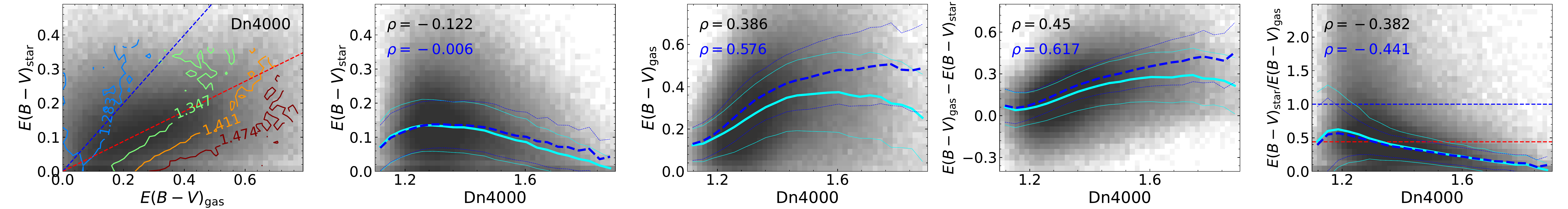}\\
    \plotone{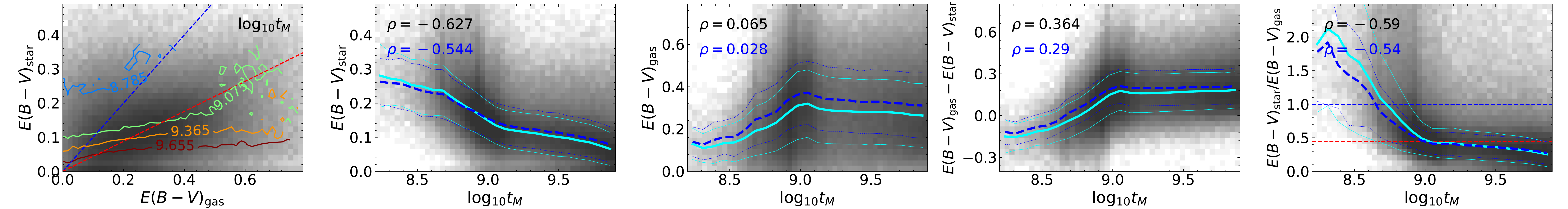}\\
    \plotone{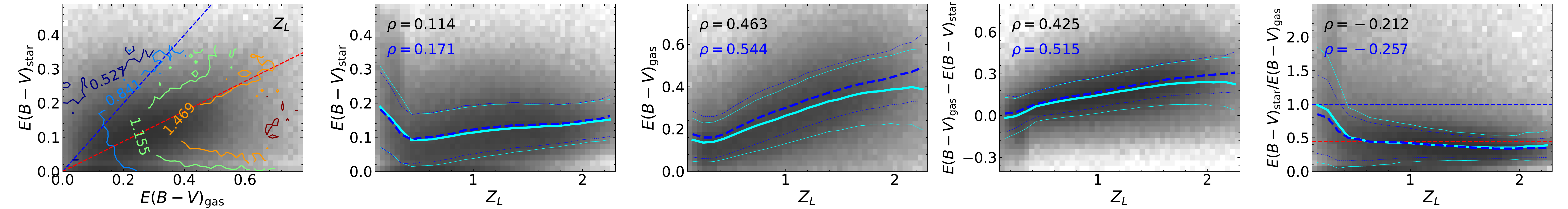}\\
    \plotone{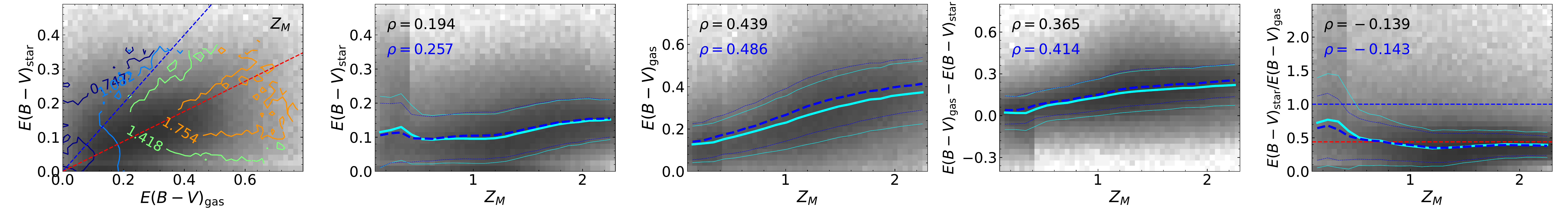}\\
    \plotone{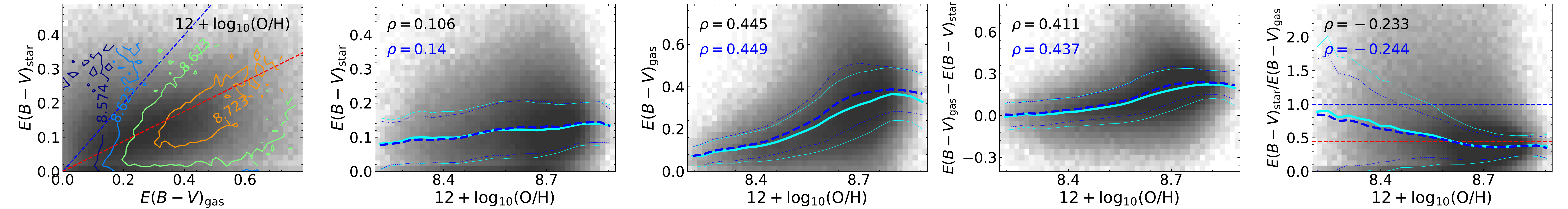}
    \caption{Correlations of \ebvstar\ and \ebvgas\ on 
    regional/global properties (panels from top to bottom):
    \logten\sigmamass, \logten\haew, sSFR,
    $D_n4000$, \logten$t_M$, $Z_L$, $Z_M$,
    12+\logten(O/H), \logten(\niioii), \logten(\oiiioii),
    \logten\mass, $T$-type and \ba.
    Symbols/lines are the same as in \autoref{fig:dependence_main}.
    }
    \label{fig:dependence_other}
\end{figure*}
\addtocounter{figure}{-1}

\begin{figure*}
    \epsscale{1.15}
    \plotone{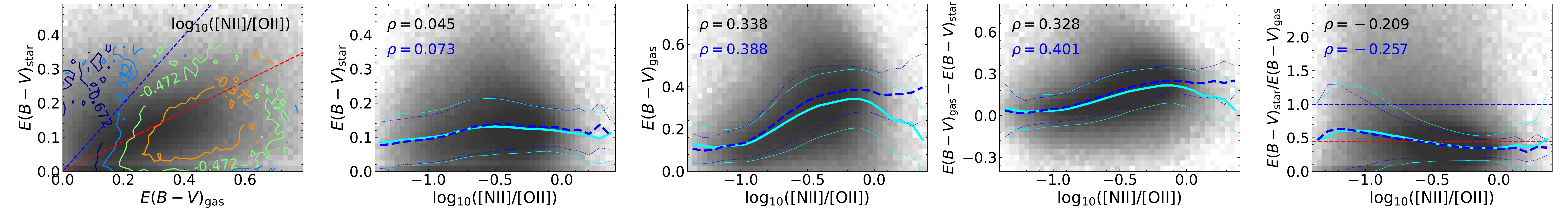}\\
    \plotone{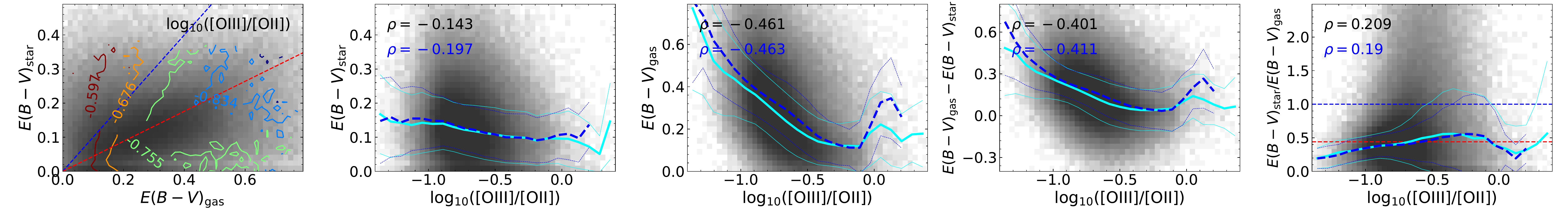}\\
    \plotone{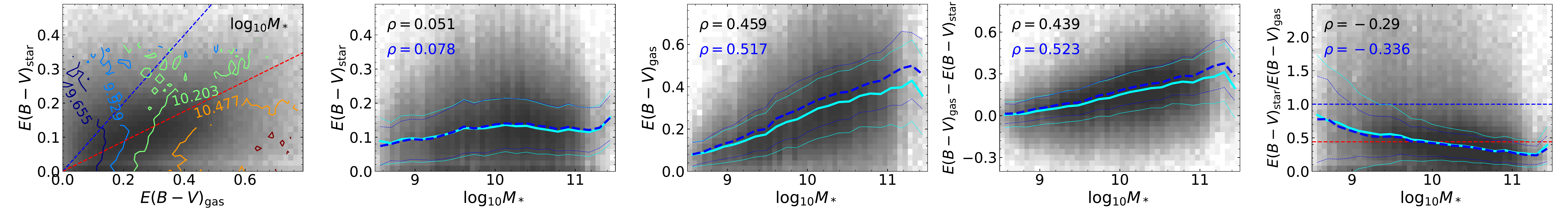}\\
    \plotone{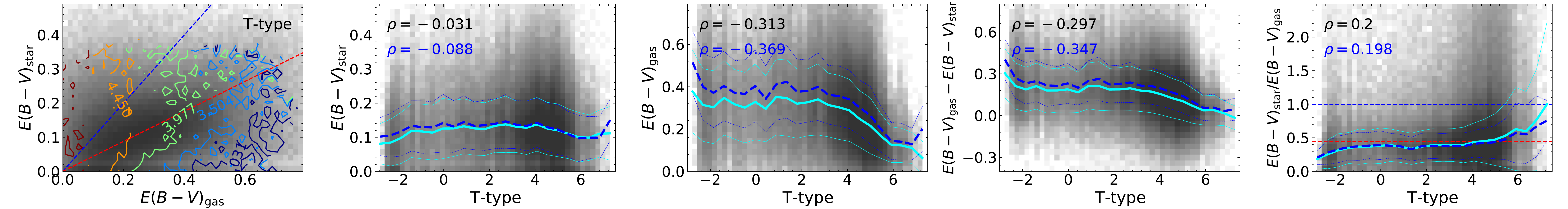}\\
    \plotone{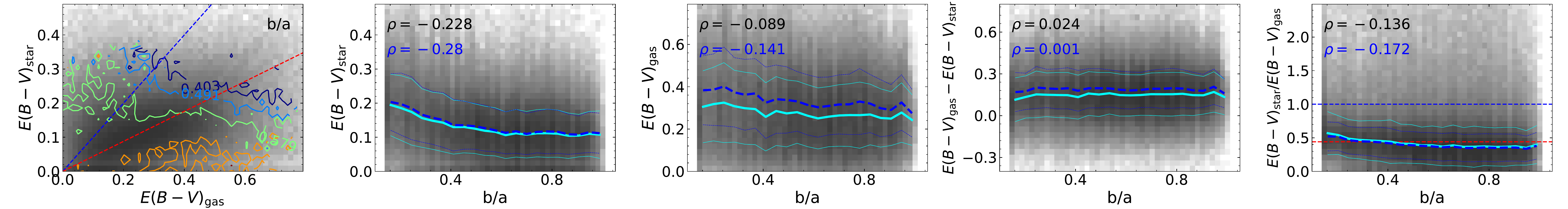}
    \caption{(Continued.)}
    % \label{fig:dependence_other}
\end{figure*}

\section{Correlations of dust attenuation parameters with regional/global properties}
\label{sec:res_phys}

In the paper we have examined the correlation of dust attenuation  
with 13 regional properties and 3 global properties. Among other properties,
\logten\age, \hasb\ and \niisii\ are found to be mostly correlated with 
the dust attenuation parameters, as shown in~\autoref{fig:dependence_main}
and discussed in detail in Section \ref{sec:res_drive}. Here we present the 
results for the rest 13 properties. Following the format of 
\autoref{fig:dependence_main}, \autoref{fig:dependence_other} shows the 
distribution of each property on the diagram of \ebvstar\ versus \ebvgas, 
as well as the correlation of the dust attenuation parameters with 
the properties. The Spearman rank correlation coefficient is indicated 
in each panel, for both the full sample and the subset of \hii\ regions
as selected by \logten\hasb$\;>39$. 
This figure, together with~\autoref{fig:dependence_main}, is 
discussed in depth in the paper. In what follows we briefly mention 
several interesting points that can be obtained from the figure. 

A dichotomy is present in the relation between the attenuation  
and \sigmamass\, in the sense that the ionized gas regions 
can be divided into two classes separated at \logten$\Sigma_\ast\sim 8$. 
This can be seen, for instance, from the top-right panel 
where \ebvratio\ decreases rapidly at \logten$\Sigma_\ast <8$
and levels off at $\sim 0.5$ when \logten$\Sigma_\ast$ exceeds $\sim8$,  
a behavior very similar to the dichotomy in the relation between 
the attenuations and \hasb\ as seen in the middle-right panel of 
\autoref{fig:dependence_main}. This similarity may be understood 
from the main sequence of star forming regions of galaxies, 
in which SFR is strongly correlated with stellar mass
\citep[e.g.,][]{2007ApJ...660L..43N,2010ApJ...721..193P,
2012ApJ...754L..29W,2016ApJ...821L..26C,2017ApJ...851L..24H}.

Our result reveal a positive correlation between \ebvratio\ and sSFR, 
with $\rho=0.274$ (the last panel in the third row).
This is consistent with many previous studies based on 
integrated spectra of galaxies \citep[e.g.,][]{2011MNRAS.417.1760W,
2014ApJ...788...86P,2019ApJ...886...28Q,2019PASJ...71....8K}, 
but is different from that of \citet{2020ApJ...888...88L}, who found 
a negative correlation with $\rho \sim -0.24$.  
In our case, the attenuation parameters depend on \haew\ and 
sSFR in a similar way, which is expected as \haew\ quantifies 
the strength of the \ha\ luminosity (essentially the SFR) 
relative to the stellar continuum luminosity (roughly proportional to $\Sigma_\ast$). 
For regions with \logten\haew$\;\ga 1$ or sSFR$\;\ga -10$, \ebvdelta\ 
is roughly constant at around zero and \ebvratio\ is close to the value 
of 0.44, with weak dependence on \haew\ or sSFR.
At lower \haew\ and sSFR, both attenuation parameters are positively 
correlated with the two properties, with \ebvdelta$\;>0$ and \ebvratio$\;<0.44$. 

Overall, all the dust attenuation parameters are weakly correlated with 
stellar metallicity, except the most metal-poor regions which appear 
to have higher \ebvstar\ and \ebvratio. 
The age $t_M$ shows similar behaviors to \age\ in its correlations 
with the attenuation parameters, but the correlation is much weeker 
for $t_M\ga 1$Gyr. The absence of a strong correlation at $t_M\ga 1$Gyr
is particularly remarkable for \ebvgas. This result may be understood
from the fact that $t_M$ is dominated by low-mass old stars
which contribute little to dust-related processes. 
As an empirical indicator of mean stellar age, the 4000\AA\ break 
is also correlated with the attenuation parameters, 
as expected. Similar to $t_M$, $D_n4000$ shows no correlation 
with \ebvgas\ in old regions with $D_n4000\ga 1.5$, 
which may be explained by the know fact that $D_n4000$ is sensitive 
only to stellar populations younger than 1-2Gyr \citep[e.g.,][]{2003MNRAS.341...33K}.

The gas-phase metallicity \metgas\ is estimated with the O3N2 estimator.
The line ratio of \niioii\ has also been frequently used to estimate 
the gas-phase metallicity, particularly when contribution from DIG 
becomes important \citep{2017MNRAS.466.3217Z}. Both parameters 
show weak correlation with \ebvstar, positive correlation with \ebvgas\ 
and \ebvdelta, and negative correlation with \ebvratio. 
The behaviors of \metgas\ are quite similar to \niisii, but 
with weaker correlations with all the dust attenuation parameters.
The ionization parameter \oiiioii\ correlates with all the dust attenuation 
parameters but in a different way from the metallicity parameters. 
These trends are broadly consistent with 
\citet[][see their figure 8]{2020ApJ...888...88L}.

\autoref{fig:dependence_other} also shows the correlations of the dust 
attenuation parameters of ionized gas regions with three global 
properties of their host galaxies: \mass, $T$-type and \ba. 
The \ebvstar\ shows no obvious correlation with \logten\mass\ 
and $T$-type, but decreases slightly with increasing \ba. 
The latter result is understandable. Smaller \ba\ correspond to 
more inclined disks where starlight has to 
travel through the dusty ISM with a longer path. The \ebvgas\ 
increases with stellar mass, and shows anti-correlations with 
$T$-type and \ba. The \ebvdelta\ shows slightly positive 
correlation with \logten\mass, slightly negative correlation 
with $T$-type, and weak/no correlation with \ba. The 
\ebvratio\ decreases with stellar mass. This result is 
consistent with that in \citet{2020ApJ...888...88L}, who found 
that the globally averaged \ebvratio\ is a linearly decreasing 
function of \logten\mass\ (see their figure 10). The \ebvratio\ 
shows slightly positive correlation with $T$-type and almost 
no correlation with \ba. Generally, the trends of the dust attenuation 
parameters with the global properties are consistent with the
correlations with the regional properties as described above. 

\begin{figure*}
    \centering
    \fig{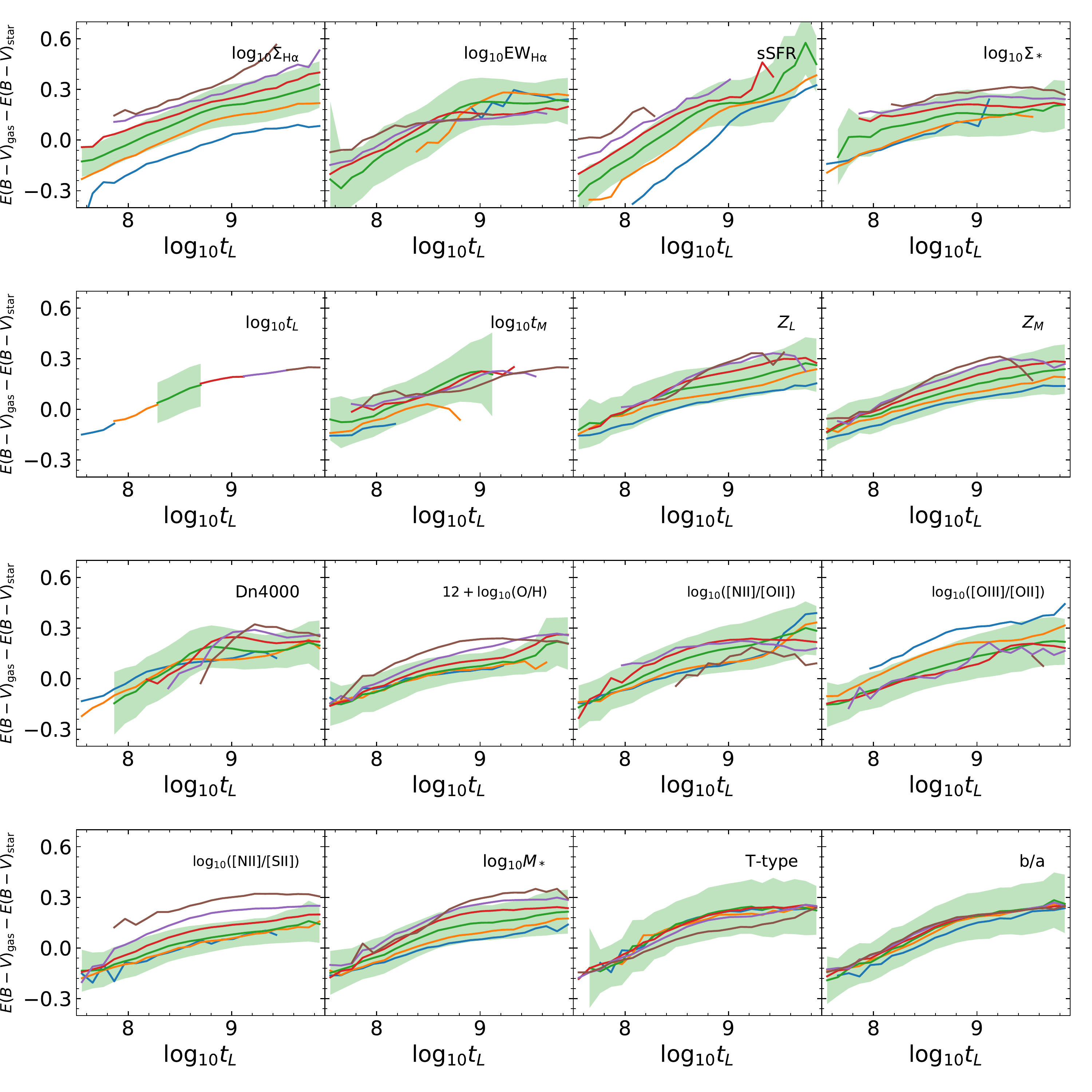}{0.48\textwidth}{(a) \ebvgas$-$\ebvstar\ versus $\log_{10}$\age}
    \fig{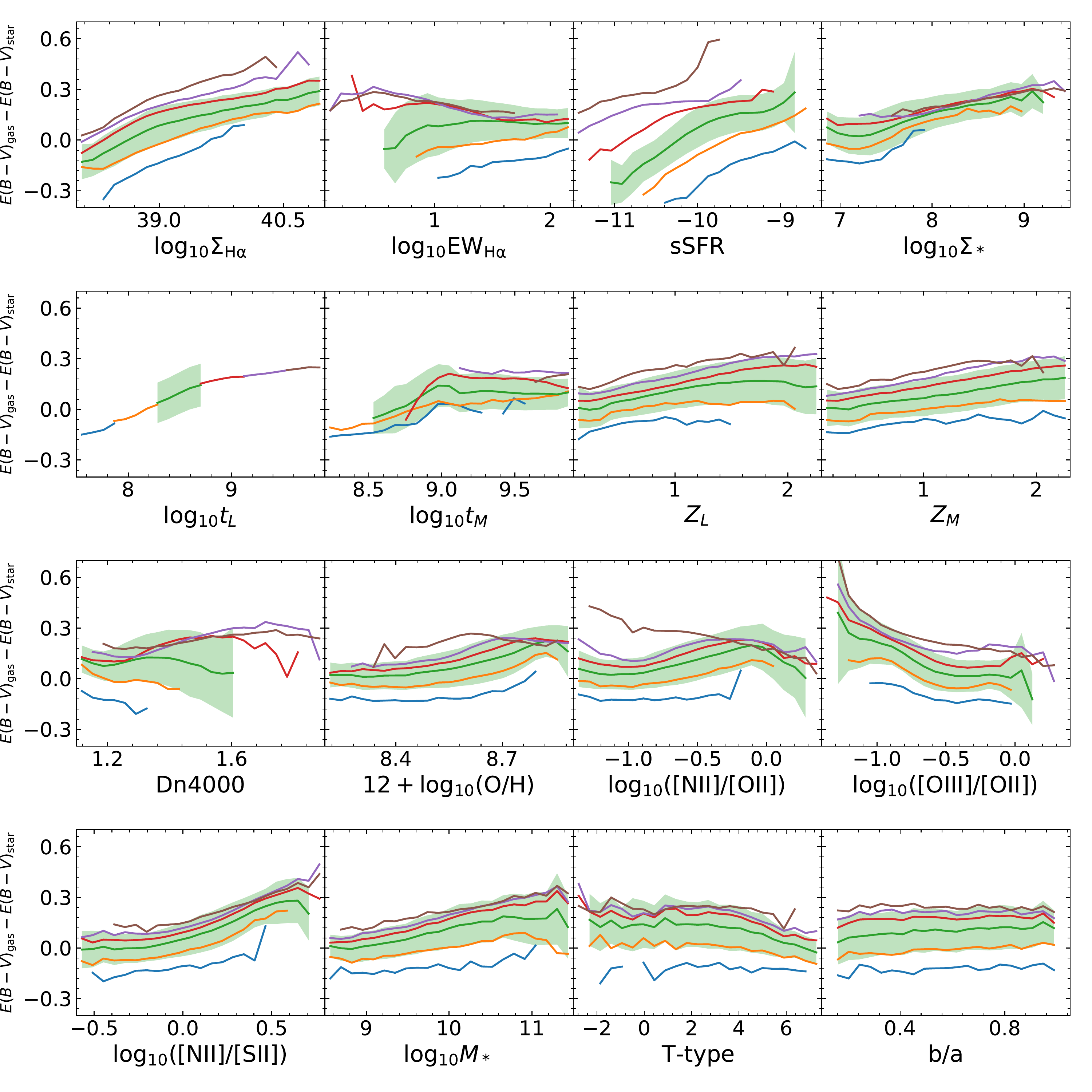}{0.48\textwidth}{(b) \ebvgas$-$\ebvstar\ versus regional/global properties}
    \caption{\ebvgas$-$\ebvstar\ as function of stellar age (\logten\age) 
    for ionized gas regions in different intervals of regional/global properties
    (panels a), and as functions of regional/global properties 
    but in different intervals of \logten\age~(panels b). Symbols/lines 
    are the same as in \autoref{fig:age_ebvstar}.}
    \label{fig:age_ebvdifference}
\end{figure*}

\begin{figure*}
    \centering
    \fig{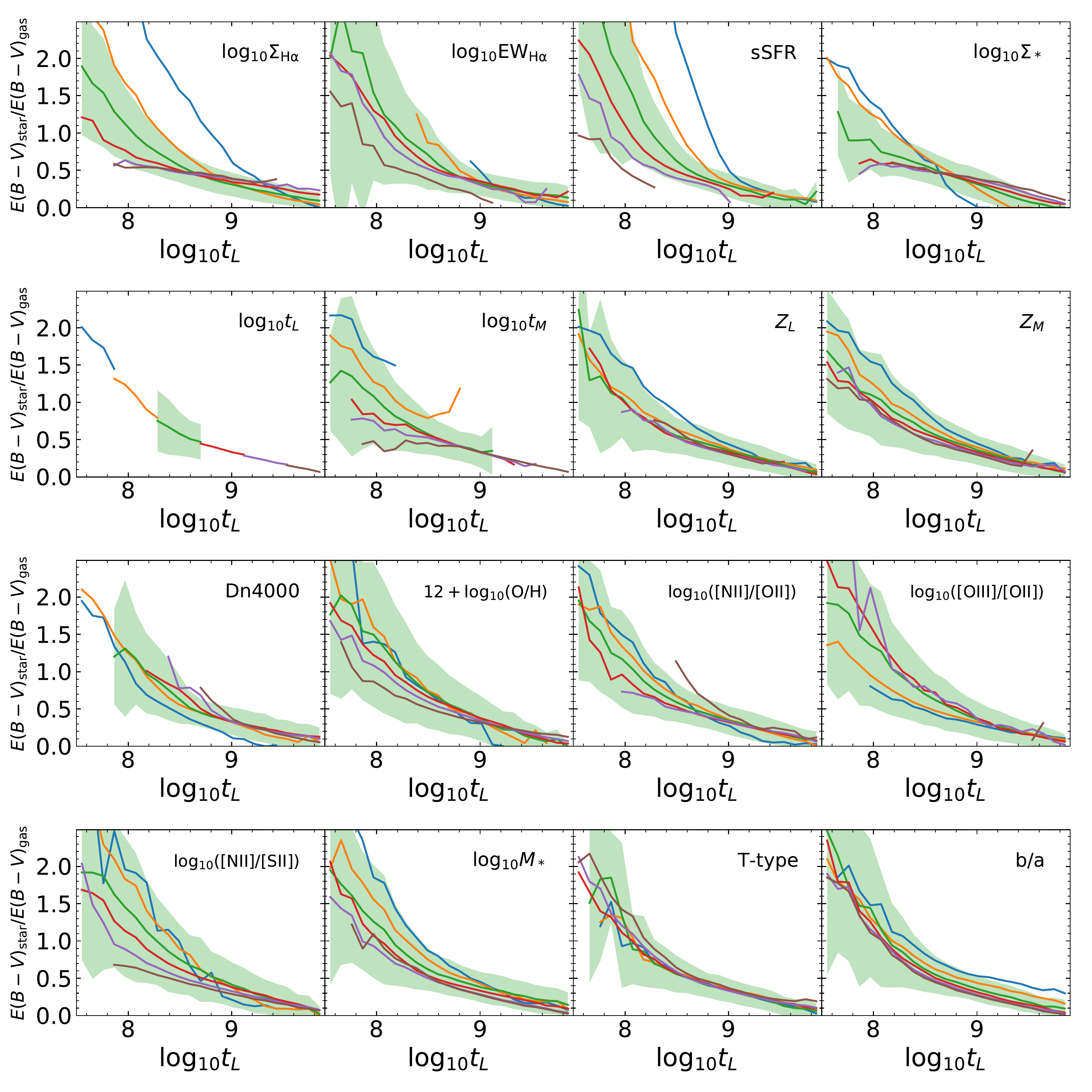}{0.48\textwidth}{(a) \ebvratio\ versus $\log_{10}$\age}
    \fig{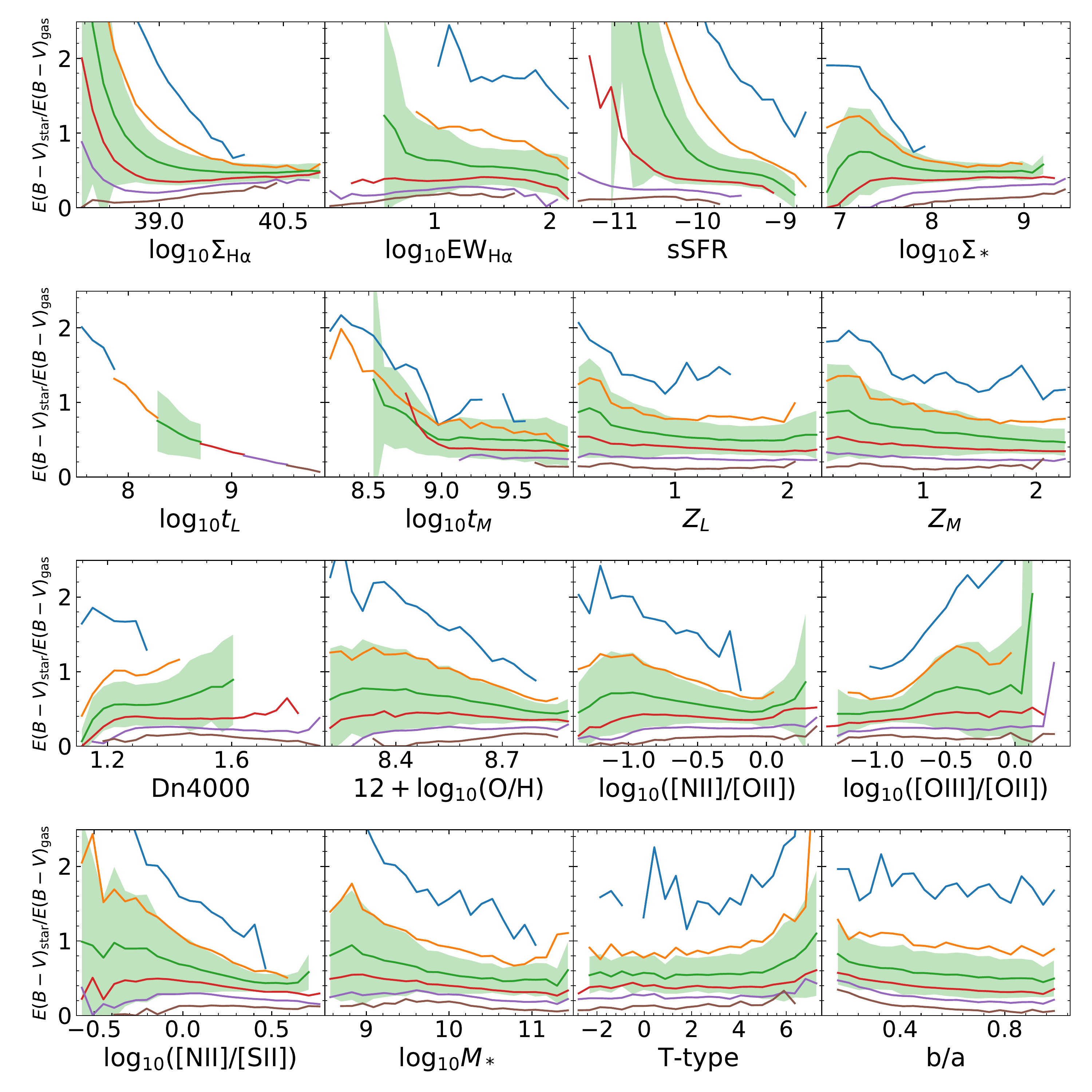}{0.48\textwidth}{(b) \ebvratio\ versus regional/global properties}
    \caption{\ebvstar/\ebvgas\ as functions of stellar age (\logten\age) 
    but in different intervals of regional/global properties (panels a), 
    and as function of regional/global properties but in different intervals 
    of \logten\age~(panels b). Symbols/lines are the same as in 
    \autoref{fig:age_ebvstar}.}
    \label{fig:age_ebvratio}
\end{figure*}

\section{Role of specific properties in driving dust attenuation}
\label{sec:res_t_L_more}

Section \ref{sec:res_t_L} examines the role of stellar age in driving 
stellar attenuation, and Section \ref{sec:role_hasb_niisii} investigates 
the role of \hasb, \niisii\ and \oiiioii\ in driving the attenuation 
in gas. In those sections, for simplicity, results are shown only 
for some of the dust attenuation parameters or regional/global properties.
Here we present the remaining results for completeness. 

\autoref{fig:age_ebvdifference} and \autoref{fig:age_ebvratio} 
show the correlations of the two dust attenuation parameters, 
\ebvdelta\ and \ebvratio\ respectively, with all the regional/global 
properties, aiming to test the dominant role of \logten\age. 
These figures follow the same format as \autoref{fig:age_ebvstar},
and provide complementary results to the analysis in Section \ref{sec:res_t_L}.
Each figure contains two sets of panels. In panel set (a), 
the dust attenuation parameter is plotted as function of 
\logten\age, but for different subsets of the ionized gas regions selected 
by different regional/global properties. In panel set (b), 
the full sample is divided into subsets according to \logten\age, 
and the dust attenuation parameter of each subset is plotted 
as function of the different regional/global properties. 
\autoref{fig:age_ebvstar} shows that stellar age is indeed 
a driving property for the stellar attenuation. The two 
figures here further show that the stellar age also plays 
an important role in driving the difference and ratio 
between the stellar and gas attenuation. This result is produced  
by the combined effect of the strong correlation of \logten\age\  
with \ebvstar\ and the moderate correlation of 
\logten\age\ with \ebvgas, as discussed in Sections \ref{sec:res_t_L}
and \ref{sec:role_hasb_niisii}.

\begin{figure*}
    \centering
    \fig{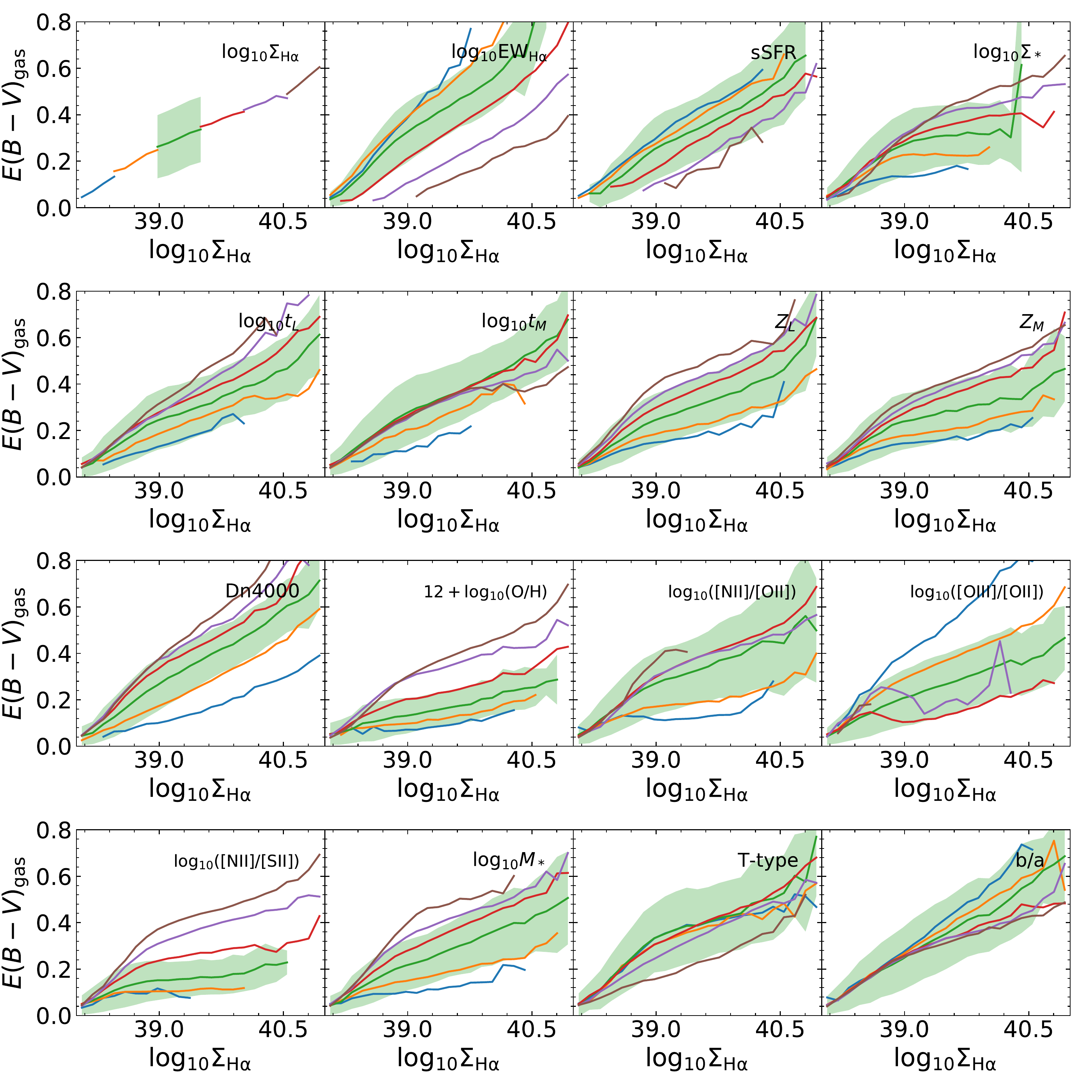}{0.48\textwidth}{(a) \ebvgas\ versus \logten\hasb}
    \fig{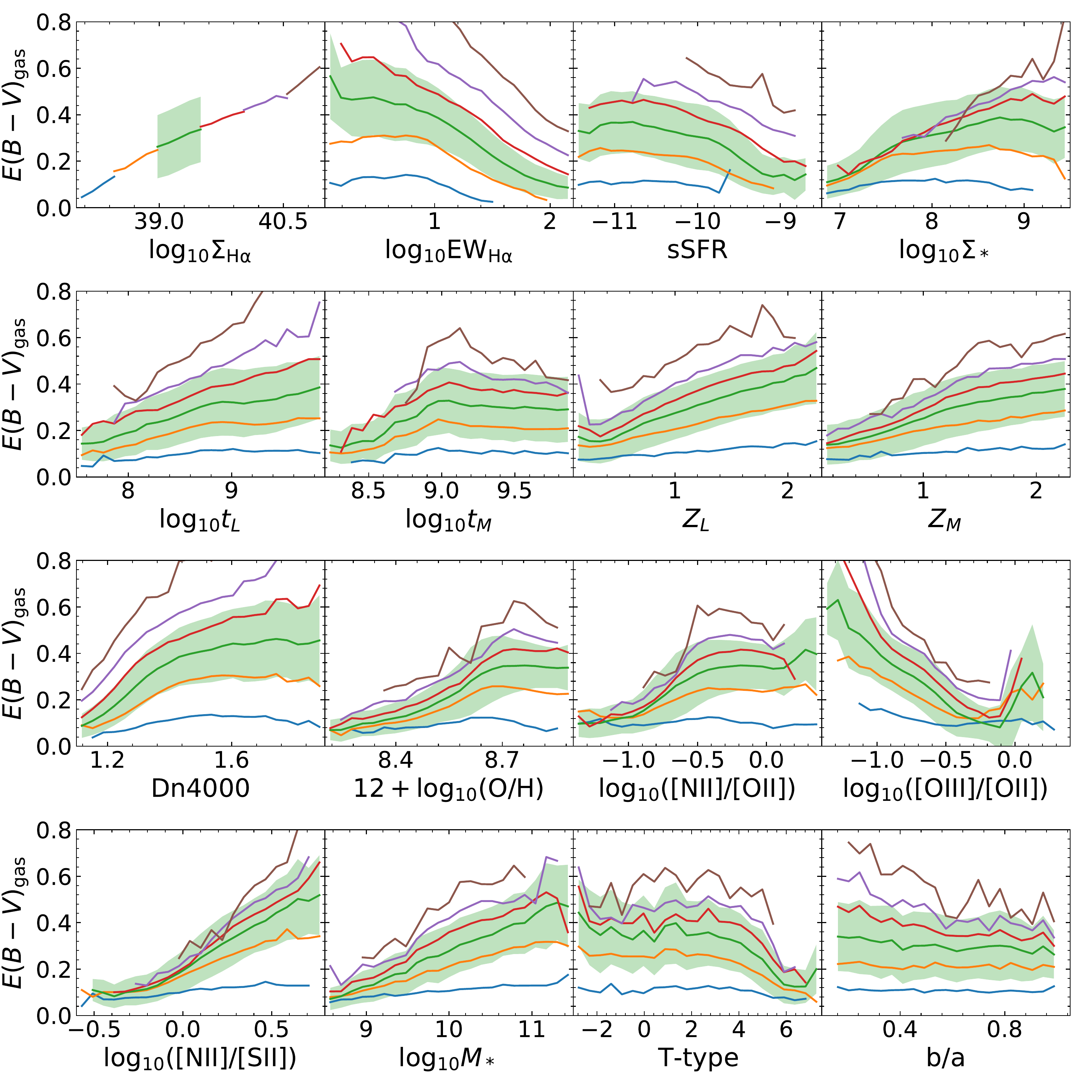}{0.48\textwidth}{(b) \ebvgas\ versus regional/global properties}
    \caption{\ebvgas\ as functions of \hasb\  
    but in different intervals of regional/global properties (panels a), 
    and as function of regional/global properties but in different intervals 
    of \hasb~(panels b). Symbols/lines are the 
    same as in \autoref{fig:age_ebvstar}.}
    \label{fig:hasb_ebvgas}
\end{figure*}

\begin{figure*}
    \centering
    \fig{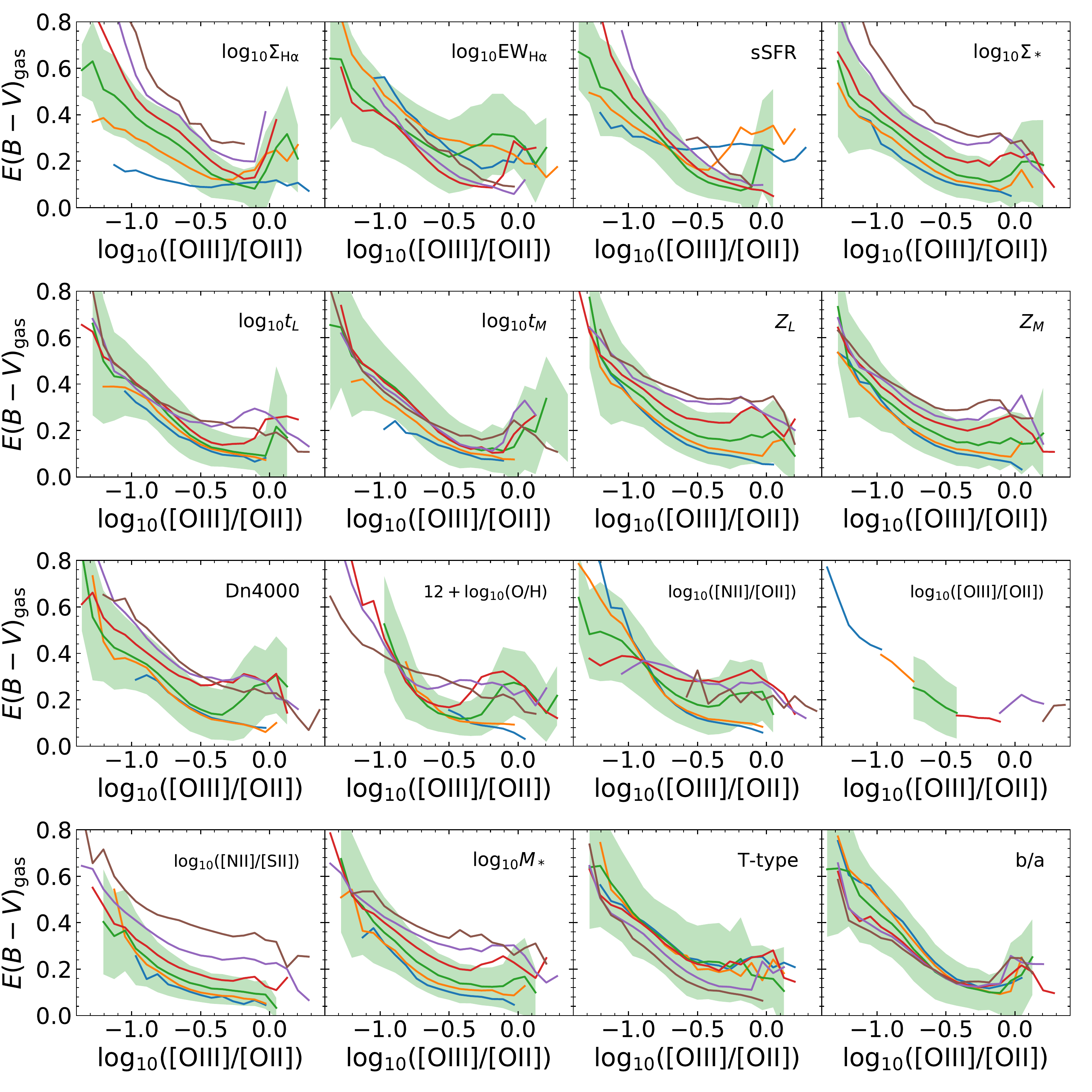}{0.48\textwidth}{(a) \ebvgas\ versus \logten\oiiioii}
    \fig{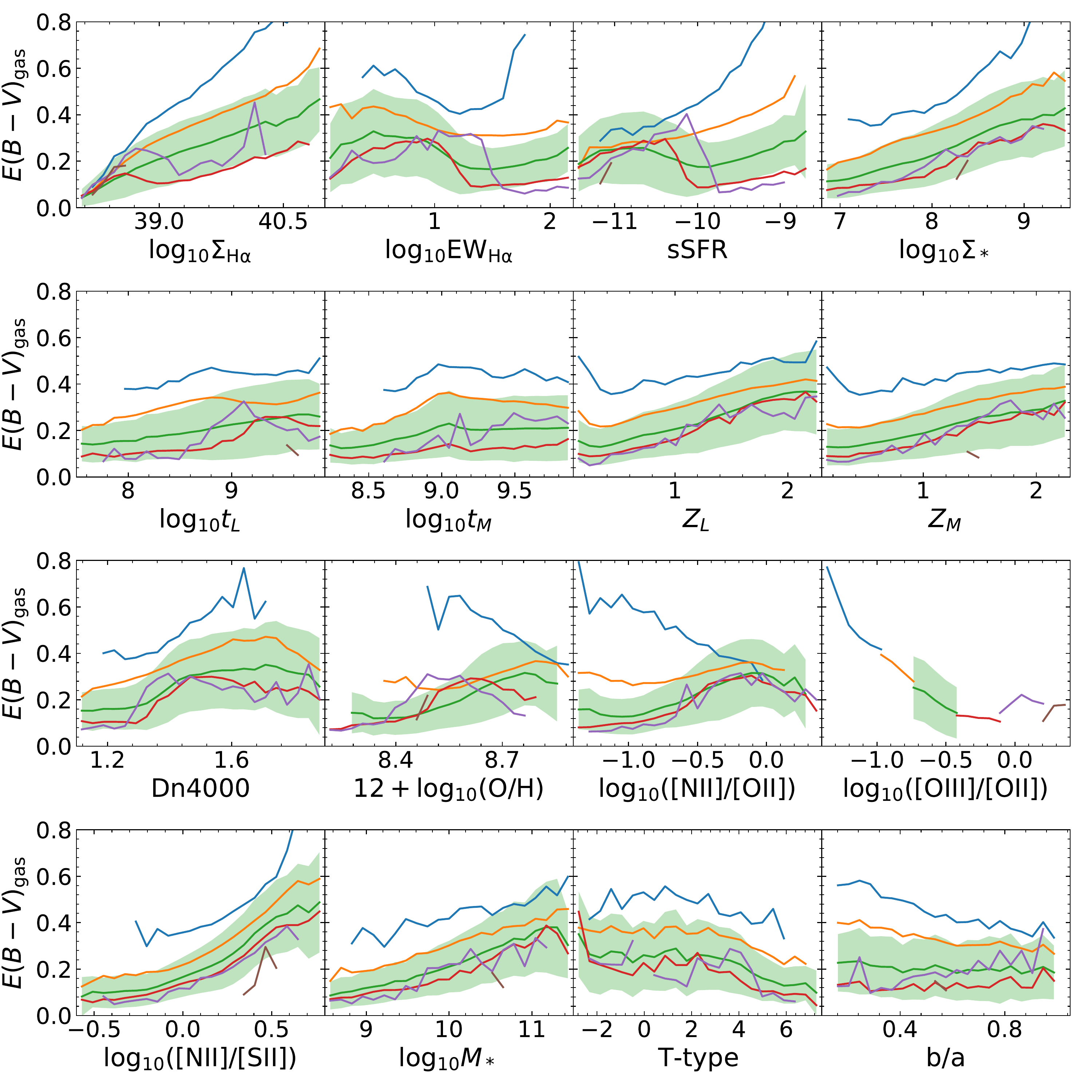}{0.48\textwidth}{(b) \ebvgas\ versus regional/global properties}
    \caption{\ebvgas\ as functions of \oiiioii\  
    but in different intervals of regional/global properties (panels a), 
    and as function of regional/global properties but in different intervals 
    of \oiiioii~(panels b). Symbols/lines are the 
    same as in \autoref{fig:age_ebvstar}.}
    \label{fig:oiiioii_ebvgas}
\end{figure*}

\autoref{fig:hasb_ebvgas} and \autoref{fig:oiiioii_ebvgas} examine the 
correlation of \ebvgas\ with the 16 regional/global properties, aiming 
to test the importance of \hasb\ and \oiiioii\ in driving the attenuation 
in gas. These figures follow the same format as \autoref{fig:niisii_ebvgas}
which tests the role of \niisii, and provide complementary results to 
the analysis in Section \ref{sec:role_hasb_niisii}. Each figure contains two 
sets of panels, with one plotting \ebvgas\ as a function of either 
\hasb\ or \oiiioii\ for subsets of ionized gas regions selected by 
different properties, and the other plotting \ebvgas\ as a function of 
different properties for subsets of ionized gas regions selected by 
either \hasb\ or \oiiioii. Comparing the two figures here with 
\autoref{fig:niisii_ebvgas}, one can find that \niisii\ shows stronger 
correlation with the gas attenuation than \hasb\ and \oiiioii. 

\bibliography{ref}{}
\bibliographystyle{aasjournal}



\end{document}